\documentclass[%
 reprint,
 superscriptaddress,preprintnumbers,
 nofootinbib,
 amsmath,
 amssymb,
 aps,
 pra
]{revtex4-2}

\usepackage[a4paper,margin=0.8in]{geometry}
\usepackage[utf8]{inputenc}
\usepackage{hyperref}
\usepackage{amsmath,amssymb,mathtools}
\usepackage{dsfont}
\usepackage{graphicx}   
\usepackage[table, dvipsnames]{xcolor}
\usepackage{colortbl}
\usepackage{url}
\usepackage[normalem]{ulem}
\usepackage{slashed}
\usepackage[capitalise, english]{cleveref}
\usepackage[noindentafter]{titlesec} 
\allowdisplaybreaks
\usepackage{comment}
\usepackage{bm}
\usepackage{booktabs}
\usepackage{nicematrix}

\usepackage{physics}
\usepackage{ragged2e}
\usepackage{multirow}
\usepackage{array}
\usepackage{makecell}
\usepackage{soul}
\usepackage{lipsum}
\usepackage{tikz}
\usepackage{tikz-feynman}
\tikzfeynmanset{compat=1.1.0}
\usepackage{siunitx}
\usepackage{xspace}
\usepackage{enumitem}
\usepackage{tikz}
\usetikzlibrary{calc,tikzmark,fit,shapes.geometric,matrix,decorations.markings,arrows.meta,decorations.pathmorphing,patterns,positioning,decorations}

    \makeatletter
    \def\CT@@do@color{%
      \global\let\CT@do@color\relax
            \@tempdima\wd\z@
            \advance\@tempdima\@tempdimb
            \advance\@tempdima\@tempdimc
    \advance\@tempdimb\tabcolsep
    \advance\@tempdimc\tabcolsep
    \advance\@tempdima2\tabcolsep
            \kern-\@tempdimb
            \leaders\vrule
                    \hskip\@tempdima\@plus  1fill
            \kern-\@tempdimc
            \hskip-\wd\z@ \@plus -1fill }
    \makeatother

\usepackage{accents}
\DeclareMathSymbol{\widetildesym}{\mathord}{largesymbols}{"65}

\def\thesubsection{\arabic{section}.\arabic{subsection}}

\def\thesection{\arabic{section}}
\titleformat*{\subsubsection}{\normalfont \small \bfseries \boldmath}
\renewcommand{\paragraph}[1]{\vspace{.3em} \indent {\bfseries \boldmath #1 ---}\xspace }
\makeatletter
    \renewcommand{\p@subsection}{}
    \renewcommand{\p@subsubsection}{}
\makeatother

\newcommand{\cC}{\mathcal{C}}
\newcommand{\cO}{\mathcal{O}}
\newcommand{\cL}{\mathcal{L}}
\newcommand{\U}{\mathrm{U}}
\newcommand{\SU}{\mathrm{SU}}

\renewcommand{\Im}{\mathop{\mathrm{Im}}}

\newcommand{\vrel}{v_\rmi{rel}}
\newcommand{\rmi}[1]{{\mbox{\scriptsize #1}}}

\definecolor{red}{rgb}{0.6,.0706,.1373}
\definecolor{blue}{rgb}{0,0.396,0.741}
\definecolor{green}{rgb}{0.25,0.6,0.2}
\colorlet{mylinkcolor}{violet}
\colorlet{mycitecolor}{violet}
\colorlet{myurlcolor}{violet}
\newcommand{\myshade}{85}

\hypersetup{
  linkcolor  = mylinkcolor!\myshade!black,
  citecolor  = mycitecolor!\myshade!black,
  urlcolor   = myurlcolor!\myshade!black,
  colorlinks = true
}

\colorlet{MFVG}{LimeGreen!70!White}
\colorlet{MFVY}{Yellow!70!White}
\newcommand{\MFVg}[1]{%
  {\setlength{\fboxsep}{1pt}\colorbox{MFVG}{$\mathstrut #1$}}%
}
\newcommand{\MFVy}[1]{%
  {\setlength{\fboxsep}{1pt}\colorbox{MFVY}{$\mathstrut #1$}}%
}

\begin{document}

\title{
 \boldmath Charting the Flavour Structure of Dark Matter
}

\author{Simone Biondini}
\email{simone.biondini@physik.uni-freiburg.de}
\affiliation{Institute of Physics, University of Freiburg, Hermann-Herder-Straße 3, 79014 Freiburg, Germany}
\author{Admir Greljo}
\email{admir.greljo@unibas.ch}
\affiliation{Department of Physics, University of Basel, Klingelbergstrasse 82, CH-4056 Basel, 
Switzerland}
\author{Xavier Ponce D\'iaz}
\email{xavier.poncediaz@unibas.ch}
\affiliation{Department of Physics, University of Basel, Klingelbergstrasse 82, CH-4056 Basel, 
Switzerland}
\author{Alessandro Valenti}
\email{alessandro.valenti@unibas.ch}
\affiliation{Department of Physics, University of Basel, Klingelbergstrasse 82, CH-4056 Basel, 
Switzerland}


\begin{abstract}
What flavour structure of $t$-channel thermal dark matter remains compatible with current flavour physics and direct detection bounds? We broadly chart the space of hypotheses using the framework of flavour symmetries and their breaking patterns. We then focus on scenarios in which the fermionic dark matter and its scalar mediator are flavour singlets, falling into the class of \emph{rank-1 flavour violation}. For two representative benchmarks, quarkphilic ($q_L$) and leptophilic ($e_R$), we perform a comprehensive phenomenological analysis, fitting the relic abundance and examining the interplay among flavour observables, direct detection, and collider searches. Our results quantify the allowed deviations from flavour-symmetric limits and assess the discovery prospects in future flavour and direct  detection experiments.
\end{abstract}

\maketitle

{
\color{black}
\tableofcontents
}

\section{Introduction} 
\label{sec:intro}

The freeze-out mechanism remains one of the most compelling paradigms explaining the observed abundance of dark matter (DM) as a thermal relic~\cite{Gondolo:1990dk, Griest:1990kh, Cirelli:2024ssz}. Through its interactions with the Standard Model (SM), DM is initially in thermal equilibrium with the primordial plasma; as the temperature falls below its mass, the equilibrium density becomes exponentially suppressed, and annihilations can no longer keep pace with the Hubble expansion, leading to freeze-out. Remarkably, particles with masses around the TeV scale and $\mathcal{O}(1)$ couplings with the SM naturally reproduce the observed DM relic density, a coincidence often referred to as \emph{the WIMP miracle}. This observation hints at a deep connection between cosmology and the energy frontier of particle physics, possibly suggesting a common origin of the electroweak scale and DM. 

Despite decades of null results from direct and indirect detection, as well as collider searches~\cite{Arcadi:2017kky, Schumann:2019eaa, Arcadi:2024ukq}, these efforts continue to motivate a thorough and systematic exploration of the thermal dark matter parameter space, including well-motivated departures from the minimal WIMP paradigm~\cite{Cirelli:2005uq, Chang:2013oia}. 
Such simplified models extend the SM by at least a dark matter candidate and a mediator. They can be broadly classified into two categories~\cite{DeSimone:2016fbz}: \emph{$t$-channel models}, featuring renormalizable DM--mediator--SM interactions~\cite{Arina:2025zpi, Garny:2015wea, Garny:2011ii, Bai:2013iqa, DiFranzo:2013vra, An:2013xka, Garny:2014waa, Kopp:2014tsa, Biondini:2025gpg}, and \emph{$s$-channel models}, in which the mediator couples linearly to SM fields~\cite{Buckley:2014fba, Abdallah:2015ter, Boveia:2016mrp, Goncalves:2016iyg, Albert:2016osu, Bell:2016ekl, Englert:2016joy, Kahlhoefer:2015bea}. In this work, we consider $t$-channel models, which are particularly interesting from a flavour physics perspective.

Indeed, among complementary probes, thermal DM scenarios are directly tested by flavour physics~\cite{Belfatto:2025ids,  Kile:2011mn, Kamenik:2011nb, Agrawal:2011ze, Agrawal:2015kje, Blanke:2017fum, Arcadi:2021glq, Acaroglu:2021qae, Arcadi:2021cwg, Demetriou:2025ewa}. TeV-scale new physics (NP) with $\mathcal{O}(1)$ couplings to SM particles must exhibit a highly non-trivial flavour structure to evade stringent bounds from flavour-changing neutral currents and CP-violating observables. This general fact is commonly referred to as the \emph{new physics flavour puzzle}~\cite{Altmannshofer:2024hmr, Isidori:2025iyu, Nir:2020jtr, Altmannshofer:2025rxc, deBlas:2025gyz}. The same concerns, therefore, apply to DM interactions with the SM, motivating the central question of this work: what is the flavour structure of $t$-channel DM that remains compatible with existing bounds?

Flavour symmetries and their breaking patterns provide a robust framework for addressing this question in a structured and systematic manner. Building on recent applications of this approach to the SMEFT~\cite{Faroughy:2020ina, Greljo:2022cah, Greljo:2025mwj, Mescia:2024rki, Kile:2014jea}, we extend it to $t$-channel DM interactions. In \cref{sec:charting}, we chart the corresponding space of hypotheses by classifying the fields and couplings, formally treated as spurions, under a set of assumed flavour symmetries. Our work generalises the well-known case of Minimal Flavour Violation~\cite{DAmbrosio:2002vsn, Batell:2011tc, Lopez-Honorez:2013wla, Agrawal:2014una} and its variants~\cite{Agrawal:2014aoa, Blanke:2017tnb, Chen:2015jkt, Acaroglu:2022hrm}. Guided by general lessons from the SMEFT, we identify particularly well-motivated candidate scenarios for dedicated studies. For the other class of simplified DM models, namely $s$-channel scenarios, only the mediator couples directly to the SM fields. As a result, the flavour structure reduces to identifying the flavour representations of linear extensions of the SM~\cite{deBlas:2017xtg} and their allowed couplings~\cite{Greljo:2023adz, Greljo:2023bdy}, with the DM field itself playing no active role in flavour observables.

In the central part of this work, we test the flavour symmetry expectations on specific cases. For concreteness, we focus on the simplest benchmark scenarios in which both the DM particle and the mediator are flavour singlets. In this setup, the DM--mediator--SM coupling is a priori an arbitrary vector in the flavour space of the corresponding SM fermion, giving rise to \emph{rank-1} flavour violation, for which we adopt the general parametrization introduced in recent $B$-physics studies~\cite{Gherardi:2019zil, Marzocca:2024hua}. We perform a thorough phenomenological study of a leptophilic benchmark in \cref{sec:leptophilic} and a quarkphilic benchmark in \cref{sec:quarkphilic}. In each case, we compute the complete set of relevant flavour observables and confront them with the relic density requirement and complementary constraints from collider searches and direct detection (DD). This allows us to quantitatively assess whether, and to what extent, viable scenarios must lie close to flavour-symmetric limits, which is often assumed rather than tested in the existing literature. We conclude in \cref{sec:conclusion} with a discussion of how these lessons generalize beyond the simplest scenarios.

\section{Charting Dark Matter with Flavour Symmetries}
\label{sec:charting}

The generic $t$-channel interaction between a SM fermion $f_{\text{SM}}$ and the dark sector fields $\chi$ and $\Phi$ can be written as
\begin{equation}\label{eq:def-t}
\mathcal{L} \supset y\, \bar f_{\text{SM}}\, \chi\, \Phi + \text{h.c.} \; .
\end{equation}
Here we have suppressed flavour indices in both the fields and the coupling $y$. The fields $\chi$ and $\Phi$ are assumed to be odd under a stabilizing $\mathcal{Z}_2$ symmetry.\footnote{DM stability can be a consequence of unbroken discrete subgroups of flavour symmetries, see e.g.~\cite{Batell:2011tc,Bishara:2015mha}.} For the purposes of this section, it is not relevant whether the actual DM particle is either $\chi$ or $\Phi$, which we will collectively denote by $X\equiv (\chi \Phi)$. Our focus is instead on the following question: what flavour structure of the coupling $y$ allows the dark sector to reproduce the observed DM relic abundance, while remaining compatible with constraints from flavour physics?

This question is particularly relevant for the $t$-channel models considered here. These scenarios favor dark sector masses in the range $0.1$–$10\,\text{TeV}$ with $\mathcal{O}(1)$ couplings to achieve the observed relic abundance (see \cite{Arina:2025zpi} for a recent review). On the other hand, precision measurements of flavour transitions among SM fermions are well known to constrain generic NP with ${\cal O}(1)$ flavour-violating couplings up to scales of ${\cal O}$(PeV) or higher \cite{deBlas:2025gyz}. Flavour-changing neutral currents (FCNC) are highly suppressed in the SM due to the non-generic structure of the Yukawa interactions, which gives rise to approximate flavour symmetries. In the quark sector, flavour violation is entirely controlled by the CKM matrix through weak interactions, leading, for example, to strong suppressions of FCNC via the GIM mechanism \cite{Glashow:1970gm}. In the lepton sector, charged lepton flavour violation is suppressed by the smallness of neutrino masses through an analogous mechanism. NP can violate these approximate flavour symmetries and generate large flavour-violating effects, making FCNC exceptionally sensitive probes of physics beyond the SM.

The challenge of accommodating TeV-scale NP while satisfying flavour constraints is a general one. It is particularly well known in the context of solutions to the SM hierarchy problem, where new states near the TeV scale are required to stabilize the Higgs mass, but would generically introduce large sources of flavour violation, for a recent example, see~\cite{Glioti:2024hye}. To address this tension, the framework of flavour symmetries was introduced to organize new physics (NP) couplings in a controlled manner, meaning that the resulting flavour structure is stable under renormalization group evolution and does not rely on fine-tuned cancellations. This allows NP at much lower scales than in flavour-anarchic scenarios. Our goal is to adapt these tools to the present context and develop a systematic classification of $t$-channel DM models that reconciles the relic abundance requirement with stringent constraints from FCNC and CP violation. The presentation is intentionally pedagogical to remain accessible to readers not accustomed to the use of flavour symmetries in BSM model building.

\subsection{Minimal Flavour Violation}
\label{sec:MFV}
When dealing with DM simplified models, the most widely studied framework in this context is Minimal Flavour Violation (MFV) \cite{Chivukula:1987py, DAmbrosio:2002vsn, Batell:2011tc, Lopez-Honorez:2013wla, Agrawal:2014una}. In MFV, the SM fields are assigned to the fundamental representations of the approximate flavour symmetry $\U(3)^5$ respected by the gauge interactions, $f_{\text{SM}} \sim \bm{3}_f$. This symmetry is broken by the dimension four Yukawa interactions, leaving only the subgroup $\U(1)_B \times \U(1)_e \times \U(1)_\mu \times \U(1)_\tau$ unbroken. Importantly, this breaking pattern leads to the special suppression properties of flavour violation discussed above. New interactions such as the one in \cref{eq:def-t} would generically spoil this structure and thus lead to very strong flavour constraints.

The impact of NP on flavour can be elegantly understood by treating the couplings that break $\U(3)^5$, namely the SM Yukawas and the new coupling $y$, as \emph{spurions}. These are non-dynamical fields that transform under $\U(3)^5$ in such a way that the full flavour symmetry is formally restored, and are then assigned fixed background values. In this language, NP couplings can avoid inducing large flavour violation in two ways. They may transform as flavour singlets, in which case they introduce no new sources of flavour breaking and can naturally be of order one. Alternatively, they may be constructed as product expansions in terms of the SM spurions, which in MFV are the Yukawa matrices themselves. In this case, flavour violation induced by NP is aligned with that of the SM and appears only as a small correction, allowing TeV-scale NP to remain compatible with current data.

Such an expansion is not always possible. The representation of the spurion $y$ under $\U(3)^5$ is fixed by the representations of $f_{\text{SM}}$ and the dark sector operator $X$, since the product of all three must contain a singlet in order for the Lagrangian to be flavour invariant. An expansion of $y$ in terms of the SM spurions,
\begin{align}
Y_{u} \sim (\bm{3}_q,\overline{\bm{3}}_{u}), \quad Y_{d} \sim (\bm{3}_q,\overline{\bm{3}}_{d}),\quad  Y_e \sim (\bm{3}_\ell,\overline{\bm{3}}_e),
\label{eq:U3spurions}
\end{align}
is possible only if the representation of $y$ appears in the decomposition of products of these spurions. When this condition is not satisfied, the breaking of $\U(3)^5$ induced by $y$ is not necessarily aligned with the SM one, and large new sources of flavour violation may be expected. 

To make this discussion concrete, we present in \cref{tab:U3} several examples of the representation of $y$ under $\U(3)^3$, the flavour group of the quark sector, as fixed by the interaction in \cref{eq:def-t} once the representations of $f_{\text{SM}}$ and $X$ are specified. In the table, entries shown in green correspond to representations that are either singlets under $\U(3)^5$ or can be expanded purely in terms of $Y_u$. In these cases, the coupling $y$ can naturally be of order one, supported by the large top Yukawa coupling, and TeV-scale dark sector masses can reproduce the observed relic abundance. Entries shown in yellow correspond to representations that can be expanded in terms of SM Yukawas but require at least one insertion of $Y_d$ or $Y_e$. This implies an additional suppression by at least one power of $y_{b,\tau} \sim 10^{-2}$, which is too small to account for the relic abundance in most of the parameter space. Finally, uncolored entries correspond to representations that cannot be constructed from SM spurions.

\begin{table}[!htb]
\centering
\renewcommand{\arraystretch}{1.3}
\begin{tabular}{c|c|c|c|c|c}
$X \backslash \bar f_{\text{SM}}$ & $\overline{\bm{3}}_q$ & $\overline{\bm{3}}_u$ & $\overline{\bm{3}}_d$ & $\overline{\bm{3}}_\ell$ & $\overline{\bm{3}}_e$ \\
\hline
$\bm{1}$ & $\bm{3}_q$& $\bm{3}_u$ & $\bm{3}_d$ &  $\bm{3}_\ell$ & $\bm{3}_e$ \\
$\bm{3}_q$ & \MFVg{\bm{1}} $\oplus$ \MFVg{\bm{8}_q}&
\MFVg{ (\bm{3}_u, \overline{\bm{3}}_q)} & \MFVy{\mathstrut (\bm{3}_d, \overline{\bm{3}}_q)}& $(\bm{3}_\ell, \overline{\bm{3}}_q)$ & $(\bm{3}_e, \overline{\bm{3}}_q)$ \\
$\overline{\bm{3}}_q$ & $\overline{\bm{3}}_q \oplus \bm{6}_q$      & $(\bm{3}_u, \bm{3}_q)$            & $(\bm{3}_d, \bm{3}_q)$ & $(\bm{3}_\ell, \bm{3}_q)$& $(\bm{3}_e, \bm{3}_q)$ \\
$\bm{3}_u$  & \MFVg{ (\bm{3}_q, \overline{\bm{3}}_u)} &\MFVg{\bm{1}} $\oplus$  \MFVg{ \bm{8}_u } & \MFVy{(\bm{3}_d, \overline{\bm{3}}_u)}&$(\bm{3}_\ell, \overline{\bm{3}}_u)$ &$(\bm{3}_e, \overline{\bm{3}}_u)$  \\
$\overline{\bm{3}}_u$ & $(\bm{3}_q, \bm{3}_u)$                 & $\overline{\bm{3}}_u \oplus \bm{6}_u$ & $(\bm{3}_d, \bm{3}_u)$ & $(\bm{3}_\ell, \bm{3}_u)$& $(\bm{3}_e, \bm{3}_u)$\\
$\bm{3}_d$       & \MFVy{(\bm{3}_q, \overline{\bm{3}}_d)} &
    \MFVy{(\bm{3}_u, \overline{\bm{3}}_d) } & \MFVg{\bm{1}} $\oplus$ \MFVy{ \bm{8}_d } & $(\bm{3}_\ell, \overline{\bm{3}}_d)$& $(\bm{3}_e, \overline{\bm{3}}_d)$\\
$\overline{\bm{3}}_d$ & $(\bm{3}_q, \bm{3}_d)$                 & $(\bm{3}_u, \bm{3}_d)$            & $\overline{\bm{3}}_d \oplus \bm{6}_d$ & $(\bm{3}_\ell, \bm{3}_d)$& $(\bm{3}_e, \bm{3}_d)$\\
$\bm{3}_\ell$       & $(\bm{3}_q, \overline{\bm{3}}_\ell)$ &
    $(\bm{3}_u, \overline{\bm{3}}_\ell) $ & $(\bm{3}_d, \overline{\bm{3}}_\ell)$ & \MFVg{\bm{1}} $\oplus$ \MFVy{\bm{8}_\ell} & \MFVy{ (\bm{3}_e, \overline{\bm{3}}_\ell)}  \\
$\overline{\bm{3}}_\ell$ & $(\bm{3}_q, \bm{3}_\ell)$ & $(\bm{3}_u, \bm{3}_\ell)$  & $(\bm{3}_d, \bm{3}_\ell)$ &  $\overline{\bm{3}}_\ell \oplus {\bm{6}}_\ell$ &$(\bm{3}_e, \bm{3}_\ell)$  \\
$\bm{3}_e$       & $(\bm{3}_q, \overline{\bm{3}}_e)$ &
    $(\bm{3}_u, \overline{\bm{3}}_e) $ & $(\bm{3}_d, \overline{\bm{3}}_e)$ &  \MFVy{(\bm{3}_\ell, \overline{\bm{3}}_e)} & \MFVg{\bm{1}} $\oplus$ \MFVy{\bm{8}_e} \\
$\overline{\bm{3}}_e$ & $(\bm{3}_q, \bm{3}_e)$ & $(\bm{3}_u, \bm{3}_e)$  & $(\bm{3}_d, \bm{3}_e)$ & $(\bm{3}_\ell, \bm{3}_e)$ & $\overline{\bm{3}}_e \oplus \bm{6}_e$ \\
\vdots & \vdots & \vdots & \vdots & \vdots & \vdots\\
\end{tabular}
\caption{Example flavour representations of the coupling $y$ in \cref{eq:def-t} under $\mathrm{U}(3)^5$ (MFV). Uncolored entries denote representations incompatible with an MFV expansion; yellow entries are MFV compatible but correspond to couplings likely too small to reproduce the observed DM relic abundance unless in specific corners of parameter space; green entries are MFV compatible and phenomenologically viable.}
\label{tab:U3}
\end{table}

\subsection{$\mathrm{U}(2)^5$ flavour symmetry}
\label{sec:U2}
By enforcing flavour universality, the MFV expansion is often too restrictive, excluding phenomenologically viable and theoretically motivated possibilities. This is particularly evident in flavour-conserving processes involving light quarks, where bounds on NP are far stronger than in the heavy-flavour sector. Relevant to our study, DD bounds are highly stringent for DM couplings to first-generation quarks~\cite{Cirelli:2024ssz}.\footnote{Another typical motivation for $\mathrm{U}(2)^5$ over MFV is Drell–Yan production at the LHC, which constrains semileptonic four-fermion interactions with first-generation couplings up to effective scales of $\mathcal{O}(10)\,\text{TeV}$~\cite{Greljo:2017vvb, Allwicher:2022gkm, Greljo:2022jac}.} On the other hand, constraints involving purely third-generation particles are significantly weaker. This motivates taking a smaller flavour symmetry group as the starting point, which, unlike MFV, permits departures from flavour universality and independent third-generation couplings.

A popular and well-motivated choice is a $\U(2)^5$ flavour symmetry, which distinguishes third-generation SM fermions from the first two generations~\cite{Barbieri:2011ci, Barbieri:2012uh, Faroughy:2020ina, Greljo:2022cah, Allwicher:2023shc}. This can be viewed as a subgroup of the full MFV symmetry, under which the SM fermions decompose as
\begin{align}
    \bm{3}_f \to \bm{2}_f \oplus \bm{1}.
\label{eq:U3toU2decomp}
\end{align}
This structure is justified by the fact that the third-generation Yukawa couplings are much larger than those of the first two generations, which can be neglected at leading order. In this limit, the full set of SM Yukawas respects an approximate $\U(2)^5$ symmetry.

Following the same principles as in MFV, the breaking of $\U(2)^5$ induced by the light-quark masses and by CKM mixings can be parametrized in terms of a set of four spurions~\cite{Fuentes-Martin:2019mun, Greljo:2022cah}
\begin{align}
\begin{gathered}
V_q \sim \bm{2}_q,\quad  \Delta_u \sim (\bm{2}_q,\overline{\bm{2}}_u), \quad \Delta_d \sim (\bm{2}_q,\overline{\bm{2}}_d), \\
\Delta_e \sim (\bm{2}_\ell,\overline{\bm{2}}_e).
\end{gathered}
\label{eq:U2spurions}
\end{align}
The decompositions of MFV spurions from \cref{eq:U3spurions} reads 
\begin{align}
\begin{gathered}
Y_u =
\begin{pmatrix}
\Delta_u & \vline & V_q \\
\hline
0 & \vline & y_t
\end{pmatrix}, 
~
Y_d =
\begin{pmatrix}
\Delta_d & \vline & V_q \\
\hline
0 & \vline & y_b
\end{pmatrix},\\
Y_e =
\begin{pmatrix}
\Delta_e & \vline & 0 \\
\hline
0 & \vline & y_\tau
\end{pmatrix}.
\end{gathered}
\label{eq:decomp1}
\end{align}
The spurions absent from the decomposition, $\bm{2}_\ell$, $\bm{2}_u$, $\bm{2}_d$, and $\bm{2}_e$, are phenomenologically constrained to be small.

The representation of the coupling $y$ is then fixed by the $\U(2)^5$ transformation properties of $f_{\text{SM}}$ and the dark sector field $X$, and can be expanded in terms of the spurions in \cref{eq:U2spurions}. We present several examples in \cref{tab:U2}, using the same color coding as in the MFV table.\footnote{While the tensor representations shown technically correspond to the non-Abelian part of the flavour group, the invariance under the $\U(1)$ factors is also implicitly assumed \cite{Greljo:2022cah}.}

A crucial observation is that the representation of $y$ under MFV in \cref{tab:U3} can be decomposed under $\U(2)^5$ in complete analogy with \cref{eq:U3toU2decomp} and \cref{eq:decomp1}. Since $\U(2)^5 \subset \U(3)^5$, this implies that some representations that are not compatible with MFV may nevertheless admit a consistent $\U(2)^5$ expansion. As a result, such models can become phenomenologically viable.

As an illustrative example, consider the case $X\sim \bm{1}$ with $y \sim \bm{3}_q$, which we will study in detail in \cref{sec:quarkphilic}, and which induces an interaction of the form $y_i \bar q_i X$ with $i=1,2,3$. Under $\U(2)^5$, this decomposes as $y\sim\bm{3}_q \to \bm{2}_q \oplus \bm{1}$. The $\bm{2}_q$ representation can be constructed from the spurions in \cref{eq:U2spurions}, thereby inducing flavour violation, albeit with sufficient suppression. The $\mathcal O(1)$ coupling associated with the singlet component involves third-generation fermions and can reproduce the observed relic abundance. Under MFV, no expansion of $y$ in terms of the spurions in \cref{eq:U3spurions} was instead possible, naively ruling out this scenario.

This example also highlights an important point concerning the consistency of the spurion expansion and the necessity of including next-to-leading-order terms. If the gauge representation of $X$ allows couplings to SM fields transforming in different representations of $\U(2)^5$, all such couplings must be included, with their sizes fixed by the appropriate powers of the SM spurions when applicable. In the example above, keeping only the coupling to $q_3$ with $y_3 \sim \bm{1} \sim \mathcal{O}(1)$ would be inconsistent. After electroweak symmetry breaking, the rotation to the mass basis induces small couplings to $q_{1,2}$ proportional to CKM matrix elements. At the same order in the $\U(2)^5$ expansion, this effect is consistently captured by including couplings of the form $(y_1 \bar q_1  + y_2 \bar q_2) X$, with $(y_1,y_2) \sim \bm{2}_q \sim V_q$.

\begin{table}[t]
\centering
\renewcommand{\arraystretch}{1.4}
\begin{NiceTabular}{@{}r@{}c|cc|cc|cc|c}
    &$X \backslash \bar f_{\text{SM}}$ & $\overline{\bm{2}}_q$ & $\bm{1}$ & $\overline{\bm{2}}_u$ & $\bm{1}$ & $\overline{\bm{2}}_d$ & $\bm{1}$ & \dots \\
    \midrule
    & $\bm{1}$ & \MFVy{\bm{2}_q} &  \MFVg{\bm{1}} & \MFVy{\bm{2}_u} &  \MFVg{\bm{1}} & \MFVy{\bm{2}_d} &  \MFVg{\bm{1}} & \dots\\
    \Block{2-1}{$\bm{3}_q \Bigg \{$} &
     $\bm{2}_q$& \MFVg{\bm{1}} $\oplus$ \MFVy{\bm{3}_q}& \MFVy{\overline{\bm{2}}_q} & \MFVy{(\bm{2}_u,\overline{\bm{2}}_q)} &  \MFVy{\overline{\bm{2}}_q} & \MFVy{(\bm{2}_d,\overline{\bm{2}}_q)} &  \MFVy{\overline{\bm{2}}_q} & \dots \\
      & $\bm{1}$ & \MFVy{\bm{2}_q} &  \MFVg{\bm{1}} & \MFVy{\bm{2}_u} &  \MFVg{\bm{1}} & \MFVy{\bm{2}_d} &  \MFVg{\bm{1}} & \dots\\
    \Block{2-1}{$\bm{3}_u \Bigg \{$} &
     $\bm{2}_u$& \MFVy{(\bm{2}_q,\overline{\bm{2}}_u)}& \MFVy{\overline{\bm{2}}_u} &  \MFVg{\bm{1}} $\oplus$ \MFVy{\bm{3}_u}& \MFVy{\overline{\bm{2}}_u} & \MFVy{(\bm{2}_d,\overline{\bm{2}}_u)} &  \MFVy{\overline{\bm{2}}_u} & \dots\\
      & $\bm{1}$ & \MFVy{\bm{2}_q} & \MFVg{\bm{1}} & \MFVy{\bm{2}_u} &  \MFVg{\bm{1}}& \MFVy{\bm{2}_d} &  \MFVg{\bm{1}} & \dots \\
      \Block{2-1}{$\bm{3}_d \Bigg \{$} &
     $\bm{2}_d$& \MFVy{(\bm{2}_q,\overline{\bm{2}}_d)}& \MFVy{\overline{\bm{2}}_d} &  \MFVy{(\bm{2}_u,\overline{\bm{2}}_d)} &  \MFVy{\overline{\bm{2}}_d} & \MFVg{\bm{1}} $\oplus$ \MFVy{\bm{3}_d}& \MFVy{\overline{\bm{2}}_d} & \dots\\
      & $\bm{1}$ & \MFVy{\bm{2}_q} & \MFVg{\bm{1}} & \MFVy{\bm{2}_u} &  \MFVg{\bm{1}}& \MFVy{\bm{2}_d} &  \MFVg{\bm{1}} & \dots\\
      &\vdots & \vdots & \vdots & \vdots & \vdots & \vdots & \vdots & $\ddots$ \\
      \CodeAfter
      \OverBrace[yshift=2pt]{1-3}{1-4}{$\overline{\bm{3}}_q$}
      \OverBrace[yshift=2pt]{1-5}{1-6}{$\overline{\bm{3}}_u$}   
      \OverBrace[yshift=2pt]{1-7}{1-8}{$\overline{\bm{3}}_d$}   
\end{NiceTabular}
\caption{Example flavour representations of the coupling $y$ in \cref{eq:def-t} under $\mathrm{U}(2)^5$. Subscripts refer to representation under the non-abelian part of $U(2)^5$. The decomposition of the associated MFV representation of $\bar{f}_\text{SM}$ and $X$ under $\U(2)^5$ is also included to highlight the difference in the spurion counting.
Color code as in Table~\ref{tab:U3}: $\mathrm{U}(2)$-compatible but likely too small for the relic abundance (yellow), $\mathrm{U}(2)$-invariant and viable (green). Leptons are omitted to avoid clutter, but their inclusion is straightforward.}
\label{tab:U2}
\end{table}

\subsection{Minimal Flavour Protection}
\label{sec:MFP}
One may wonder whether a flavour group even smaller than $\U(2)^5$, yet still providing sufficient protection against dangerous flavour violation, can be constructed. The answer is affirmative and goes under the name of Minimal Flavour Protection (MFP), in which the smallest viable flavour symmetry has been identified as~\cite{Greljo:2025mwj}
\begin{equation}
    \text{SU}(2)_{\boldsymbol{q}} \times \U(1)_{x_i}\,.
\end{equation}
In this framework, only the first two generations of the left-handed quarks $\boldsymbol{q} = (q_1, q_2)^T$ transform as a doublet of $\text{SU}(2)_{\boldsymbol{q}}$, while all remaining SM fermions are singlets and carry charges under the additional abelian symmetry $\U(1)_{x_i}$. 

As shown in~\cite{Greljo:2025mwj}, $\mathrm{SU}(2)_q$ cannot be further reduced, since the large CKM 1-2 mixing induces irreducible flavour violation that unavoidably overshoots either $K$–$\bar K$ or $D$–$\bar D$ mixing bounds, a fact confirmed in our quarkphilic benchmark in \cref{sec:quarkphilic}. In addition, a single non-universal flavour group $\U(1)_{x_i}$ suffices to control the other sectors. It can be identified with a linear combination of the remaining Cartan generators of $\U(3)^5$, \emph{i.e.}~$\U(1)_{x_i} \subset \U(1)^{14}$.\footnote{At the Lie-algebra level, $\mathfrak u(3)_f = \mathfrak{su}(3)_f \oplus \mathfrak u(1)_f$. The Cartan subalgebra is three-dimensional and may be chosen as 
$\{T_3,\,T_8,\,X_f\}$, where $T_3$ and $T_8$ are the diagonal generators of 
$\mathfrak{su}(3)_f$ and $X_f \propto \mathbf 1$ generates $\mathfrak u(1)_f$.} 

Therefore, under the MFP subgroup, the $U(3)^5$ SM fermion representations decompose as
\begin{align}
\begin{aligned}
    \bm{3}_{x_q} &\to \bm{2}_{x_{\boldsymbol{q}}} \oplus \bm{1}_{x_{q_3}} \,,\\ 
    \bm{3}_{x_{f}} &\to \bm{1}_{x_{f_1}} \oplus \bm{1}_{x_{f_2}} \oplus \bm{1}_{x_{f_3}},
\end{aligned}
\label{eq:U3toU1decomp}
\end{align}
where $f = u, \, d, \, \ell,\, e$, while the first line is for $q$. The corresponding $\U(1)$ charges are reported as subscripts. The spurions required to reproduce the observed SM masses and mixings consist of three $\SU(2)_q$ doublets and a singlet:
\begin{align}
    \boldsymbol{V}\sim \bm{2}_{x_V}, \quad  \boldsymbol{W} \sim \bm{2}_{x_W}, \quad \boldsymbol{U}\sim \bm{2}_{x_U}, \quad \boldsymbol{z}\sim \bm{1}_{x_z}.
    \label{eq:MFPspurions}
\end{align}
The SM charges $x_{f_i}$ and $x_{q_3}$ are expressed in terms of spurion charges (see Table~I of~\cite{Greljo:2025mwj}) such that \cref{eq:decomp1} admits the following leading-order form,
\begin{align}
\begin{aligned}
\Delta_u &\sim
\begin{pmatrix}
\boldsymbol{V} \boldsymbol{z}^2 & \vline & \boldsymbol{W}
\end{pmatrix},
~\Delta_d \sim
\begin{pmatrix}
\boldsymbol{V} \boldsymbol{z}^2 & \vline & \widetilde{ \boldsymbol{V}} \boldsymbol{z}^\ast
\end{pmatrix},
\\
\Delta_e &\sim
\begin{pmatrix}
\boldsymbol{z}^3 & \vline & 0 \\
\hline
0   & \vline & \boldsymbol{z}^2
\end{pmatrix}, ~
V_q^{(d)} \sim \boldsymbol{U}\, \boldsymbol{z}^\ast, ~
y_b, y_\tau \sim \boldsymbol{z},
\end{aligned}
\label{eq:MFPstruc}
\end{align}
correctly reproducing the observed SM flavour hierarchies for all spurions of $\mathcal{O}(0.01)$.
In addition, the spurion charges can be chosen such that the interaction and mass bases are sufficiently aligned, and that direct spurion insertions yield adequate suppression of SMEFT operators. This was explicitly demonstrated in~\cite{Greljo:2025mwj} for the benchmark charge assignment $\{x_V,x_W,x_U,x_z\}=\{1,-2,-3,7\}$. In general, structural requirements imposed by phenomenological constraints lead to relations among the flavour charges that must always be satisfied. For instance, the charges of $q_3$ and $u_3$ must coincide in order to reproduce the large top Yukawa coupling.

As in MFV and $\mathrm{U}(2)^5$, the MFP symmetry–spurion framework allows one to systematically construct the allowed flavour structures of the coupling $y$. In \cref{tab:MFP} we present several representative examples of the representation of $y$ with the usual colour coding. These illustrate when the flavour symmetry is further reduced from $\U(3)^5$ and $\U(2)^5$ down to the MFP framework, that new scenarios with sufficient suppression of flavour-violation arise.

A final, important point is that, for appropriate charge assignments of the leptons under $\U(1)_{x_i}$, the MFP framework accidentally respects a larger symmetry in the SM charged lepton Yukawa, namely 
\begin{equation}
    \U(1)_e \times \U(1)_\mu \times \U(1)_\tau,
\end{equation}
 up to a very good approximation. By choosing the charge of $X$ in the $t$-channel interaction in \cref{eq:def-t} such that its coupling to a specific lepton flavour is allowed without spurion insertions (and therefore of $\mathcal{O}(1)$), couplings to the other lepton flavours \emph{necessarily} require insertions of the spurions in \cref{eq:MFPspurions}. As a result, these couplings are small, strongly suppressing charged-lepton flavour violation, which would otherwise impose severe constraints.

In the presence of multiple $t$-channel interactions, $\mathcal{L} \supset y^k\bar f_{\rm SM} X^k$, a flavour-protected scenario compatible with DM relic abundance, barring fine-tuned cancellations, requires each coupling $y^k$ to transform as a flavour singlet, $y^k \sim \bm{1}_0$, for all $k$.
As an illustrative example, consider a triplet of leptophilic DM $\chi^k$ where $k=1,2,3$ and interaction
\begin{equation}
    \mathcal{L} \supset y_{i k} \bar e_i \chi_k \phi\,.
\end{equation}
In the formulation of dark MFV~\cite{Agrawal:2014aoa}, $y_{i k}$ is a new spurion of $\mathrm{U}(3)_e \times \mathrm{U}(3)_\chi$. Our goal here is to identify the flavour-safe texture of this spurion. Starting from MFV in \cref{tab:U3}, and identifying $\mathrm{U}(3)_\chi = \mathrm{U}(3)_e$, the flavour-singlet coupling $y_{ik}=y \delta_{ik}$ implies universal interactions governed by a single parameter.\footnote{Identifying $\mathrm{U}(3)_\chi =\U(3)_\ell$ realizes $y_{ik} = (Y_e^\dagger)_{ik}$, which is flavour-protected but in generic parameter space leads to DM overabundance due to small $y_\tau$. 
} Moving to $\mathrm{U}(2)^5$, the DM triplet would decompose into $\tau$-philic singlet and $e\mu$-philic doublet. Two possibilities with $\mathcal{O}(1)$ couplings arise: a universal coupling in the $1$–$2$ sector with a distinct coupling to the third family, or three independent $\tau$-philic couplings. Other possibilities also exist in which at least one coupling is $\mathcal{O}(1)$, while the others are spurion suppressed. In such cases, the $\chi^k$ components with suppressed couplings must be heavier and decay into the component with the $\mathcal{O}(1)$ coupling. Importantly, these decay channels respect the spurion expansion according to the representations of $\chi^k$, consistent with the suppression of flavour violation. This is not restrictive, as even very small couplings suffice to ensure that heavier states decay into the lightest DM state within phenomenologically acceptable timescales.

Finally, within MFP, the DM triplet decomposes into three singlets, each of which may carry one of the $x_{e_i}$ quantum numbers. This results in ten distinct flavour-protected scenarios with $\mathcal{O}(1)$ couplings, allowing all $\chi^k$ components to contribute sizably to coannihilations. Again, there are several other viable scenarios in which at least one coupling is $\mathcal{O}(1)$, while the others are controlled by spurions.

\begin{table}[t]
\centering
\renewcommand{\arraystretch}{1.4}
\begin{NiceTabular}{r@{}c|cc|ccc|c}
    &$X \backslash \bar f_{\text{SM}}$ & $\overline{\bm{2}}_q$ & $\overline{\bm{1}}_{q_3}$ & $\overline{\bm{1}}_{u_1}$ & $\overline{\bm{1}}_{u_2}$ & $\overline{\bm{1}}_{u_3}$  & \dots \\
    \midrule
    \Block{2-1}{$\bm{3}_q \Bigg \{$}  &
     $\bm{2}_q$& \MFVg{\bm{1}} $\oplus$ \MFVy{\bm{3}_0}& \MFVy{\overline{\bm{2}}_{-\bm{q}+q_3}} & \MFVy{\overline{\bm{2}}_{-\bm{q}+u_1}} &  \MFVy{\overline{\bm{2}}_{-\bm{q}+u_2}} & \MFVy{\overline{\bm{2}}_{-\bm{q}+u_3}} &  \dots\\
      & $\bm{1}_{q_3}$ & \MFVy{\bm{2}_{\bm{q}-q_3}} &  \MFVg{\bm{1}} & \MFVy{\bm{1}_{u_1-q_3}} &  \MFVy{\bm{1}_{u_2-q_3}} & \MFVg{\bm{1}} &  \dots\\
    \Block{3-1}{$\bm{3}_u \Bigg \{$} &
     $\bm{1}_{u_1}$& \MFVy{\bm{2}_{\bm{q}-u_1}}& \MFVy{\bm{1}_{q_3-u_1}} &  \MFVg{\bm{1}} & \MFVy{\bm{1}_{u_2-u_1}} & \MFVy{\bm{1}_{u_3-u_1}} &  \dots\\
      & $\bm{1}_{u_2}$ & \MFVy{\bm{2}_{\bm{q}-u_2}} & \MFVy{\bm{1}_{q_3-u_2}} & \MFVy{\bm{1}_{u_1-u_2}} &  \MFVg{\bm{1}}& \MFVy{\bm{1}_{u_3-u_2}} &  \dots\\
      & $\bm{1}_{u_3}$ & \MFVy{\bm{2}_{\bm{q}-u_3}} & \MFVg{\bm{1}} & \MFVy{\bm{1}_{u_1-u_3}} &  \MFVy{\bm{1}_{u_2-u_3}}& \MFVg{\bm{1}} &  \dots \\
      &\vdots & \vdots & \vdots & \vdots & \vdots & \vdots & $\ddots$ \\
      \CodeAfter
      \OverBrace[yshift=2pt]{1-3}{1-4}{$\overline{\bm{3}}_q$}
      \OverBrace[yshift=2pt]{1-5}{1-7}{$\overline{\bm{3}}_u$}
\end{NiceTabular}
\caption{Example flavour representations of the coupling $y$ in \cref{eq:def-t} under MFP. The subscripts denote the $\U(1)_{x_i}$ charges, while the number in bold identifies the $\SU(2)_{\bm{q}}$ representation. The decomposition of the associated MFV representation of $\bar{f}_\text{SM}$ and $X$ under MFP is also included to highlight the difference in the spurion counting. Color code as in Table~\ref{tab:U3}: $\mathrm{U}(2)$-compatible but likely too small for the relic abundance (yellow), $\mathrm{U}(2)$-invariant and viable (green). The table can be straightforwardly extended to other cases.}
\label{tab:MFP}
\end{table}

\begin{figure}[ht!]
    \centering    \includegraphics[width=0.9\linewidth]{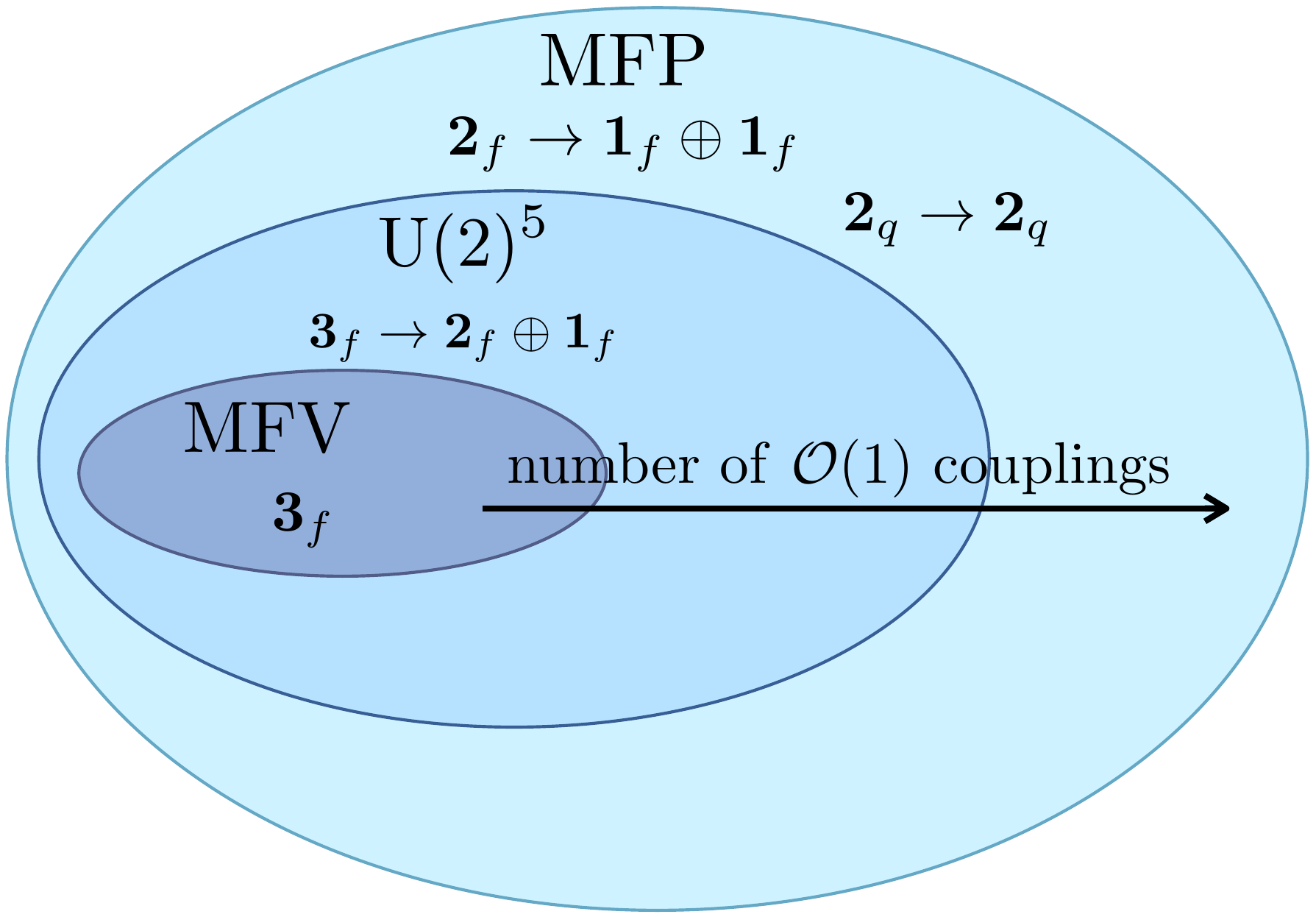}
    \caption{Schematic illustration of the space of flavour-protected $t$-channel DM scenarios in different flavour symmetry frameworks.}
    \label{fig:MFV_to_MFP}
\end{figure}

\paragraph{Summary} Lowering the assumed flavour symmetry from MFV to $\mathrm{U}(2)^5$, and further to the MFP framework, enlarges the space of flavour-protected scenarios for the coupling in \cref{eq:def-t}. MFV enforces flavour universality up to Yukawa breaking, $\mathrm{U}(2)^5$ allows controlled departures associated with the third generation, and MFP admits a broader class of non-universal structures, all while maintaining controlled suppression of flavour violation through symmetry and spurion power counting. The resulting hierarchy of symmetries, $\mathrm{MFP} \subset \mathrm{U}(2)^5 \subset \mathrm{MFV}$, corresponds to an \emph{inverted} hierarchy in the size of the allowed parameter space, as illustrated schematically in \cref{fig:MFV_to_MFP}.

In what follows, we concentrate on two minimal benchmark configurations discussed in \cref{sec:leptophilic} and \cref{sec:quarkphilic}, which already exhibit rich and distinctive phenomenology and illustrate the general applicability of the framework developed above. The primary goal is to test these expectations for viable flavour structures in the simplest scenarios, leaving a systematic investigation for future studies.

\section{Leptophilic Dark Matter} 
\label{sec:leptophilic}

In this section, we focus on a concrete leptophilic scenario in which DM is a complete singlet fermion $\chi$  (Dirac or Majorana) and the mediator is a single scalar field $\Phi$ carrying the same SM quantum numbers as $e_R$. The portal interaction is given by
\begin{align}
\mathcal{L} \supset -y_i \, \bar e_{R,i} \chi \Phi + \textrm{h.c.}.
\label{eq:leptophilicL}
\end{align}
The three components of the coupling vector $\bm{y} \sim \bm{3}_e$ of $\U(3)_e$ break the exact SM lepton flavour symmetry $\U(1)_e \times \U(1)_\mu \times \U(1)_\tau$ and $\mathrm{U}(1)_{\chi + \Phi}$ down to a single generalized lepton number $\mathrm{U}(1)_L$. These global rotations can be used to remove any complex phase in the vector, implying CP invariance. Hence, we parametrize the components of $\bm{y}$ as coordinates on a sphere of radius $y \equiv \norm{ \bm{y}}$, similarly to \cite{Gherardi:2019zil, Marzocca:2024hua}. This parametrization requires two angles, whose orientation is, in principle, arbitrary. A convenient choice is to use the polar angle $\theta_\tau$ to quantify the alignment toward a $\tau$-only coupling and the azimuthal angle $\phi_{e\mu}$ to describe the orientation within the $(e \mu)$ plane:
\begin{align}
\label{eq:angles_lept}
\bm{y} = y  \, ( \sin \theta_\tau \cos \phi_{e\mu} , \sin \theta_\tau  \sin \phi_{e\mu} , \cos \theta_\tau) .
\end{align}
This choice is motivated by the fact that bounds involving $\tau$'s are generally weaker than those involving electrons or muons, as will be shown later, and with this parametrization, the degree of $\tau$ alignment becomes immediately apparent. However, this choice is by no means unique. The other two permutations are explored in Appendix~\ref{app:LFV_directions}, where we also quantify the degree of $\mu$- and $e$-alignment, respectively.

\begin{figure}
    \centering
    \includegraphics[width=\linewidth]{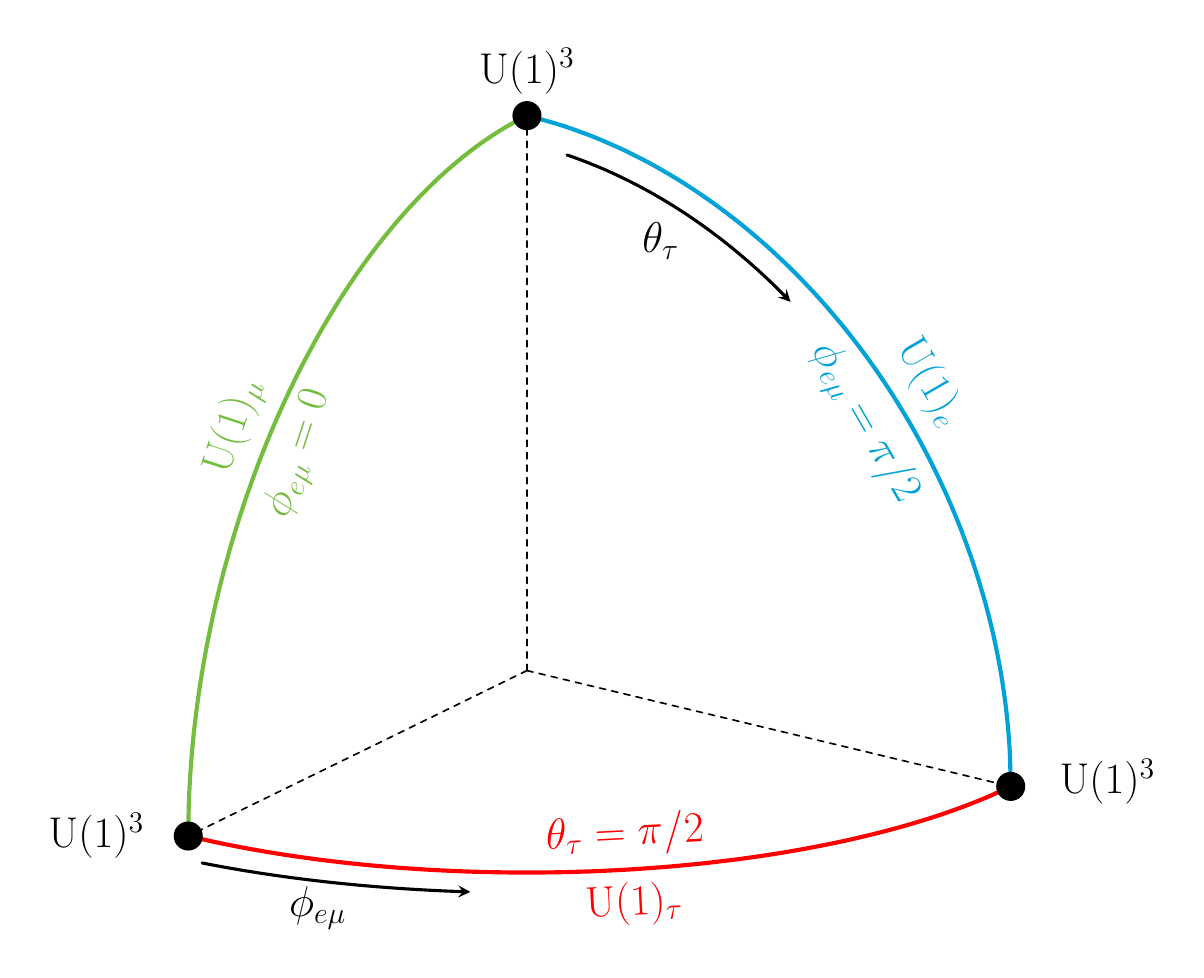}
    \caption{Visual representation of the parametrization on the octant of the sphere in terms of the polar angle $\theta_\tau$ and azimuthal angle $\phi_{e\mu}$, and corresponding symmetric limits.}
    \label{fig:sphere}
\end{figure}

The geometry of the parameter space is that of an octant of a sphere, $S^2/(Z_2)^3$, reflecting that the phases in $y$ are unphysical; for this reason, we can restrict $\theta_\tau \in [0,\pi/2]$ and $\phi_{e\mu} \in [0,\pi/2]$.\footnote{This implies that bounds on $\theta_\tau, \phi_{e\mu}$ are symmetric in shifts of $\pi$ and with respect to reflections along integer multiples of $\pi/2$.} It is illustrated in Fig.~\ref{fig:sphere}. Along the ``meridian'' $\phi_{e\mu}=0$, the coupling to muons vanishes, corresponding to a $\U(1)_\mu$-symmetric limit, while $\phi_{e\mu}=\pi/2$ gives no coupling to electrons, i.e. a $\U(1)_e$-symmetric limit. The ``parallel'' at $\theta_{\tau}=\pi/2$ defines the $\U(1)_\tau$-symmetric case. Finally, points naively realizing a $\U(1)^2$ limit (such as $\theta_{\tau}=0 \leftrightarrow \U(1)_e \times \U(1)_\mu$) automatically imply full $\U(1)^3$ conservation due to the overall $\U(1)_L$ preserved by the interaction in \cref{eq:leptophilicL}, as already discussed in \cref{sec:charting}. This reflects the trivial fact that coupling to a single flavour does not induce any flavour-violating transition.

The various flavour symmetry scenarios introduced in \cref{sec:charting} can be smoothly explored using this parametrization. Within MFV, the vector $\bm{y} \sim \bm{3}_e$ does not admit an expansion in terms of SM spurions (see the first row of \cref{tab:U3}). Hence, there is no flavour-symmetric parametrization.
Within $\mathrm{U}(2)^5$, the vector $\bm{y}$ can be decomposed as $\bm{y} \sim \bm{2}_e \oplus \bm{1}$. The singlet component, parametrizing the coupling to $\tau$, can consistently be of $\mathcal{O}(1)$. At the same time, in the minimal-breaking scenario, there is no spurion associated with the $\bm{2}_e$ representation. This structure realises the $\mathrm{U}(1)^3$ limit, represented by the upper corner of \cref{fig:sphere}.

Finally, within MFP one may couple at $\mathcal{O}(1)$ to any single flavour (electron, muon, or tau) at the time, thereby realising all three corners of \cref{fig:sphere}. Couplings to the other flavours, corresponding to departures from a given corner, must then be constructed using the spurions in \cref{eq:MFPspurions}. With the general charge parametrisation introduced in \cite{Greljo:2025mwj}, the remaining components of the vector $y_i$ are expected to be suppressed by at least two powers of the spurions. For instance, if one chooses to couple to tau with $\mathcal{O}(1)$ coupling, the parametrization in \cref{eq:angles_lept} leads to the expectation $\theta_{\tau} \lesssim 10^{-4}$. Our results show that, if this is the only TeV-scale new-physics sector, the charge assignment of \cite{Greljo:2025mwj} may be relaxed, allowing larger $\theta_\tau$.

\subsection{Relic Abundance}
For Yukawa couplings $y_i$ of order unity, DM candidates in $t$-channel models can be regarded as WIMP-like particles. As such, they are expected to be in thermal equilibrium in the early universe for $T \gg M_\chi$. Their relic abundance is therefore determined by the thermal freeze-out mechanism~\cite{Gondolo:1990dk, Griest:1990kh}, where the key quantity is the thermally averaged annihilation cross section of DM pairs into lighter SM states. 

Despite its apparent simplicity, thermal freeze-out receives an important modification in $t$-channel models due to the presence of the mediator $\Phi$, which effectively belongs to the dark sector. When the mediator mass is sufficiently close to the DM mass, more precisely, when the relative mass splitting $\delta_\mathrm{DM} \equiv (M_\Phi-M_\chi)/M_\chi$ becomes $\delta_\mathrm{DM} \lesssim 0.3$ \cite{Griest:1990kh, Baker:2015qna},  the mediator remains abundant in the thermal bath during DM freeze-out.\footnote{The relative mass splitting needed to significantly affect the DM relic density depends non-trivially on the interplay between the Yukawa coupling $y_i$ and SM gauge couplings.} Since the mediator ultimately decays into DM and chemical equilibrium is maintained within the dark sector, coannihilation processes involving fermion–scalar initial states, as well as scalar pair annihilations, must be included to obtain an accurate prediction of the relic abundance.

Coannihilations play an important role in $t$-channel models~\cite{Garny:2015wea, Biondini:2018pwp, Garny:2018ali, Becker:2022iso, Biondini:2018ovz, Biondini:2019int, Bollig:2021psb, Arina:2025zpi}, as well as in generic next-to-minimal WIMP scenarios~\cite{Griest:1990kh, Edsjo:1997bg, Ellis:2014ipa, Biondini:2017ufr, Tsai:2019eqi}. Complementary bounds from DD and collider searches constrain the $\chi\chi \to \mathrm{SM}\,\mathrm{SM}$ annihilation rate via crossing symmetry, 
which may imply that this process is insufficient to deplete the primordial abundance, leading to DM overproduction. Entering the coannihilation regime introduces additional efficient channels for depleting the dark sector population, thereby opening cosmologically viable regions of parameter space.

The effects of the coannihilating particle can still be captured by a single Boltzmann equation \cite{Griest:1990kh,Edsjo:1997bg}
\begin{equation}
    \dot{n} = \langle \sigma_{\textrm{eff}} \, \vrel \rangle (n^2 -n_{\textrm{eq}}^2) \, ,
    \label{BE_start}
\end{equation}
where $n$ stands for the total number density of the dark sector. The effective cross section is a combination of the various annihilation processes that involve the dark sector states, namely 
\begin{eqnarray}
   \sigma_{\textrm{eff}} \, \vrel   = \frac{1}{(\sum_k n_k^{\textrm{eq}})^2}\sum_{i,j} n_i^{\textrm{eq}}n_j^{\textrm{eq}}\sigma_{ij}  \vrel \, ,
   \label{eff_cross_coann}
\end{eqnarray}
where $i,j$ run over the particles $\chi$ and $\Phi$ and their complex conjugates. The behavior of the effective cross section in \cref{eff_cross_coann}, and its dependence on the coannihilation channels, can be summarized qualitatively as follows \cite{Garny:2015wea} (see \cite{Baker:2015qna,Biondini:2025gpg} for a more detailed expression)
\begin{equation}
   \sigma_{\textrm{eff}} \, \vrel  \sim \sigma_{\chi \chi} \vrel +  \sigma_{\chi \Phi} \vrel \, R +   \sigma_{\Phi \Phi} \vrel \, R^2 \, ,
   \label{Eff_cross_coann}
\end{equation}
where $R = n_\Phi^{\textrm{eq}}/ n_\chi^{\textrm{eq}} \propto e^{-(M_\Phi-M_\chi)/T}$ is a Boltzmann suppression factor for the
processes involving the coannihilating particle $\Phi$, penalizing larger mass splittings. 

The cross sections up to order $\mathcal{O}(\vrel^2)$ in the non-relativistic regime relevant for freeze out have been obtained earlier in the literature, see e.g.~\cite{Garny:2015wea, Kopp:2014tsa, Biondini:2022ggt}. Here we briefly summarize the behaviour of the dark fermion pair annihilation cross sections $\sigma_{\chi \chi} \vrel$ and $\sigma_{\chi \bar{\chi}} \vrel$ for the Majorana and Dirac case respectively, which is controlled by the Yukawa coupling. We drop the subscript indicating right-handed chirality of the leptons when writing the relevant processes to avoid clutter.
 More specifically, the process $\chi \chi \to e \bar{e}$ receives a helicity suppressed $s$-wave contribution for Majorana DM, so the velocity suppressed $p$-wave can in fact dominate at freeze-out.\footnote{We have checked that for right-handed leptons, radiative processes with a photon or a $Z$ boson in the final state, $\chi \chi \to e \bar e \gamma(Z)$, which lift the helicity suppression, remain largely subdominant compared to the $\mathcal{O}(\vrel^2)$ term of the $2 \to 2$ cross section at freeze out. We have instead included the corresponding radiative process for the quarkphilic scenario, which accounts for corrections of order 10\%. Cross sections are taken from \cite{Garny:2015wea,Ibarra:2014qma}.} This is indeed the case for the leptophilic model, where the ratio $(m_\tau/M_\chi)^2 \simeq 2 \times 10^{-3}$ is largest for the smallest DM mass considered in this work ($M_\chi = 45$ GeV) and is quite smaller than $\langle \vrel^2 \rangle \simeq 4 \times 10^{-2}$ at the freeze-out. In contrast, the Dirac case features an $s$-wave contribution to $\sigma_{\chi \bar{\chi}} \vrel$ that is not helicity suppressed. Therefore, the coannihilation effects are generally more relevant for the Majorana option.

The flavour structure of the Yukawa interaction implies that nine annihilation processes must be considered, namely $\chi \chi \to e_i \bar e_j$ with $i,j = e, \mu, \tau$, and the individual cross sections depend on the mixing angles $\theta_\tau$ and $\phi_{e\mu}$. However, upon neglecting the lepton masses, which is well justified in the phenomenologically relevant regime $m_{\tau}/M_\chi \ll 1$, the total annihilation rate becomes flavour independent. Indeed, after summing over all final states, $\sum_{i,j} \sigma \vrel (\chi \chi \to e_i \bar e_j)$, the result no longer depends on the flavour mixing angles. We have checked this for all processes involving the Yukawa interaction, namely also for $\chi \Phi \to \gamma e_i$, $\chi \Phi \to Z e_i$, $\Phi \Phi^\dagger \to e_i \bar{e}_j$ and $\Phi \Phi \to e_i e_j$. The same applies for the Dirac DM option.

Thermal freeze-out involves non-relativistic particle annihilations. When the annihilating states interact through gauge bosons or scalars, mediating long-range forces, Sommerfeld enhancement~\cite{Hisano:2004ds, Iengo:2009ni, Feng:2010zp} and bound-state effects~\cite{Detmold:2014qqa, vonHarling:2014kha, Petraki:2015hla} can significantly modify the annihilation cross sections and affect the relic density prediction. 
In our work, we include Sommerfeld factors and bound state effects for mediator pair annihilations, namely $\Phi \Phi^\dagger \to \textrm{SM} \, \textrm{SM}'$. For Majorana DM, the process $\Phi \Phi \to e e$ and its conjugate also contribute, which depends solely on the Yukawa coupling. The Sommerfeld factors enhance or suppress the perturbative cross section depending on whether the potential is attractive (for $\Phi \Phi^\dagger$) or repulsive (for $\Phi \Phi$ and $\Phi^\dagger \Phi^\dagger$). The formation of bound states and their decays provides an additional channel for depleting the heavy scalars. Our treatment of near-threshold effects follows the implementations described 
in~\cite{Petraki:2015hla,Bollig:2021psb,Biondini:2022ggt}, consistently accounting for the electroweak crossover that sets the temperature-dependent masses of the 
electroweak gauge bosons~\cite{Biondini:2022ggt,Biondini:2025ihi}.Since the scalar mediator couples to the SM photon and $Z$ boson with a typical strength $\alpha_{\mathrm{ew}} \sim 10^{-2}$, the resulting Sommerfeld and bound-state corrections remain modest, typically at most $\mathcal{O}(10\%)$ level in the relic density prediction. These effects become more pronounced in the quarkphilic scenario, as discussed in \cref{sec:quarkphilic}.

It is worth mentioning that we have neglected couplings of the scalar mediator with the Higgs boson in our analysis, focusing instead on the Yukawa interaction in \cref{eq:def-t}. In general, including such couplings could affect the relic density through additional annihilation channels of the mediator, as well as direct detection and collider searches when the scalar couplings are of order unity (see e.g.~\cite{Biondini:2018ovz,Biondini:2018pwp,Harz:2019rro,Biondini:2025gpg,Olgoso:2025jot}), and could also impact the nature of the electroweak phase transition \cite{Biondini:2022ggt,Liu:2021mhn}.

In our analysis, we compute the DM relic density and restrict the 
parameter space to reproduce the observed value,
$\Omega_{\mathrm{DM}} h^2 = 0.1200 \pm 0.0012$~\cite{Planck:2018vyg}. 
The resulting parameter space defines targets for the complementary 
experimental searches discussed below.

\subsection{Direct (and Indirect) Detection}
\label{sec:leptoDD}

Bounds from DD of DM scattering on nuclei arise in the leptophilic context only at the one-loop level. They are directly linked to the requirement that $\chi$ constitutes the DM of the universe, and therefore they technically apply only along the relic density contours and for a fixed direction of the coupling vector $\bm{y}$. In $t$-channel models, the contribution to spin-independent and spin-dependent scattering originates from loop-induced penguin diagrams involving the Higgs boson, the $Z$, or a photon~\cite{Ibarra:2015fqa, Hisano:2018bpz, Arcadi:2023imv}. The dominant one comes from photon penguins, giving rise to the magnetic dipole, charge radius, and anapole operators. At the same time, those involving the Higgs or $Z$ are suppressed by the small lepton masses~\cite{Kopp:2014tsa, Ibarra:2015fqa, Ibarra:2024mpq}. Consequently, the resulting bounds from the magnetic dipole moment, which dominate for larger masses, depend primarily on the overall magnitude $y = \norm{\bm{y}}$ and are nearly independent of the angular parameters.

The Majorana case is special because many of these diagrams sum up to zero as a consequence of the absence of a DM vector current. In this situation, one must rely solely on the anapole moment, which leads to weaker, lepton flavour-dependent bounds~\cite{Kayser:1983wm, Radescu:1985wf, Kopp:2014tsa, Garny:2015wea, Ibarra:2024mpq}.
Technically, this conclusion is specific to the case of a single DM particle. If multiple flavours of $\chi$ were present, the corresponding vector currents would no longer vanish identically, thereby allowing, in principle, contributions from the other operators. However, for this to occur, the mass splitting between the DM flavours must be smaller than the kinetic energy of non-relativistic DM while still producing a detectable nuclear recoil. This requirement renders the region of parameter space in which this possibility applies somewhat fine-tuned~\cite{Bramante:2016rdh}.

Concretely, to derive the experimental limits, we employ the expressions for the differential nucleus cross-section given in \cite{Kopp:2014tsa, Ibarra:2024mpq}, and follow the procedure outlined in Appendix D of \cite{Hisano:2018bpz} due to the non-trivial dependence on the recoil energy of the various contributions. We use the latest results from the Lux–Zeplin (LZ) experiment \cite{LZ:2024zvo} at 90\% CL, and the projections for DARWIN from \cite{DARWIN:2016hyl}. The DD constraints are evaluated along relic density contours by fixing the Yukawa coupling as $y = y(M_\chi,\,  \delta_\text{DM})$, ensuring self-consistent bounds. The resulting limits are shown in the left panels of \cref{fig:Majorana_RH_lept} and \cref{fig:Dirac_RH_lept} for the Majorana and Dirac cases, respectively, and will be discussed in \cref{sec:interplay_leptophilic}.

Indirect detection probes DM through present day annihilations in astrophysical environments, most notably via $\gamma$-ray observations of dwarf spheroidal galaxies, searches for monochromatic $\gamma$-ray lines from the Galactic centre, and measurements of the antiproton flux \cite{Garny:2015wea, Kopp:2014tsa, Garny:2018icg, Cirelli:2013hv}. As reviewed in \cite{Garny:2015wea, Arina:2025zpi} and confirmed by recent analyses of $t$-channel models~\cite{Olgoso:2025jot, ElHedri:2016onc}, current indirect detection limits are generally weak and do not compete with present or projected DD sensitivities for this class of scenarios.

In leptophilic models, the phenomenology depends strongly on the nature of the DM. For Dirac DM, annihilation into lepton pairs proceeds through an unsuppressed $s$-wave and indirect searches constrain masses below $\mathcal{O}(100)~\text{GeV}$ \cite{DeLaTorreLuque:2023fyg}, well below the DD reach discussed in Sec.~\ref{sec:interplay_leptophilic}. For Majorana DM, instead, the $s$-wave is helicity suppressed by $(m_e/M_\chi)^2$ and the $p$-wave by $\vrel^2$ with $\vrel \approx 10^{-3}$, leading to highly suppressed annihilation rates today~\cite{Bringmann:2012ez,Kopp:2014tsa}. Radiative annihilation $\chi\chi \to e\bar e\gamma$ and loop-induced $\chi\chi \to \gamma\gamma$ lift this suppression and are constrained by $\gamma$-line searches from Fermi-LAT and HESS \cite{Fermi-LAT:2022byn, McDaniel:2023bju, DeLaTorreLuque:2023fyg, HESS:2018cbt}. Using the cross sections from \cite{Garny:2011ii, Garny:2015wea}, we find no constraints from sharp spectral features in the parameter space consistent with the observed relic abundance.

\subsection{Collider Searches}

The mediator $\Phi$ carries the same quantum numbers as an MSSM right-handed slepton~\cite{Martin:1997ns}. Therefore, direct searches for supersymmetric sleptons at LEP and the LHC can be directly applied to this scenario~\cite{ATLAS:2019lff, ATLAS:2019lng, ATLAS:2022hbt, CMS:2025ttk, DELPHI:2003uqw, ATLAS:2024fub, CMS:2019zmn}.\footnote{For a comprehensive review of $t$-channel DM searches at LHC, we refer the reader to \cite{Arina:2025zpi}.} These searches are largely independent of the Dirac or Majorana nature of DM. At both colliders, sleptons are predominantly pair-produced through electroweak Drell–Yan processes. Most searches target the decay of a slepton into a charged lepton and a neutralino, corresponding precisely to our process $\Phi \to \ell_i \chi$. These results can thus be straightforwardly reinterpreted, as they depend only mildly on the size of the Yukawa coupling provided that their decay is prompt. Such searches are particularly effective when the mass splitting between $\Phi$ and $\chi$ is large (tens of GeV) but quickly lose sensitivity in the compressed regime, where the decay products become too soft to be reconstructed efficiently. In that region, the most robust limit arises from measurements of the $Z$-boson width, which impose $m_{\Phi} > m_Z/2$. It is plausible that recasting searches designed for compressed chargino–neutralino systems, e.g.,~\cite{OPAL:2002lje}, could extend the coverage of this region in the slepton-like case, but a dedicated study is beyond the scope of this work.

LEP bounds require some additional care due to the tree-level coupling of $\Phi$ to electrons. In this case, $\Phi$ can also be produced via $t$-channel exchange in $e^+e^-$ collisions, which can dominate over the electroweak production if the Yukawa coupling $y$ exceeds the gauge coupling of $\Phi$. Because of the different chiral structures involved, a complete cancellation between the $s$- and $t$-channel diagrams cannot occur, so the limits derived here considering only the $s$-channel can somehow be regarded as conservative.

Finally, the radiative return process $e^+ e^- \to \chi \chi \gamma$, mediated by $t$-channel $\Phi$ exchange, could also provide additional constraints. These bounds depend strongly on the magnitude of the Yukawa coupling $y$. Adapting the results of Ref.~\cite{Kopp:2014tsa} to our setup, we verified that when evaluated along the relic-density contours $\bigl(y = y(M_\chi, \delta_\text{DM})\bigr)$ such measurements do not impose any new constraints in the parameter space of the model. Hence, we do not include them in the plots presented in \cref{sec:interplay_leptophilic}.

\subsection{Flavour Observables}
\label{sec:lep_flavour}

\begin{figure*}[!htb]
    \centering
    \includegraphics[width=\linewidth]{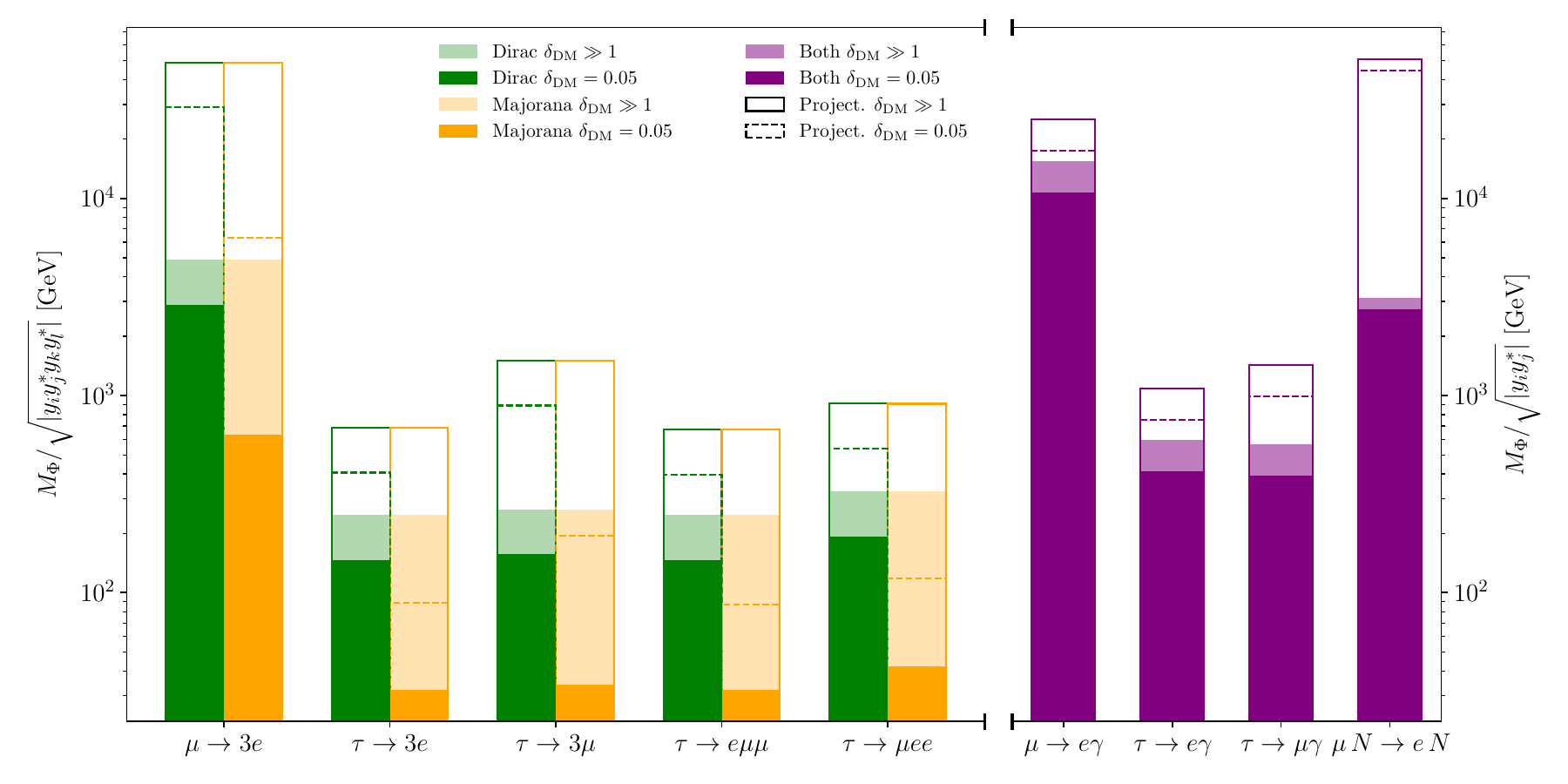}
    \caption{\textbf{Leptophilic DM}.
    Bounds on the mediator mass $M_\Phi$ from lepton flavour-violating decays, with the dependence on the couplings $y_i$ factored out. On the left-hand side of the plot are the three-body decays, and on the right are the radiative decays. The different shades of the colour correspond to different values of $\delta_\text{DM}= (M_\Phi-M_\chi)/M_\chi$, while dashed (solid) lines denote future projections for $\delta_{\rm{DM}} = 0.05$ ($\delta_{\rm{DM}}\gg 1$). The Dirac and Majorana cases are shown separately on the left. Note that $\delta_\text{DM} \gg 1$ corresponds to the limit $M_\chi/M_\Phi \ll 1$ while keeping $M_\Phi$ fixed.}
    \label{fig:LFV_scale_bound}
\end{figure*}

Lepton-flavour-violating (LFV) processes provide powerful null tests of the 
SM. Any observed signal would constitute unambiguous evidence 
for NP, while negative searches translate into stringent constraints 
on new mass scales or interactions. Among the many possible probes~\cite{Calibbi:2017uvl, Calibbi:2022ddo, Fernandez-Martinez:2024bxg, Greljo:2025ljr}, 
the most constraining are $\mu\to e\gamma$, $\mu \to e$ conversion in nuclei, and $\mu\to 3e$ decays. These observables dominate current limits in the 
charged-lepton sector and are primary targets of upcoming experiments, 
notably MEG~II~\cite{MEGII:2025gzr}, Mu2e~\cite{Bernstein:2019fyh}, COMET~\cite{Moritsu:2022lem}, and Mu3e~\cite{Blondel:2013ia}. Tau decays are currently less constraining for comparable couplings but probe complementary regions of parameter space and are expected to improve with ongoing and future experiments, such as Belle~II~\cite{Belle-II:2018jsg}, STCF~\cite{Achasov:2023gey}, and FCC-ee~\cite{FCC:2025lpp}.

In the $t$-channel DM scenario considered here, all these processes arise only at the loop level and are further suppressed by a numerically small loop function; explicit expressions are collected in \cref{app:matching,app:FV_observables}. Despite these suppression factors, LFV observables provide stringent constraints on the parameter space of interest. In \cref{fig:LFV_scale_bound}, we present the resulting bounds on the mediator mass from $\mu \to e$ conversion, three-body decays, and radiative transitions, factoring out the dependence on the flavour structure.

The different shades denote bounds corresponding to two choices of the mass splitting between the DM particle and the mediator, shown separately for the Dirac (green) and Majorana (orange) scenarios. Dashed lines indicate future projections. Near mass degeneracy leads to a pronounced suppression in the Majorana case, while the effect is considerably milder for Dirac DM. This behaviour originates from cancellations among box diagrams in the Majorana scenario, whose loop function vanishes in the exact degeneracy limit (see \cref{app:matching} and \cref{fig:loop_functions}). In contrast, radiative decays and $\mu \to e$ conversion are governed by the same loop function in both Dirac and Majorana cases and exhibit only a mild suppression as the mass splitting decreases.

Radiative decays $\mu \to e\gamma$, $\tau \to e\gamma$, and $\tau \to \mu\gamma$, with current limits set by MEG~II~\cite{MEGII:2023ltw}, BaBar~\cite{BaBar:2009hkt} and Belle~\cite{Belle:2021ysv}, provide the most stringent constraints for couplings $y \sim \mathcal{O}(1)$, despite the chirality suppression inherent to these processes. Together with $\mu \to e$ conversion in nuclei, these observables probe the same parametric combination of couplings $\sim y_i y_j^\ast / M_\Phi^2$. Future sensitivities for $\mu \to e\gamma$ and $\mu \to 3e$ are taken from the MEG~II and Mu3e projections discussed above, while the expected reach of COMET~II~\cite{COMET:2018auw} and Mu2e~\cite{Mu2e:2014fns} will significantly improve the sensitivity to $\mu \to e$ conversion. Notably, the latter could surpass the constraints from radiative muon decays in the future.

Three-body decays receive contributions from two distinct parametric structures. In addition to the same $\sim y_i y_j^\ast / M_\Phi^2$ combination arising from dipole and penguin insertions, they also receive box diagram contributions scaling as $\sim y_i y_j^\ast y_k y_l^\ast / M_\Phi^2$ (see \cref{app:FV_observables}). For $y\lesssim g_\text{SM}$ the $\sim y^2/M_\Phi^2$ contribution dominates the three-body amplitude, but it is always more strongly constrained by $\mu \to e$ conversion and radiative decays than by the three-body channels themselves, both for current bounds and for future projections. As a consequence, three-body decays become phenomenologically relevant only once the box contribution dominates, i.e. when $y\gtrsim g_\text{SM}$, for which they effectively probe the quartic coupling structure. Accordingly, in \cref{fig:LFV_scale_bound} we show three-body constraints exclusively on the $\sim y^4 / M_\Phi^2$ combination, using future sensitivities for $\tau$ channels from~\cite{deBlas:2025gyz}, which include projections from Belle~II~\cite{Belle-II:2018jsg}, STCF~\cite{Achasov:2023gey}, and FCC-ee~\cite{FCC:2025lpp}.

\subsection{Interplay and Summary}
\label{sec:interplay_leptophilic}

In this section, we combine the aspects discussed separately in the previous sections, addressing the Majorana and Dirac cases separately due to their distinct features. The main results are presented in \cref{fig:Majorana_RH_lept} and \cref{fig:Dirac_RH_lept} for the Majorana and Dirac scenarios, respectively. 

The left panels of these figures show constraints from collider searches and DD experiments, together with the parameter space compatible with the observed DM relic abundance. In particular, the grey shaded regions indicate portions of parameter space where reproducing the observed relic abundance requires non-perturbative couplings ($y > \sqrt{4\pi}$)~\cite{Cahill-Rowley:2015aea}, or where the DM relic density falls below the observed value, corresponding to thermal underproduction starting around $y \lesssim g_\text{SM}$ (more specifically the hypercharge coupling of the SM). This is due to gauge-driven coannihilations processes $\Phi \Phi^\dagger \to \textrm{SM} \,  \textrm{SM}'$, which do not depend on $y$, and that efficiently depletes the DM whenever the dark fermion and mediator are in chemical equilibrium. In this situation, the correct relic abundance can be obtained only along the line in the $(m_\chi, \delta_\text{DM})$ plane denoting the boundary of the underproduction region.\footnote{There exists a lower limit below which thermal contact between the mediator and DM is no longer maintained, $y \approx 10^{-7}$. In this regime, coannihilations may still occur without chemical equilibrium between the dark sector states \cite{Garny:2017rxs, DAgnolo:2017dbv}, a possibility we do not consider in this work.} Interestingly, near this boundary, the relic abundance becomes essentially 
independent of the couplings $y_i$, which can be small. In this regime, 
complementary constraints from flavour and DD require no 
particular symmetry alignment and allow for an anarchic flavour structure, unlike when $y \gtrsim g_\text{SM}$. 

The right panels of \cref{fig:Majorana_RH_lept} and \cref{fig:Dirac_RH_lept} show the exploration of the various flavour directions using the basis defined in \cref{eq:angles_lept} in the $(\phi_{e \mu},\,\theta_\tau)$ plane for a benchmark point. Using the same colour coding as a reference, the corresponding flavour-symmetric trajectories and points on the sphere shown in Fig.~\ref{fig:sphere} are also represented.
In the plots, a specific benchmark point in $(M_\chi,\,\delta_{\rm DM},y)$ is chosen along the relic abundance contours and highlighted in the left panels by a black star. This benchmark is motivated by its location at the edge of the excluded region and by the non-trivial interplay among the various experimental probes, which we now discuss in detail.

\subsubsection{Majorana Dark Matter}

\begin{figure*}[!htb]
    \centering
    \includegraphics[width=\linewidth]{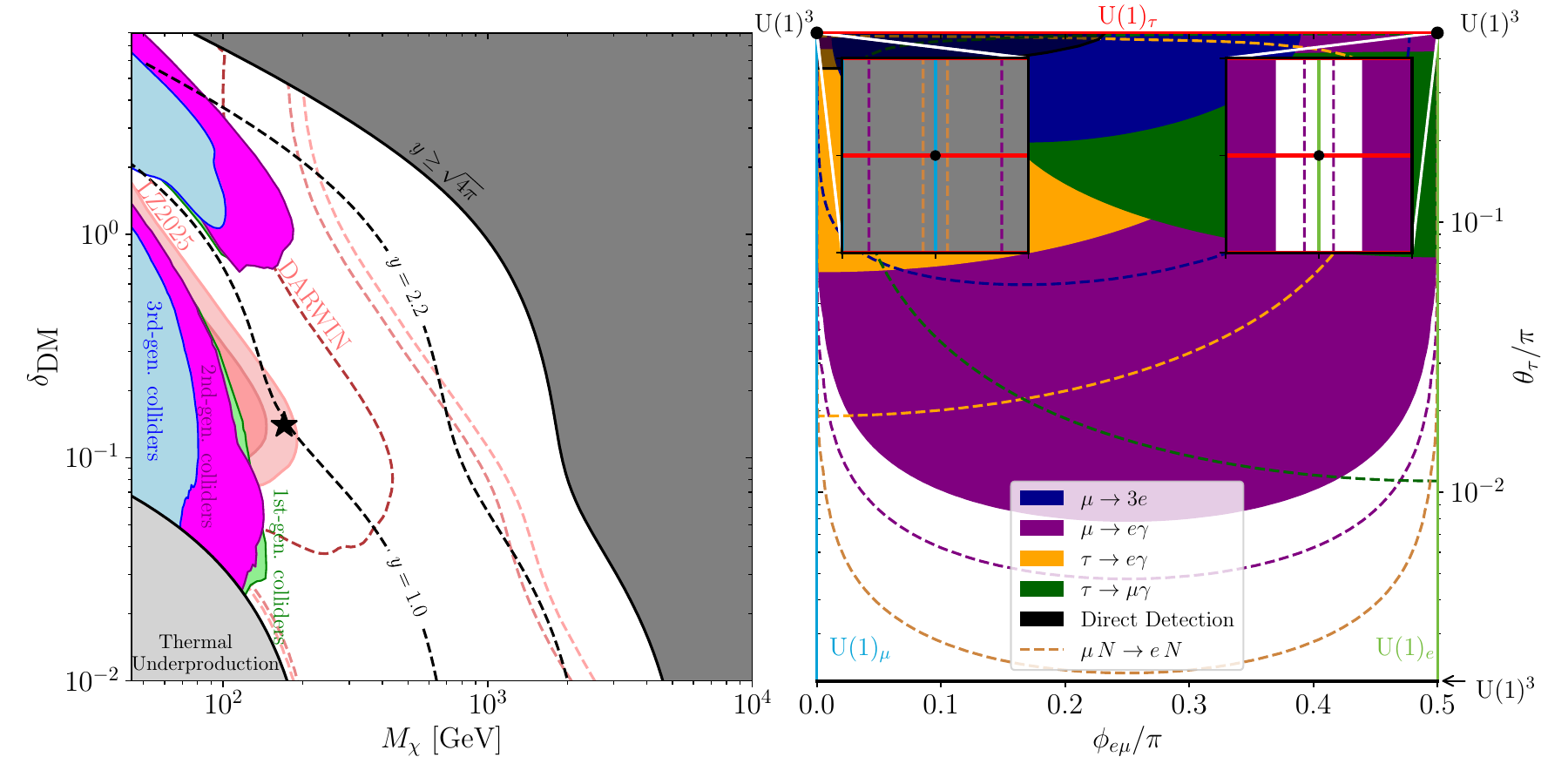}
    \caption{
    \textbf{Leptophilic Majorana DM}. \textit{Left panel:} Constraints in the $(M_\chi,\delta_{\rm DM})$ plane, with $\delta_{\rm DM}\equiv M_\Phi/M_\chi-1$. Shown are 90\% CL bounds from collider searches (green, purple, and blue, corresponding to couplings to first-, second-, and third-generation leptons, respectively), DD (pink, with darker shades indicating couplings to heavier lepton generations; only first- and second-generation bounds are visible), and relic abundance considerations; representative contours for $y=1$ and $y=2.2$ are shown. The DD limits are evaluated along relic-density contours with varying couplings. Gray regions indicate parameter space where the coupling required to reproduce the observed relic abundance is non-perturbative, or where the correct abundance cannot be achieved via thermal freeze-out due to overdepletion. Dashed lines denote projected sensitivities from DARWIN.
    \textit{Right panel}: Constraints from flavour physics and DD in the $(\phi_{e\mu},\,\theta_\tau)$ plane, evaluated at the benchmark point indicated by the black star in the left panel ($M_\chi \simeq 170 \text{ GeV},\delta_\text{DM} \simeq 0.14, y \simeq 0.98$). Dashed lines denote future projected sensitivities (see \cref{sec:lep_flavour}). The boxes in the upper-left and upper-right corners show zoomed-in views of the parameter space, illustrating that flavour constraints become ineffective in these regions and that, at the benchmark point, DD excludes the $U(1)_\mu$ limit but not the $U(1)_e$ limit. This region of parameter space is best visualised in \cref{fig:basis_reparametrization}. See \cref{sec:interplay_leptophilic} for details.
    }
    \label{fig:Majorana_RH_lept}
\end{figure*}

In the Majorana case, bounds from DD are very weak, as they arise exclusively from the anapole contribution. Consequently, most of the parameter space evades current direct-detection bounds. However, future DARWIN sensitivity is expected to cut deeply into it, as shown in the left panel of \cref{fig:Majorana_RH_lept}. The model is therefore currently mainly probed by collider searches, which set a lower bound on the DM mass of $\mathcal{O}(100)$\,GeV, depending on the splitting, and by flavour constraints. The latter are exhibited in the right panel of \cref{fig:Majorana_RH_lept} for a benchmark point in parameter space where the DM abundance matches the observed value for $y \simeq 0.98$, with $M_\chi \simeq 170$ GeV and $\delta_\text{DM} \simeq 0.14$.\footnote{Although in this benchmark $M_{\chi,\Phi}$ are at the EW scale, the SMEFT calculations presented in the \cref{app:FV_observables} are still valid, as discussed there.}

\begin{figure*}[!htb]
    \centering
    \includegraphics[width=\linewidth]{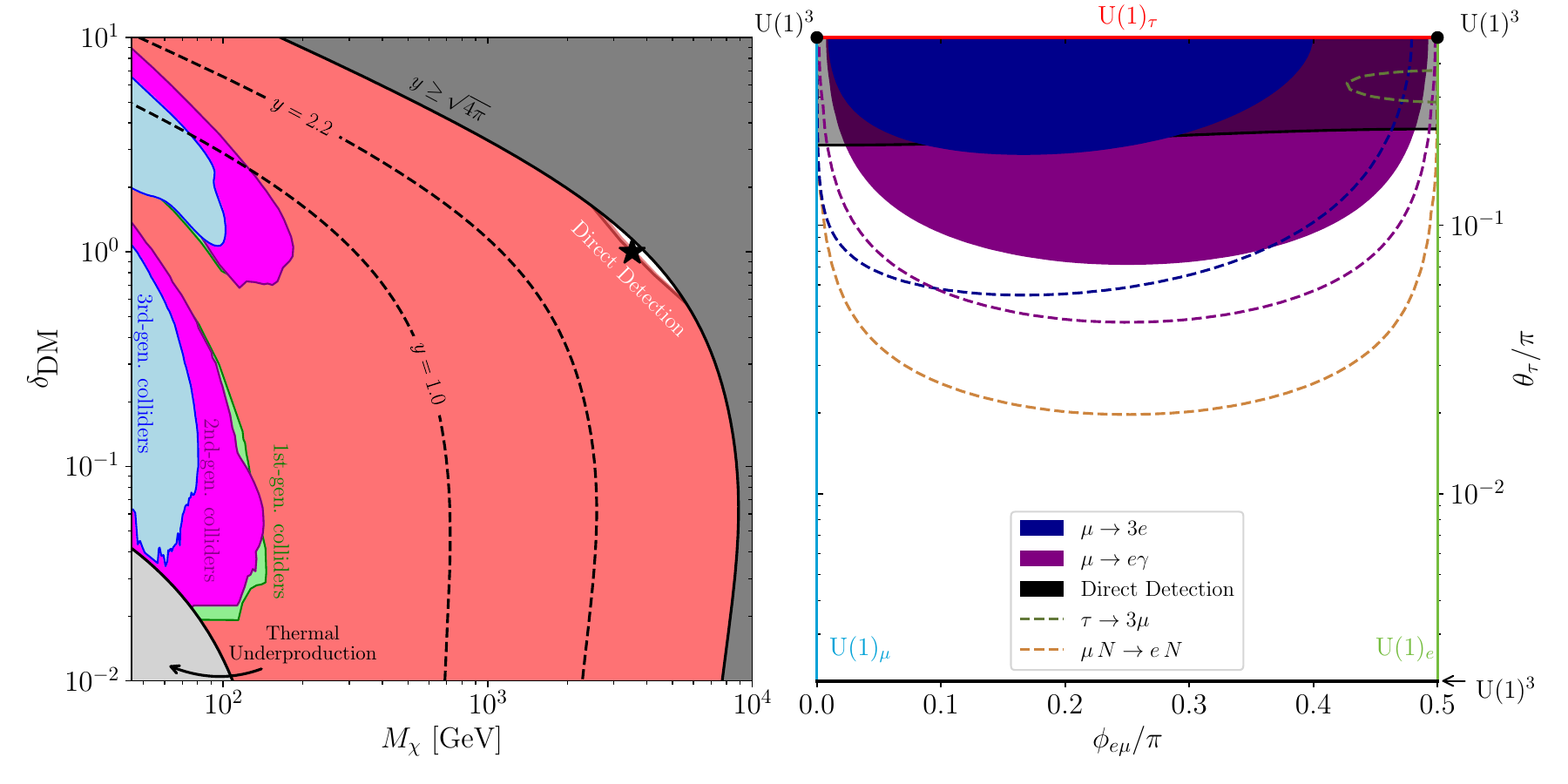}
    \caption{\textbf{Leptophilic Dirac DM}. \textit{Left panel:} Constraints in the $(M_\chi,\delta_{\rm DM})$ plane, with $\delta_{\rm DM}\equiv M_\Phi/M_\chi-1$. Shown are 90\% CL bounds from collider searches (green, purple, and blue, corresponding to couplings to first-, second-, and third-generation leptons, respectively), DD (pink, with darker shades indicating couplings to heavier lepton generations; here the bound is essentially the same for all generations), and relic abundance considerations; representative contours for $y=1$ and $y=2.2$ are shown. The DD limits are evaluated along relic-density contours with varying couplings. Gray regions indicate parameter space where the coupling required to reproduce the observed relic abundance is non-perturbative, or where the correct abundance cannot be achieved via thermal freeze-out due to overdepletion.
    \textit{Right panel}: Constraints from flavour physics and DD in the $(\phi_{e\mu},\,\theta_\tau)$ plane, evaluated at the benchmark point indicated by the black star in the left panel ($M_\chi \simeq 3.5 \text{ TeV},\delta_\text{DM} \simeq 1, y \simeq 3.35 $). Dashed lines denote future projected sensitivities discussed in \cref{sec:lep_flavour}.  See \cref{sec:interplay_leptophilic} for details.
    }
    \label{fig:Dirac_RH_lept}
\end{figure*}

We observe in the right panel that bounds from flavour transitions vanish as one approaches the flavour-symmetric limits (see \cref{fig:sphere}) that forbid the corresponding flavour-violating decay. For instance, $\mu \to e\gamma$ vanishes at $\phi_{e\mu} = 0,\, \pi/2$, where $\mathrm{U}(1)_{\mu}$ and $\mathrm{U}(1)_{e}$ are respectively restored, and analogous behaviour is found for the other transitions. Note that this particular choice of angles selects $\tau$ flavour over $\mu$ and $e$, as the point at the north pole of the sphere in \cref{fig:sphere} is special: for $\theta_\tau = 0$, one gets $\tau$ alignment independently of $\phi_{e\mu} \in [0, 2\pi)$. However, this asymmetry is merely an artefact of the chosen parametrisation; the other two bases are plotted in App.~\ref{app:LFV_directions}, zooming on the other two corners of \cref{fig:sphere}. In most of the parameter space, the dominant constraint arises from $\mu\to e\gamma$, shown in purple.
As a result, for a generic azimuthal angle, \cref{fig:Majorana_RH_lept} shows that $\tau$ alignment must be satisfied at the $10^{-2}$ level. In contrast, \cref{fig:basis_reparametrization} implies significantly stronger bounds for $\mu$- or $e$-alignment, at the level of $10^{-4}$ for the benchmark point considered. As a reminder, the maximal misalignment naturally expected in MFP is $\mathcal{O}(10^{-4})$, as discussed at the beginning of \cref{sec:leptophilic}, indicating that current and near-future experiments are beginning to probe the motivated region of parameter space.

We again remark that as the dark sector spectrum becomes increasingly degenerate, the three-body decays are suppressed, as shown in \cref{fig:loop_functions}. This suppression weakens the corresponding bounds, making the radiative decays the only relevant constraints (see \cref{fig:LFV_scale_bound}). Since radiative decays, both in the Dirac and Majorana DM cases, are only mildly suppressed, and leptonic $t$-channel DM requires relatively low masses, the process $\mu \to e\gamma$ always needs some degree of flavour protection. If the model is aligned in either the $\mu$- or $e$-flavour direction, $\theta_\tau$ may still remain relatively unconstrained, at least in the more degenerate regimes. This region could be probed by future Belle-II searches for $\tau \to \ell\gamma$ decays~\cite{Belle-II:2018jsg, Banerjee:2022xuw, deBlas:2025gyz}, which are expected to explore the parameter space with $\theta_\tau \sim \mathcal{O}(1)$ for both scenarios. Conversely, the model may be aligned in the $\theta_\tau \to 0$ limit, where the full SM lepton flavour symmetry is restored. In this limit, future experimental sensitivities are expected to improve by roughly an order of magnitude in the branching ratio, translating to a marginally stronger bound of $\theta_\tau /  \pi \lesssim 10^{-2}$ for future MEG II projections~\cite{MEGII:2018kmf}.

Finally, this benchmark highlights the complementarity between flavour physics probes and DD. In particular, DD constrains flavour non-universal but flavour conserving directions compatible with $\U(1)_e\times \U(1)_\mu \times \U(1)_\tau$ symmetric case. The universality is broken by lepton masses as explained in \cref{sec:leptoDD}. In contrast, flavour observables are sensitive to the $\mathrm{U}(1)^3$-breaking effects. To show this, in Fig.~\ref{fig:Majorana_RH_lept} we have zoomed in on the two respective $\mathrm{U}(1)^3$ symmetric points corresponding to coupling to the first or second family only. In this case, DD can completely exclude the $e$-specific scenario while imposing no constraint on the $\mu$-specific case; see also \cref{fig:basis_reparametrization}.

\subsubsection{Dirac Dark Matter}

The most recent experimental DD limits basically exclude the case of Dirac DM, as shown in the left panel of \cref{fig:Dirac_RH_lept}. The bounds here are stronger because the DM vector current no longer vanishes, leading to the presence of magnetic dipole and charge-radius operators as discussed in \cref{sec:leptoDD}. These generate significant spin-independent contributions to DM–nucleus scattering, ruling out the entire parameter space except for a small region around the benchmark point at the edge of perturbativity. This conclusion is flavour-independent. For the benchmark point, DD is shown in gray in the right panel of \cref{fig:Dirac_RH_lept}. Notably, flavour constraints continue to require a degree of alignment with symmetry limits to avoid the $\mu\to e\gamma$ constraint.

\section{Quarkphilic Dark Matter} 
\label{sec:quarkphilic}

As a next example, we study a minimal quarkphilic scenario with flavour-singlet $\chi$ and $\Phi$ coupled to the left-handed quark doublets:
\begin{align}
\mathcal{L} \supset -y_i \,  \bar q_{L,i} \chi \Phi  + \textrm{h.c.}.
\label{eq:quarkphilicL}
\end{align}
Since $\bm{y} \sim \bm{3}_q$ of $\U(3)_q$, as before, there is no MFV expansion associated with this coupling (see \cref{tab:U3}), which \textit{a priori} hints at $O(1)$ flavour violation. However, it can be decomposed as $\bm{3}=  \bm{2} \oplus \bm{1}$ under $\mathrm{SU}(2)_q$, in the U(2)$^5$ or MFP framework. Specifically, we can express the expansion according to Table~\ref{tab:U2} in terms of the minimal $\U(2)_q \subset \mathrm{U}(2)^5$-breaking spurions as 
\begin{align}
\bm{y} \sim (a V_q,b)
\label{eq:yquark_U2}
\end{align}
with $\norm{V_q}=O(V_{cb})$ defined in \cref{eq:U2spurions}, and $a,b$ are $O(1)$ couplings. After the diagonalization of the SM quark mass matrices, this generically results in the structure 
$
\bm{y}\sim (\lambda_c^3,\lambda_c^2, 1) 
$
with $\lambda_c$ the Cabibbo angle (see also \cite{Gherardi:2019zil, Marzocca:2024hua}). The same or more suppressed structure could also be realized in the MFP case, depending on the specific $\mathrm{U(1)}_{x_i}$ charge assignments.

To test these flavour symmetry hypotheses, we adopt an agnostic angular parametrisation, analogous to that introduced in the leptophilic case:
\begin{align}
\label{eq:quark_angles}
\bm{y} = y \,  (\sin \theta  \cos \phi\, e^{i\alpha_1}, \sin \theta \sin \phi \,  e^{i\alpha_2}, \cos \theta).
\end{align}
This time we employ the full angular range on the sphere, with $\theta \in [0,\pi)$ and $\phi \in [0,2\pi)$. The parameters $\alpha_{1,2} \in [-\pi/2,\pi/2]$ represent two irreducible physical CP-violating phases, providing the first qualitative difference with respect to the leptophilic model.
Fixing $y = \norm{\bm{y}}$, the expansion \cref{eq:yquark_U2} implies, generically, that $\cos \phi \sim \lambda_c$ and $\sin \theta \sim \lambda_c^2$, which showcases the typical pattern associated to the $\U(2)$ limit. 

We choose to define the angular parametrization in \cref{eq:quark_angles} in the down-aligned mass basis, namely in the basis in which the down quark Yukawa matrix is diagonal, which can always be achieved~\cite{Greljo:2022cah}. In this basis, the vector $\bm{y}$ in \cref{eq:quark_angles} should be identified with the coupling to down quarks. The corresponding couplings to up quarks are then fixed by the CKM matrix. Explicitly, after diagonalization of the SM Yukawas, one finds
\begin{align}
\begin{aligned}
    \bm{y}_{\text{down}} &= \bm{y}, \\
    \bm{y}_{\text{up}} &= V_{\text{CKM}} \, \bm{y}.
\end{aligned}
\label{eq:quark_angles_mass}
\end{align}
As a consequence, the interactions with up and down quarks are generically misaligned. This misalignment has important phenomenological implications, as we will see later. It also reflects the fact that, once again, only an $\SU(2)_q$ symmetry on the first two generations provides a meaningful organizing principle for left-handed quark couplings. Indeed, alignment along any of the light generations, corresponding to a $\U(1)$ symmetry associated with light left-handed quark flavour numbers, is not well motivated: already in the SM such symmetries are significantly broken by the Cabibbo angle, see \cref{eq:quark_angles_mass}.
In contrast, in the leptophilic scenario of \cref{sec:leptophilic}, assigning individual lepton flavour numbers is consistent, since treating neutrinos as effectively massless is a good approximation in experiments probing charged-lepton flavour violation.

\subsection{Relic Abundance}

The discussion of thermal freeze-out for DM interacting with quarks shares many similarities with the leptophilic scenario. As in that case, one must track the abundances of both the DM and the mediator. The main difference is that the scalar $\Phi$ now carries full SM gauge charges, introducing several additional annihilation channels for the dark sector states.

A first qualitative difference is that quarks in the final state of the annihilation processes cannot generally be treated as massless. This is particularly relevant for the top quark, whose mass is non-negligible when the DM and mediator masses are of order $\mathcal{O}(1)$ TeV. We anticipate that, for typical DM masses compatible with the observed relic density and still allowed by DD, collider searches, and flavour observables, the ratio $m_t/M_\chi$ is in practice relatively small. Nevertheless, it is instructive to highlight that in the quarkphilic case the DM relic abundance becomes \emph{flavour-dependent}, and in particular depends non-trivially on the mixing angle $\theta$.

For illustration, we focus on the Majorana DM pair annihilation process $\chi \chi \to q_i \bar q_j$, with $i, j = 1, 2, 3$ (we again drop the subscript indicating chirality of quarks to avoid clutter.). The corresponding cross sections can be organized into a symmetric $3 \times 3$ matrix with six independent entries. Considering only the massive top quark, the $2 \times 2$ block associated with the first and second generations can be computed in the massless limit, where the leading nonvanishing contribution arises from the $p$ wave at order $\vrel^2$. In contrast, when a third-generation quark, namely its up-type component, enters the final state, the cross section acquires a nonvanishing $s$-wave contribution.
Summing over quark generations yields the following result
\begin{align}
 &   \sum_{i,j=1}^{3} \sigma \vrel (\chi \chi \to q_i \bar q_j)
= 2 \sin^4 \theta \,  \sigma_{\chi \chi}^{m_q=0} \vrel \nonumber
\\
& ~~~+ 2  \cos^2 \theta \, \sin^2 \theta \left(   \sigma_{\chi \chi}^{m_q=0} \vrel  + \sigma^{m_q}_{\chi \chi} \vrel \right)
\nonumber
\\
& ~~~+  \cos^4 \theta \, \left(   \sigma_{\chi \chi}^{m_q=0} \vrel  + \sigma^{2 m_q}_{\chi \chi} \vrel \right) \, .
\label{cross_flavour_chi_chi_quark}
\end{align}
The first line of eq.~\eqref{cross_flavour_chi_chi_quark} corresponds to the first two generations, and the overall factor of two accounts for the $\SU(2)_\text{L}$ multiplicity. The second line refers to the non-diagonal entries involving quarks of the first, second, and third generations; here the overall factor of two counts the two possible quark pairs in the final state, $q_i \bar q_3$ and $q_3 \bar q_i$. Finally, the third line accounts for the third generation, where one quark is treated as massless (the bottom quark) and the other as massive (the top quark). Accordingly, $\sigma_{\chi \chi}^{m_q=0} \vrel$, $\sigma_{\chi \chi}^{m_q} \vrel$, and $\sigma_{\chi \chi}^{2 m_q} \vrel$ denote the cross sections with zero, one, or two massive quarks in the final state, respectively. Their explicit expressions are given in the Appendix \ref{app:Cross_section_DM}. In the massless limit, the sum of the cross sections in \cref{cross_flavour_chi_chi_quark} reduces to the flavour-blind result found in the leptophilic scenario up to a color factor (see Eqs.~\eqref{sigma_chi_chi_mass0_Maj} and \eqref{sigma_chi_chi_mass0_Dir} for explicit expressions). Accounting for the flavour dependence of the cross sections leads to a modest but noticeable effect on the relic density contours, as illustrated for the Majorana DM scenario in the left panel of \cref{fig:Majorana_LH_quark} for $\theta = 0$ and $\theta = \pi/2$. A dependence on the quark mass is also induced in the process $\Phi \Phi \to q_i q_j$ in the Majorana dark matter scenario, whose expression is given in Appendix~\ref{app:Cross_section_DM}. For the remaining cross sections, results are available in earlier studies \cite{Garny:2015wea,Ibarra:2015nca,Baker:2015qna,Biondini:2018ovz}.  

The second aspect that requires particular care in the quarkphilic scenario is the treatment of near-threshold effects. Due to the interaction of the mediator with QCD gluons, Sommerfeld and bound-state effects are more pronounced than in the leptophilic scenario. Coannihilations of colored particles have been extensively studied in the literature \cite{DeSimone:2014qkh, Ellis:2014ipa, Ibarra:2015nca, ElHedri:2017nny}, and more recently $t$-channel models have been employed as a test bed to scrutinize bound-state effects and their impact on the relic density of the accompanying DM particle~\cite{Garny:2018ali, Biondini:2018pwp, Biondini:2018ovz, Binder:2023ckj, Garny:2021qsr,Becker:2022iso}. The Sommerfeld factors are largely dominated by QCD-induced interactions, with electroweak contributions typically providing only a subleading correction in the parameter space relevant to the models considered here.\footnote{In general, whenever the mediator interacts with QCD gluons, the numerical importance of QCD-assisted coannihilations over the corresponding electroweak contributions can be appreciated by inspecting the mediator--mediator annihilation cross sections, which scale as $g_{\mathrm{SM}}^4$ \cite{Bodwin:1994jh,Vairo:2003gh}. Moreover, Sommerfeld enhancements and bound-state effects become increasingly important for larger values of the coupling constants that induce the long-range interactions \cite{vonHarling:2014kha,Petraki:2015hla,Hisano:2004ds}.
} 

Scalar pair annihilations can occur in a color singlet, color octet, or color sextet configuration (the latter being relevant only for the Majorana DM option), with an attractive potential in the singlet case and repulsive potentials in the other two cases. The Sommerfeld factors multiply the corresponding color-decomposed annihilation channels; e.g.~see~\cite{ElHedri:2016onc, Biondini:2018pwp}. Bound state formation can occur via two main processes, namely radiative emission of a gluon~\cite{Mitridate:2017izz,Harz:2018csl, Biondini:2023zcz} and scattering with light plasma constituents~\cite{Biondini:2018pwp, Binder:2019erp, Biondini:2025jvp}. Despite exhibiting a larger rate \cite{Biondini:2018pwp,Binder:2019erp}, the latter has been found to give a quite small correction to the DM energy density \cite{Biondini:2025jvp}. 

To go beyond the minimal inclusion of bound-state effects, namely considering only the ground state, we also include excited states. Their impact can be sizeable~\cite{Garny:2021qsr, Binder:2021vfo, Binder:2023ckj, Biondini:2023zcz}, particularly depending on the number of excited states included. In this work, we consider four additional states: the $2S$ state and the $2P_m$ states with $m = -1, 0, 1$. While further improvements are possible by including more excited states, this quickly becomes numerically demanding and lies beyond the scope of the present study.\footnote{A very recent work has proposed a numerical tool and public code for incorporating bound-state effects \cite{Becker:2025vgq}, including the possibility of accounting for excited states.} For the quarkphilic case with Majorana DM, neglecting Sommerfeld and bound-state effects leads to order-one differences in the relic density prediction. For instance, for $y = 1.0$ and $\delta_{\mathrm{DM}} = 0.2\,(0.05)$, one finds corrections of about 50\% (up to a factor of $\sim 2.5$) in the DM mass range relevant for this work.

\subsection{Collider Searches}

Similarly to the leptophilic case, collider searches for squarks decaying into SM quarks and neutralinos can be directly applied here. We therefore employ the results from LEP and, most importantly, from the LHC for searches of light-flavour squarks, sbottoms, and stops~\cite{ATLAS:2024lda, ATLAS:2017bfj, CMS:2019ybf, CMS:2019zmd, ATLAS:2020syg, ATLAS:2021kxv, ATLAS:2020dsf, ATLAS:2020xzu, ATLAS:2024rcx, ATLAS:2021yij, ATLAS:2021yij, CMS:2019ybf, CMS:2025ttk}. The light-flavour squarks searches employ untagged jets, so these searches are eventually only sensitive to  the angle $\theta$.
Since the dominant production mechanism is QCD Drell–Yan pair production, we rescale the experimental limits for several squark flavours to the number of flavours relevant for our setup by adjusting the production cross section, using the cross-sections in Ref.~\cite{Dorsner:2018ynv}. As in the leptophilic case, there also exists a $t$-channel diagram for production through couplings to first generation quarks; however, this contribution is subleading for $y \lesssim g_s$, and a complete cancellation with the $s$-channel process is anyway not possible, so we neglect it. This renders our bounds somewhat conservative.

\begin{figure*}[t]
    \centering
    \includegraphics[width=\linewidth]{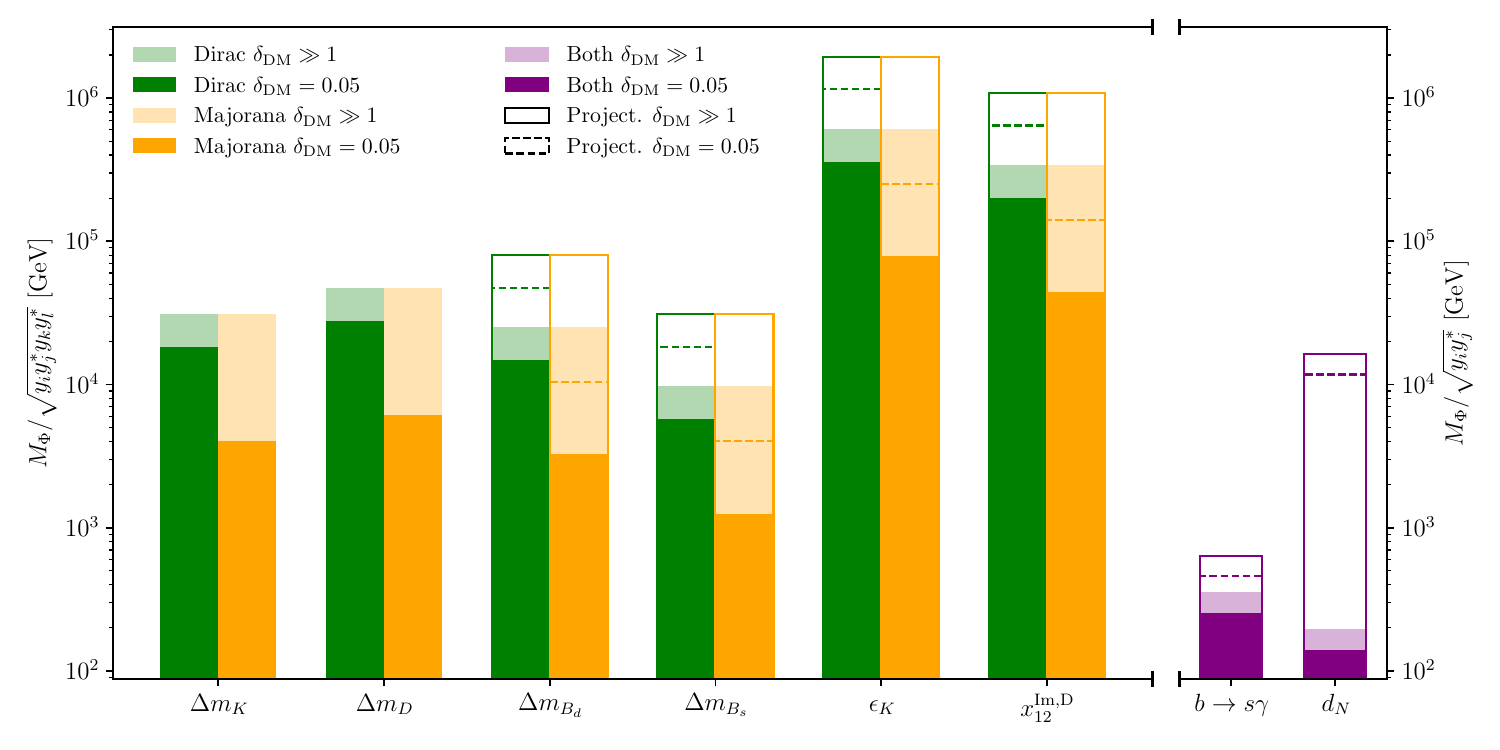}
    \caption{\textbf{Quarkphilic DM}. Bounds on the mediator mass $M_\Phi$ from quark-flavour-violating processes, with the dependence on the couplings $y_i$ factored out. On the left hand-side of the plot we report meson-mixing observables, where the bounds refer to the real part of the flavour coefficients for $\Delta m_{K,\, D}$, to the absolute value for $\Delta m_{B_d, B_s}$ and the imaginary for $\epsilon_K$ and $x^{\mathrm{Im}, D}_{12}$. Different shades of the colours correspond to different values of $\delta_\text{DM} = (M_\Phi-M_\chi)/M_\Phi$. Note that $\delta_\text{DM} \gg 1$ corresponds to the limit $M_\chi/M_\Phi \ll 1$ while keeping $M_\Phi$ fixed. Shown on the right are the constraint from $b \to s \gamma$ and $d_N$. Dashed (solid) lines correspond to future projections for $\delta_{\rm{DM}} = 0.05$ ($\delta_{\rm{DM}}\gg 1$) following~\cite{deBlas:2025gyz}.}
    \label{fig:QFV_scale}
\end{figure*}

\subsection{Direct (and Indirect) Detection}
\label{sec:quark_DD}

DD constraints depend on the structure of the coupling vector $\bm{y}$. Couplings to first-generation quarks ($u,\, d$) allow for tree-level scattering off nuclei, leading to very strong constraints. For couplings to third-quark generation, the leading one-loop contributions arise from photon, $Z$, and Higgs penguin diagrams, as in the leptophilic case, together with new box diagrams that generate effective couplings to gluons~\cite{Ibarra:2015fqa}. Since the $Z$ penguin contributions to DM-nucleus scattering scale as $(m_q/M_\chi)^2$, these effects become particularly relevant for couplings to the top quark. 

Regarding the second quark generation, an often overlooked feature in literature is the misalignment implied by \cref{eq:quark_angles_mass}. This offset has important consequences for DD, as one cannot realise a pure single-generation alignment. In particular, choosing second family alignment, $(\phi, \theta)=(\pi/2,\pi/2)$, leads to up quark couplings of the form
\begin{equation}
\bm y_{\text{up}} \sim (\lambda_c, 1, \lambda_c^2)\,, 
\end{equation}
which contains a sizeable first-generation component. This induces a tree-level contribution to DD, leading to very strong constraints on second-family alignment. An analogous discussion, with the roles of up and down quarks interchanged, would apply if one instead chose a basis in which the up quark Yukawa matrix is diagonal.

In the Majorana case, several of the diagrams cancel due to the vanishing DM vector-current $\bar \chi \gamma^\mu \chi = 0$ and magnetic dipole $\bar \chi \sigma^{\mu\nu} \chi = 0$, resulting in weaker bounds than in the Dirac scenario. In this case, the leading constraints are due to the effective DM-gluon couplings, except for couplings to first-generation quarks, where the spin-dependent cross section sets strong bounds~\cite{Arcadi:2023imv}. With third-family coupling, the Higgs exchange is enhanced due to the large top mass; see, for instance, Ref.~\cite{Biondini:2018ovz}.

To obtain the limits, we take the results of \cite{Gondolo:2013wwa, Ibarra:2015fqa} for the analytical expressions of the various contributions, and once again follow the procedure of \cite{Hisano:2018bpz}, using the most recent results from LZ \cite{LZ:2024zvo} and the DARWIN projections \cite{DARWIN:2016hyl}. In the Majorana case, the vanishing of vector currents simplifies the procedure, as the dependence on the recoil energy essentially reduces to that employed by the experimental collaborations. As a result, we can directly use the bounds on the spin-independent and spin-dependent nuclear cross sections.

Finally, indirect detection probes DM through present-day annihilations, 
notably via $\gamma$-ray observations of dwarf spheroidal galaxies and the 
Galactic centre, as well as measurements of the antiproton flux ~\cite{Garny:2015wea, Kopp:2014tsa, Garny:2018icg, Cirelli:2013hv}. For $t$-channel 
scenarios, current indirect detection bounds are generally weak and remain 
subdominant to present and projected DD sensitivities~\cite{Garny:2015wea, Arina:2025zpi, Olgoso:2025jot, ElHedri:2016onc}.

\subsection{Flavour Observables}
\label{sec:}

Quark flavour-changing neutral currents are sensitive probes of NP. Unlike lepton flavour violation, they are suppressed but non-vanishing in the SM, reflecting the approximate nature of quark flavour symmetries. Neutral meson mixings, in particular, are measured with extraordinary precision, making them among the most sensitive tests of NP. Their interpretation, therefore, relies on accurate SM predictions, which are often limited by hadronic uncertainties. In $B$-meson systems, these arise mainly from lattice determinations of hadronic matrix elements, while in kaon observables such as $\epsilon_K$, additional uncertainties from CKM parameters are relevant.

In our $t$-channel DM scenario, meson mixing is generated at one loop via box diagrams. \cref{fig:QFV_scale} shows the resulting bounds on the mediator mass $M_\Phi$, with the dependence on the couplings $y_i$ factored out, as derived from various flavour observables; details of the extraction are provided in \cref{app:FV_observables}. Different observables probe different components of the flavour coefficients: $\epsilon_K$ and $x_{12}^{\mathrm{Im},D}$ constrain the imaginary parts of the $\bm{y}$ vector, $\Delta m_{K,D}$ are sensitive to the real parts, while $\Delta m_{B_q}$ and $\phi_q$ depend on both through interference effects. Since the bound from the latter observables are numerically similar, we report in \cref{fig:QFV_scale} only the one from $\Delta m_{B_q}$. In contrast to the LFV case shown in Fig.~\ref{fig:LFV_scale_bound}, bounds from flavour-violating dipole transitions such as $b\to s\gamma$ are significantly weaker than those induced by four-quark operators. The dependence on the mass splitting is shown for two benchmark scenarios. In the Majorana case, constraints from neutral meson mixing become weaker in the small mass-splitting limit, as implied by the behavior of the loop functions shown in \cref{fig:loop_functions}. For future projections, we use Fig.~5.15 of Ref.~\cite{deBlas:2025gyz} to estimate the improvement factor on the effective scale shown in \cref{fig:QFV_scale} for the observables considered, based on the anticipated full datasets in flavor physics, including FCC-ee measurements in the $b$- and $\tau$-sectors. We add a note of caution that progress in lattice calculations, which is difficult to anticipate, may significantly alter these projections.

Other $\Delta F = 1$ observables beyond dipole transitions are not relevant for our analysis. The reason is that the contributions to the Wilson coefficient $C_{10}$ cancel once all diagrams are consistently included as a consequence of gauge invariance, see Eq.~(5.17) of~\cite{Davighi:2023evx}. As a result, potentially stringent probes such as $B_s \to \mu^+ \mu^-$ do not receive a sizeable contribution. Other flavour observables yield only subleading constraints compared to those already imposed by $b \to s \gamma$, and are therefore not reported. Note that in our framework, we cannot fit $P_5'$ because of constraints due to other observables; see \cref{app:FV_observables} for more details. 

\begin{figure*}[tb!]
    \centering
    \includegraphics[width=\linewidth]{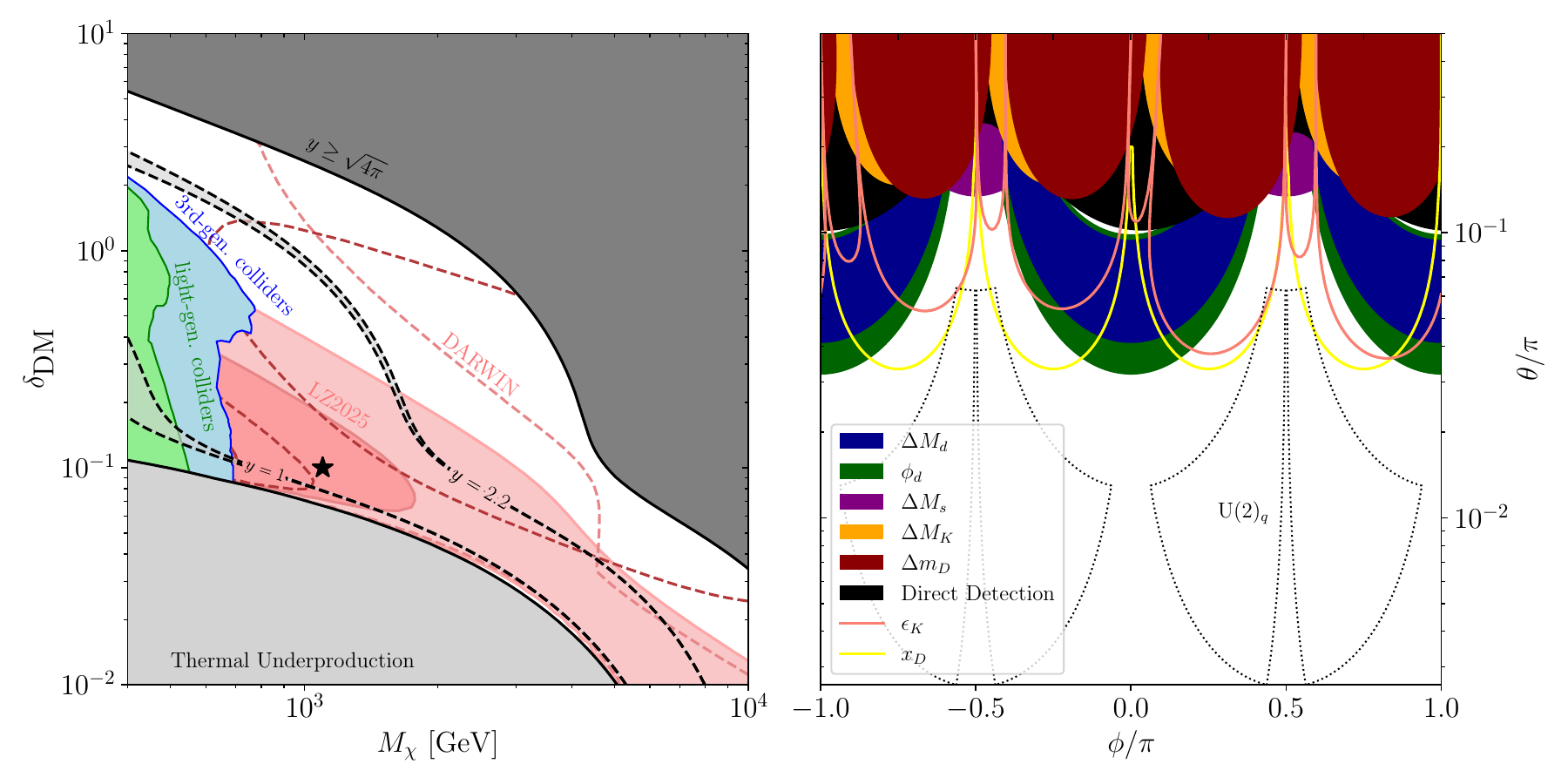}
    \caption{
    \textbf{Quarkphilic Majorana DM}. \emph{Left panel}: Constraints in the $(M_\chi,\delta_{\rm DM})$ plane, with $\delta_{\rm DM}\equiv M_\Phi/M_\chi-1$. Shown are 90\% CL bounds from collider searches (green and blue, corresponding to couplings to light and third-generation quarks, respectively), DD (pink, with darker shades indicating couplings to heavier quark generations), and relic abundance considerations; representative contours for $y=1$ and $y=2.2$ are shown. The DD limits are evaluated along relic-density contours with varying couplings. Gray regions indicate parameter space where the coupling required to reproduce the observed relic abundance is non-perturbative, or where the correct abundance cannot be achieved via thermal freeze-out due to overdepletion. Dashed lines denote projected sensitivities from DARWIN. \emph{Right panel}: Constraints from flavour physics and DD in the $(\phi, \theta)$ plane, corresponding to the benchmark point indicated by a black star in the left panel ($M_\chi \simeq 1.1 \text{ TeV},  \delta_\text{DM} \simeq 0.1, y \simeq 1.38$). The dotted-black contours encompass regions compatible with a $\U(2)_q$ spurion expansion, using $a,b \in [\lambda_c,5]$  in \cref{eq:yquark_U2}, as in Ref.~\cite{Gherardi:2019zil}.
    See \cref{sec:majoranaquark} for details. }
    \label{fig:Majorana_LH_quark}
\end{figure*}

Finally, we comment on bounds from EDMs. These are also sourced from the (chromo-)dipole operators. The rotation to the mass basis induces an electric dipole of up, down, and strange quarks of order $d_q/e\sim m_q \lambda_C \Im[y_1 y_2^*]/16\pi^2 M_\Phi^2$ and similarly for the chromo-electric dipole, see \cref{app:FV_observables}. With our specific down-aligned choice \cref{eq:quark_angles_mass}, only $d_u$ is induced, and leads to the weak constraint $M_\Phi/\sqrt{|\text{Im}y_1 y_2^*|} \gtrsim 200$ GeV at best.
Even for ${\cal O}(1)$ CP-violating phases, the current bound is weaker compared to all the other ones we discussed. However, future projections on EDMs, in particular of the proton, are expected to improve the bound on the overall scale by almost two orders of magnitude \cite{Alarcon:2022ero, n2EDM:2021yah, pEDM:2022ytu, deBlas:2025gyz}. Such an improvement could bring the resulting constraint close to that from other flavour observables, as shown in~\cref{fig:QFV_scale}.

\subsection{Interplay and Summary}

Our main results are presented in \cref{fig:Majorana_LH_quark} and \cref{fig:Dirac_LH_quark} for Majorana and Dirac DM, respectively. The plots follow the same conventions as in the leptophilic case: the left panels show constraints from relic abundance, DD, and collider searches in the $(M_\chi,\delta_{\rm DM})$ plane. The gray region in the left panels corresponding to thermal underproduction is more extended in this case, since in the coannihilation regime the strong interactions of the mediator dilute the DM abundance very efficiently. Bounds from collider searches are also stronger than in the leptophilic case, excluding colored mediator masses $m_\Phi \lesssim 600$ GeV. The right panels show flavour bounds in the $(\phi,\, \theta)$ plane for the specific benchmark point marked by a star in the left panels, that we discuss in detail in the following.

\subsubsection{Majorana Dark Matter}
\label{sec:majoranaquark}

DD and flavour observables provide complementary but mutually reinforcing constraints. As shown in the left panel of Fig.~\ref{fig:Majorana_LH_quark}, DD efficiently excludes large regions of parameter space associated with first- and, to a lesser extent, second-generation couplings, while remaining largely insensitive to a dominantly third-generation alignment. The right panel demonstrates that flavour observables independently point to the same conclusion: generic viable scenarios are driven towards a predominantly third-family structure, allowing only small misalignments. Future improvements in DD sensitivity will play an important role; see DARWIN projections in the left panel.

The right panel of Fig.~\ref{fig:Majorana_LH_quark} focuses on a benchmark point $ M_\chi = 1.1\,\text{TeV},
\delta_{\rm DM} = 0.1, 
y = 1.4$ indicated by a star in the left panel, for which DD already requires a degree of third-family alignment.
Flavour observables confirm the need for a coupling structure aligned predominantly with third-generation quarks, with different transitions probing complementary regions of parameter space. Kaon and $D$-meson mixing constrain angles around $\phi \simeq \pi/4$, where first- and second-generation transitions are maximized, while transitions involving the third generation are most strongly constrained near $\theta \simeq \pi/4$. The strongest limits from $B_d$ mixing arise close to $\phi = 0$, whereas $B_s$ mixing peaks around $\phi = \pi/2$.

\begin{figure*}[th]
    \centering
    \includegraphics[width=\linewidth]{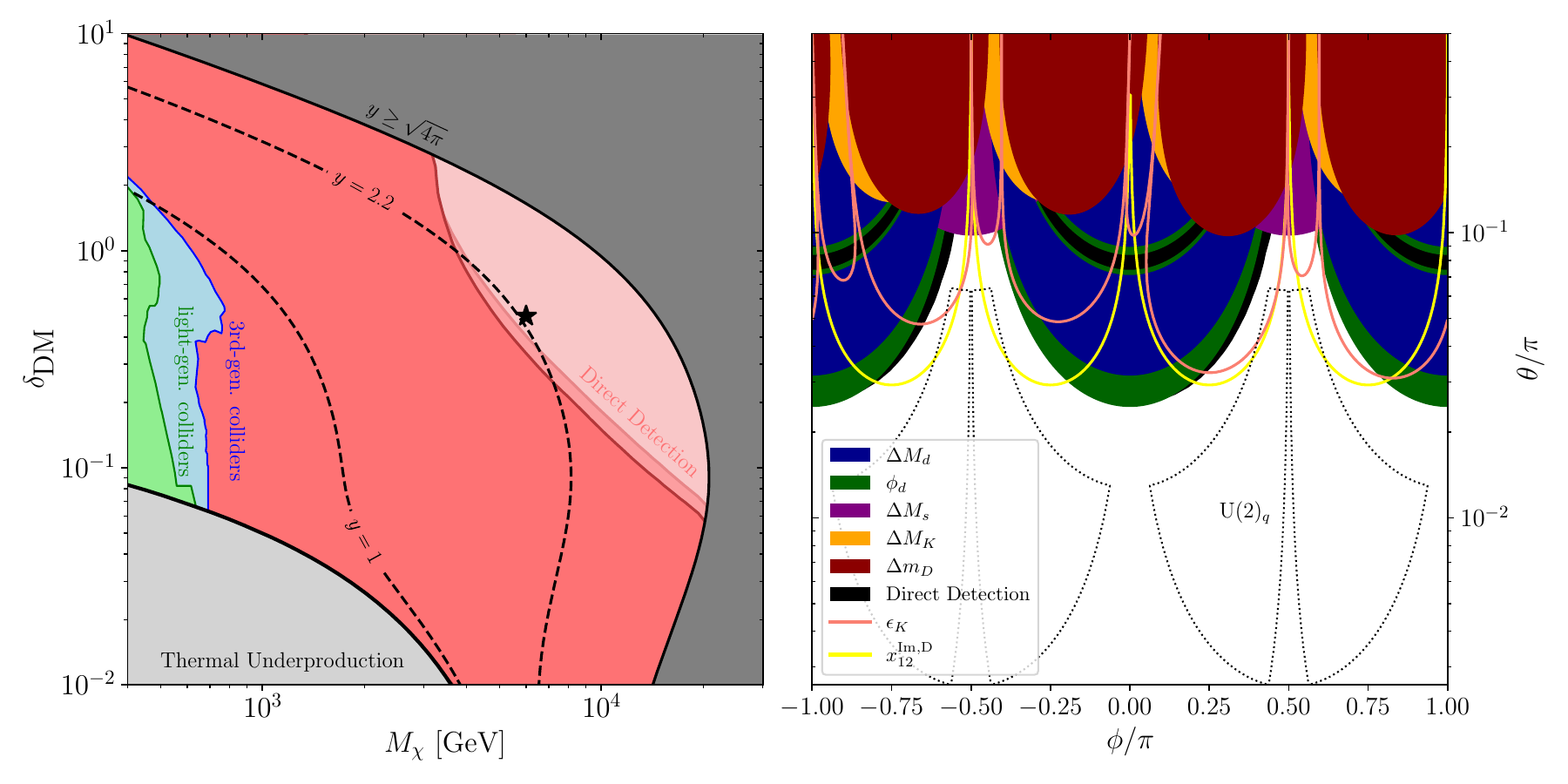}
    \caption{
    \textbf{Quarkphilic Dirac DM}. \emph{Left panel}: Constraints in the $(M_\chi,\delta_{\rm DM})$ plane, with $\delta_{\rm DM}\equiv M_\Phi/M_\chi-1$. Shown are 90\% CL bounds from collider searches (green and blue, corresponding to couplings to light and third-generation quarks, respectively), DD (pink, with darker shades indicating couplings to heavier quark generations), and relic abundance considerations; representative contours for $y=1$ and $y=2.2$ are shown. The DD limits are evaluated along relic-density contours with varying couplings. Gray regions indicate parameter space where the coupling required to reproduce the observed relic abundance is non-perturbative, or where the correct abundance cannot be achieved via thermal freeze-out due to overdepletion. \emph{Right panel}: Constraints from flavour physics and DD in the $(\theta,\phi)$ plane, corresponding to the benchmark point indicated by a black star in the left panel ($M_\chi \simeq 6 \text{ TeV},  \delta_\text{DM} \simeq 0.5, y \simeq 2.26$). The dotted-black contours encompass regions compatible with a $\U(2)_q$ spurion expansion. See \cref{sec:diracquark} for details.
    }
    \label{fig:Dirac_LH_quark}
\end{figure*}

A crucial point is that the $K$- and $D$-meson constraints are not aligned. This follows from the CKM rotation between the up and down quark sectors, \cref{eq:quark_angles_mass}. As a consequence, the $\U(1)$ limits corresponding to exact alignment with the first or second generation are not viable in the quark sector, since the inevitable rotations proportional to the Cabibbo angle induce excessively large contributions to kaon and $D$-meson mixing. This resembles what happened also for DD in \cref{sec:quark_DD}.\footnote{Instead, for DM coupled to right-handed $u_R$ or $d_R$, the MFP framework of \cref{sec:MFP} suggests that the corresponding $\U(1)$ limits could be viable, similarly to the leptophilic case, since it would correspond to removing either up- or down-type of meson-mixing observables in \cref{fig:Majorana_LH_quark,fig:Dirac_LH_quark}.}

In the right panel of Fig.~\ref{fig:Majorana_LH_quark}, we also highlight the $U(2)_q \subset \U(2)^5$ region compatible with the SM spurion expansion \cref{eq:yquark_U2}, allowing for $\mathcal{O}(1)$ variations, as defined in Ref.~\cite{Marzocca:2024hua}. Most of this region remains allowed, demonstrating that a $\U(2)^5$ flavour symmetry is compatible with the $t$-channel DM framework. The same conclusion holds in the MFP framework, with third-family misalignment being equally or more suppressed, depending on the $\U(1)_{x_i}$ charge assignment. This behaviour is generic across the parameter space, except in the degenerate limit $\delta_{\rm DM} \ll 1$, where loop-function suppressions of the relevant SMEFT coefficients permit larger $\U(2)_q$ breaking from the minimal spurion scenario. This contrasts with leptophilic $t$-channel models, where dipole-induced decays dominate and remain unsuppressed in the degenerate limit.

Finally, we assess the impact of CP-violating phases. Choosing $\alpha_2 - \alpha_1 = \pi/4$, which maximises contributions to light-meson observables through $\mathrm{Im}(M_{12}) \propto \sin[2(\alpha_2 - \alpha_1)]$, leads to a substantial strengthening of the bounds. This is consistent with Fig.~\ref{fig:QFV_scale}, where CP-violating constraints are roughly an order of magnitude stronger than CP-conserving ones. Nevertheless, the plot demonstrates that the $\U(2)_q$ flavour-spurion structure is sufficient, without the need to impose an additional CP symmetry.

Overall, the case of Majorana DM with a left-handed quarkphilic mediator represents an interesting scenario, exhibiting a non-trivial interplay between DD and flavour bounds. Our results confirm the expectation that a certain degree of alignment with the third generation is required in most of the parameter space, which can be naturally embedded in a $\U(2)^5$ or MFP flavour framework. 

\subsubsection{Dirac Dark Matter}
\label{sec:diracquark}

The quarkphilic Dirac case is significantly more constrained. As shown in \cref{fig:Dirac_LH_quark}, DD bounds are extremely strong in this scenario, to the extent that models with couplings to first-generation quarks are essentially excluded. Alignment towards the heavier generations is also strongly constrained, due to the enhanced $Z$-penguin and dipole contributions to DM-nucleus scattering in the Dirac case.
As a result, only a small region of parameter space remains viable, corresponding to large couplings close to the perturbativity bound ($y \gtrsim 2$). Such large couplings are prone to the development of Landau poles at relatively low scales~\cite{Olgoso:2025jot}, suggesting the need for a low-scale completion of the model.

The flavour bounds are shown in the right panel of \cref{fig:Dirac_LH_quark} for a benchmark point $M_\chi = 1.1~\text{TeV}$, $\delta_{\rm DM} = 0.1$, and $y = 1.38$. The considerations are similar to those in the Majorana case: flavour observables require a degree of alignment towards the third generation, as naturally realized in $\U(2)^5$ or MFP frameworks. A difference at small mass splittings, originating from the fact that for Dirac DM the loop functions entering meson mixing remain non-vanishing even in the degenerate limit, is phenomenologically irrelevant here, since this region of parameter space is already excluded by DD constraints.

In summary, the case of Dirac DM with a left-handed quarkphilic mediator is almost excluded by DD, except for scenarios with couplings close to the non-perturbative regime.

\section{Conclusion} 
\label{sec:conclusion}

In this work, we have examined thermal DM in $t$-channel models through the lens of flavour physics. These scenarios constitute a well-motivated and minimal extension of the WIMP paradigm, featuring renormalisable interactions between DM, a mediator, and SM fermions. Reproducing the observed relic abundance naturally points to TeV-scale masses and $\mathcal{O}(1)$ couplings. As a result, such models generically induce flavour- and CP-violating effects, bringing them under the stringent scrutiny of precision flavour observables. Ensuring consistency with existing flavour bounds, therefore, emerges as a central consistency requirement for $t$-channel DM.

We addressed this question systematically by adopting the framework of flavour symmetries and spurion expansions presented in \cref{sec:charting}. By classifying the DM and mediator fields and their couplings under different flavour symmetry hypotheses, we provided a structured map of viable flavour scenarios for $t$-channel models. This generalises the logic of MFV, provides a clear dictionary for identifying which flavour structures can realistically be consistent with existing bounds, and establishes a basis for future comprehensive studies of flavour in dark sectors. In practice, the flavour-singlet entries in \cref{tab:U3}, \cref{tab:U2}, and \cref{tab:MFP} serve as bottom-up simplified-model benchmarks for further investigation. As summarised in \cref{fig:MFV_to_MFP}, progressively reducing the flavour symmetry,
\begin{equation}
    \text{MFV} \;\to\; \U(2)^5 \;\to\; \text{MFP}\,,
\end{equation}
opens up a substantially broader class of viable $t$-channel DM scenarios.

To quantitatively test these symmetry expectations, we focused on simple yet representative benchmark scenarios in which both the DM particle and the mediator are flavour singlets. This choice leads to \emph{rank-1 flavour violation} in the couplings, parameterised by angular variables in \cref{eq:angles_lept} and \cref{eq:quark_angles}, and admitting a convenient geometric representation on the sphere shown in \cref{fig:sphere}. Symmetry limits then correspond to specific points or trajectories on this sphere. Rather than imposing these limits a priori, we performed an agnostic phenomenological analysis, using the full set of available experimental constraints to determine how closely viable models must align with flavour symmetry directions. We studied two concrete realisations: interactions with right-handed charged leptons in \cref{sec:leptophilic} and with left-handed quarks in \cref{sec:quarkphilic}, considering both Dirac and Majorana DM.

In the leptophilic scenario, Majorana DM remains largely viable and mostly unexplored, as illustrated by the left panel of \cref{fig:Majorana_RH_lept}, where sizable regions of parameter space survive all current constraints. As anticipated, the allowed flavour directions are organised by approximate lepton flavour number symmetries: viable scenarios require alignment with a single lepton flavour, with couplings to the third generation being somewhat less constrained, as shown in the right panel of \cref{fig:Majorana_RH_lept}. By contrast, the Dirac leptophilic case, summarised in \cref{fig:Dirac_RH_lept}, is essentially excluded by DD for all flavour directions. This conclusion relies critically on the latest results from LUX–ZEPLIN~\cite{LZ:2024zvo}.

In the quarkphilic scenario, we observe a rich interplay between flavour observables and DD. For Majorana DM, summarised in \cref{fig:Majorana_LH_quark}, a sizeable region of parameter space remains viable. As anticipated, flavour constraints demand alignment with the third quark generation, pointing to an underlying $\SU(2)_q$ flavour symmetry. Frameworks such as $\U(2)^5$ or MFP naturally realise this alignment, efficiently suppressing flavour violation through small spurions while still reproducing the observed relic abundance, as illustrated in the right panel of \cref{fig:Majorana_LH_quark}. By contrast, the Dirac quarkphilic scenario is far more constrained as shown in \cref{fig:Dirac_LH_quark}: DD limits are particularly stringent and require both a highly restrictive flavour structure, again well captured by an $\SU(2)_q$ symmetry, and a large third-generation coupling close to the perturbativity limit. 

For both scenarios considered, and in particular for the quarkphilic case, the inclusion of Sommerfeld enhancements and bound-state effects is essential to obtain an accurate determination of the relic density when the mass splitting between the DM and the mediator is moderate or small.

Overall, our results highlight a clear picture: $t$-channel DM models generically require a non-trivial flavour structure to evade flavour bounds, and flavour symmetries presented in \cref{sec:charting} provide a well-motivated guiding principle to identify viable scenarios. At the same time, DD experiments play a crucial complementary role by probing the flavour-conserving components of the same couplings. In this sense, flavour physics points to the allowed directions in flavour space, while DD constrains their overall size. 

With the progress in DD experiments, the landscape of $t$-channel fermionic DM models is coming into sharper focus: Dirac DM has already been excluded or pushed to the brink of viability, thereby shifting attention to Majorana DM in the next generation of experiments. Precision flavour measurements at present and future colliders~\cite{deBlas:2025gyz} will continue to probe ever smaller departures from flavour symmetry limits. At the same time, collider searches at the HL-LHC and electroweak precision measurements at FCC-ee~\cite{Olgoso:2025jot} will provide complementary probes. Together, this multi-front experimental programme will test the viability of thermal DM at the TeV scale.

\section*{Acknowledgments}
We thank Daniel Naredo-Tuero, Pablo Olgoso, and Stefan Vogl for helpful discussions. This work was supported by the program “Swiss
High Energy Physics for the FCC” (CHEF). \\

\appendix

\onecolumngrid
\renewcommand{\thesubsection}{\thesection\arabic{subsection}}

\section{SMEFT Matching and Flavour Observables}

In this section, we report the relevant one-loop matching expressions of the benchmark models to the effective operators inducing quark and lepton flavour observables, as well as the formulae of the observables considered in this analysis.

\subsection{Matching to the SMEFT}
\label{app:matching}

\begin{figure}[ht]
    \centering
    \begin{tikzpicture}
    \begin{feynman}
    \def\xscale{0.25} 
    \def\yscale{0.3} 
    \vertex (i0) at (0*\xscale, 0*\yscale) {\(\bar f_{L/R}\)};
    \vertex (i1) at (0*\xscale, 6*\yscale) {\(f_{L/R}\)};
    \vertex (i2) at (20*\xscale, 0*\yscale) {\(\bar f_{L/R}\)};
    \vertex (i3) at (20*\xscale, 6*\yscale) {\(f_{L/R}\)};
    \vertex (i4) at (6*\xscale, 6*\yscale);
    \vertex (i5) at (14*\xscale, 6*\yscale);
    \vertex (i6) at (6*\xscale, 0*\yscale);
    \vertex (i7) at (14*\xscale, 0*\yscale);
    \diagram* {
    (i1) -- [solid] (i4) -- [solid, edge label = \( \chi\)] (i6) -- [solid] (i0),
    (i2) -- [solid] (i7) -- [solid, edge label = \(\chi\)] (i5) -- [solid] (i3),
    (i4) -- [scalar, edge label = \(\Phi\)] (i5),
    (i6) -- [scalar, edge label = \(\Phi\)] (i7)
    };
    \end{feynman}
    \end{tikzpicture} $\quad$
    \begin{tikzpicture}
    \begin{feynman}
    \def\xscale{0.25} 
    \def\yscale{0.3} 
    \vertex (i0) at (0*\xscale, 0*\yscale) {\( f_R\)};
    \vertex (i1) at (20*\xscale, 0*\yscale) {\(\bar f_L\)};
    \vertex (i2) at (14*\xscale, 6*\yscale);
    \vertex (i3) at (17*\xscale, 5*\yscale);
    \vertex (i4) at (10*\xscale, 3*\yscale);
    \vertex (i8) at (11*\xscale, 4*\yscale);
    \vertex (i5) at (6*\xscale, 0*\yscale);
    \vertex (i6) at (14*\xscale, 0*\yscale);
    \vertex (i7) at (17*\xscale, 0*\yscale);
    \diagram* {
    (i0) -- [solid] (i5) -- [solid, edge label = \(\chi\)] (i6) -- [solid] (i7) -- [solid] (i1),
    (i8) -- [photon, edge label = \(\gamma\)] (i2),
    (i4) -- [scalar, quarter left] (i6),
    (i5) -- [scalar, quarter left, edge label = \(\Phi\)] (i4),
    (i7) -- [scalar, edge label' = \(H\)] (i3)
    };
    \end{feynman}
    \end{tikzpicture} $\quad$
    \begin{tikzpicture}
    \begin{feynman}
    \def\xscale{0.25} 
    \def\yscale{0.3} 
    \vertex (i0) at (0*\xscale, 0*\yscale) {\(f\)};
    \vertex (i1) at (20*\xscale, 8*\yscale) {\(f/H\)};
    \vertex (i2) at (20*\xscale, 5*\yscale) {\(f/H\)};
    \vertex (i3) at (20*\xscale, 0*\yscale) {\(f\)};
    \vertex (i4) at (14*\xscale, 6*\yscale);
    \vertex (i5) at (10*\xscale, 3*\yscale);
    \vertex (i6) at (6*\xscale, 0*\yscale);
    \vertex (i7) at (14*\xscale, 0*\yscale);
    \diagram* {
    (i0) -- [solid] (i6) -- [solid, edge label = \(\chi\)] (i7) -- [solid] (i3),
    (i1) -- [solid] (i4) -- [solid] (i2),
    (i5) -- [photon, edge label = \(V_\mu\)] (i4),
    (i5) -- [scalar, quarter left, edge label = \(\Phi\)] (i7),
    (i6) -- [scalar, quarter left, edge label = \(\Phi\)] (i5)
    };
    \end{feynman}
    \end{tikzpicture}
    \caption{Representative Feynman diagrams generating the four-fermion and $\mathcal{O}_{He}$ operators (box and penguins, left and bottom) and the dipole operators (right) for quarks and leptons. Diagrams were drawn using \texttt{FeynCraft}~\cite{Gaunt:2025peq}.}
    \label{fig:FeynmanDiagrams}
\end{figure}
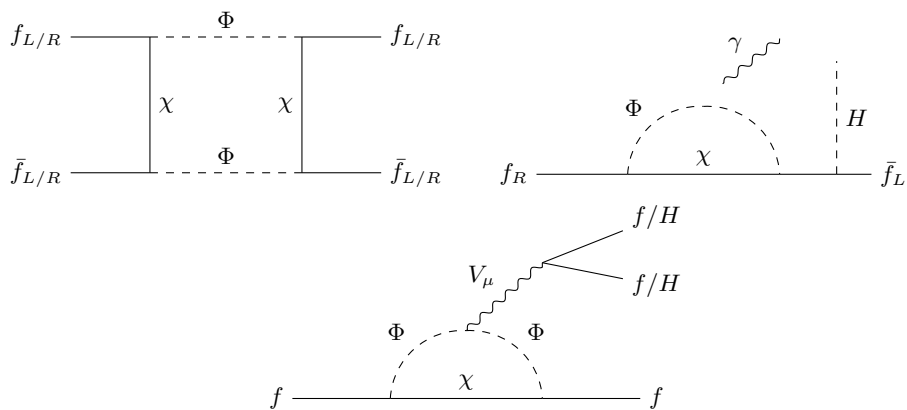
\begin{figure}[hbt]
    \centering
\includegraphics[width=0.8\linewidth]{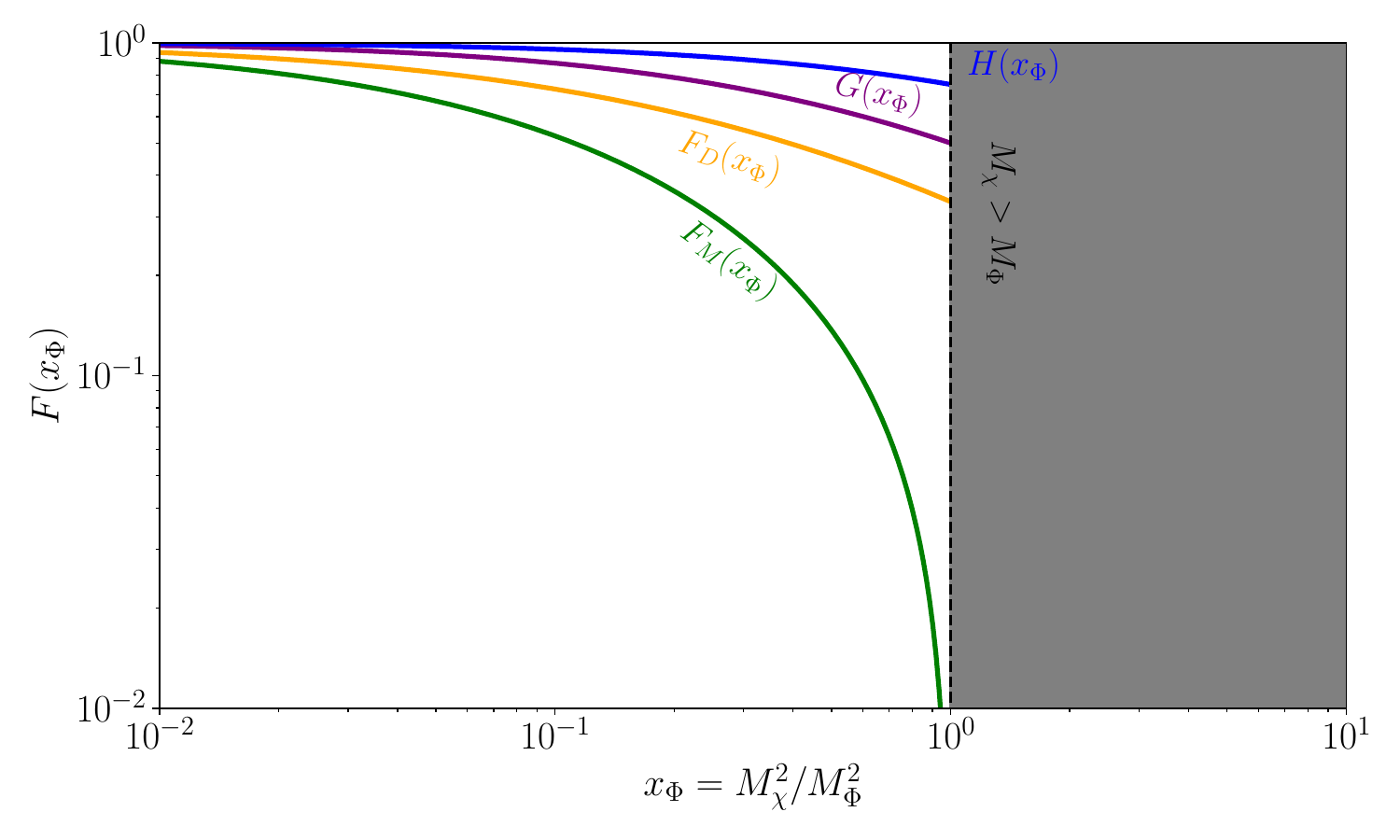}
    \caption{Loop functions for the 4-fermion operators and dipoles, in terms of the mass ratio $x_\Phi \equiv M_\chi^2/M_\Phi^2 = (1 + \delta_{\rm DM})^{-2}$.}
    \label{fig:loop_functions}
\end{figure}

After integrating out at one-loop the heavy mediator and DM candidate, we end with the following relevant subset of SMEFT operators for the leptophilic scenario
    \begin{align}
    \begin{aligned}
    \cL_{\mathrm{EFT}}^{e} &=\cC_{eB}\cO_{eB} + \cC_{He}\cO_{He} +\cC_{ee}\cO_{ee}  + \cC_{\ell e}\cO_{\ell e} + \cC_{qe}\cO_{q e} + \cC_{eu}\cO_{e u} + \cC_{ed}\cO_{e d}  \\
    &= \left[\cC_{eB}\right]_{ij} \bar{\ell}_{i}H \sigma_{\mu\nu} e_{j}  B^{\mu \nu } + \left[\cC_{He}\right]_{ij} \bar{e}_{i}  \gamma^\mu e_{j}(H^\dagger i\overleftrightarrow{D}_\mu H)\\
    &+\left[\cC_{ee}\right]_{ijkl} \bar{e}_{i}  \gamma^\mu e_{j}\bar{e}_{k}  \gamma_\mu  e_{l}
    +\left[\cC_{\ell e }\right]_{ijkl} \bar{\ell}_{i}  \gamma^\mu \ell_{j}\bar{e}_{k}  \gamma_\mu  e_{l} \\
    &+\left[\cC_{qe}\right]_{ijkl} \bar{q}_{i}  \gamma^\mu q_{j}\bar{e}_{k}  \gamma_\mu  e_{l}
    +\left[\cC_{eu}\right]_{ijkl} \bar{e}_{i}  \gamma^\mu e_{j}\bar{u}_{k}  \gamma_\mu  u_{l}
    +\left[\cC_{ed}\right]_{ijkl} \bar{e}_{i}  \gamma^\mu e_{j}\bar{d}_{k}  \gamma_\mu  d_{l},
    \end{aligned}
    \label{eq:lept_EFT}
\end{align}
while for the quarkphilic scenario
\begin{align}
\begin{aligned}
    \cL_{\mathrm{EFT}}^{q} &=\cC_{d(u)B}\cO_{d(u)B}+\cC_{d(u)W}\cO_{d(u)W}+\cC_{d(u)G}\cO_{d(u)G}+\cC_{qq}\cO_{qq} ^{(1)}\\
    &= \left[\cC_{d(u)B}\right]_{ij} \bar{q}_{i}H \sigma_{\mu\nu}  d_{j}  B^{\mu \nu }+\left[\cC_{d(u)W}\right]_{ij} \bar{q}_{i}H \sigma_{\mu\nu} \tau^a d_{j}  W^{a \, \mu \nu } \\
    &+ \left[\cC_{d(u)G}\right]_{ij} \bar{q}_{i}H \sigma_{\mu\nu} T^A d_{j}  G^{A\, \mu \nu }
    +\left[\cC_{qq}\right]_{ijkl} \bar{q}_{i}  \gamma^\mu q_{j}\bar{q}_{k}  \gamma_\mu q_{l} \, .
\end{aligned}
    \label{eq:quark_EFT}
\end{align}
Note that we do not include any $\Delta F=1$ operators beside dipoles; see the end of \cref{app:FV_observables} for the explanation.
In Fig.~\ref{fig:FeynmanDiagrams}, we show the corresponding diagrams that generate the four-fermion, penguin, and dipole operators. 
Starting from \cref{eq:leptophilicL,eq:quarkphilicL}, we perform the matching onto the corresponding Wilson coefficients using \texttt{Matchete}~\cite{Fuentes-Martin:2022jrf}. For the leptophilic scenario, the result can be written as
\begin{align}
    &\left[C_{eB}\right]_{ij} =  \frac{g'}{384 \pi^2 M_\Phi^2}  y_i y_k^*Y_{kj} G(x_\Phi),
    \quad 
[C_{He}]_{ij} = -\frac{g^{\prime 2}}{576\pi^2 M_\Phi^2} y_i y_j ^* H(x_\Phi),
\\
& [C_{ee}]_{ijkl} = \frac{1}{64 \pi^2 M_\Phi^{2}} \left[ 
-\frac{1}{2}y_{i}y_j^* y_{k}  y_{l}^*  F(x_\Phi)  + \frac{g^{\prime 2}}{9} \left(\delta_{i j} y_k y_l ^* + \delta_{i l} y_j ^* y_k \right)  H(x_\Phi)
\right],
\\
& [C_{\ell e}]_{ijkl} = \frac{g^{\prime 2}}{576 \pi^2 M_\Phi^{2}}  \delta_{i j} y_k y_l ^* H(x_\Phi),
\quad
[C_{qe}]_{ijkl} = -\frac{g^{\prime 2}}{1728 \pi^2 M_\Phi^{2}}  \delta_{i j} y_k y_l ^* H(x_\Phi),\\
& [C_{eu}]_{ijkl} = -\frac{g^{\prime 2}}{432 \pi^2 M_\Phi^{2}}  \delta_{k l} y_i y_j^* H(x_\Phi),
\quad
[C_{ed}]_{ijkl} = \frac{g^{\prime 2}}{864 \pi^2 M_\Phi^{2}}  \delta_{kl} y_i y_j ^* H(x_\Phi),
\end{align}
while for the quarkphilic scenario
\begin{align}
    [C_{qq}^{(1)}]_{ijkl} &=-\frac{1}{128\pi^2 M_\Phi^{2}}\,
y_{i} y_j^* y_{k} y _{l}^* F(x_\Phi), \quad 
[C_{u(d)B}]_{ij} = -\frac{g'}{2304 \pi^2 M_\Phi^2} y_i y^*_k Y_{u(d), kj} G(x_\Phi) ,  \\
[C_{u(d)W}]_{ij} &= -\frac{g}{768 \pi^2 M_\Phi^2} y_i y^*_k Y_{u(d), kj} G(x_\Phi), \quad  [C_{u(d)G}]_{ij} = -\frac{g_s}{384\pi^2 M_\Phi^2} y_i y^*_k Y_{u(d), kj} G(x_\Phi),    
\end{align}
with $x_\Phi = M_\chi^2/m_\Phi ^2$, and the loop functions given by
\begin{align}
\label{eq:loop_dirac}
 G(x_\Phi)&= \frac{1}{(1-x_\Phi)^4} \left[\left(1-x_\Phi\right)\left(1-x_\Phi(5+2x_\Phi)\right)-6x_\Phi^2 \log{x_\Phi}\right] \quad \textrm{(for dipoles)} , \\
\label{eq:loop_dipole}
 H(x_\Phi)&= \frac{1}{2(1-x_\Phi)^4} \left[\left(1-x_\Phi\right)\left(
 2-x_\Phi(7-11 x_\Phi)\right)+6x_\Phi^3 \log{x_\Phi}\right] \quad \textrm{(for penguin)} , \\
\label{eq:loop_mix}
   F_D(x_\Phi) &= \frac{1}{\left(1-x_\Phi \right)^{3}}\left[1-x_\Phi^{2}+2x_\Phi\log{x_\Phi}\right]
 \quad \textrm{(for box, $\chi$ Dirac)} , \\
 \label{eq:loop_majorana}
 F_M(x_\Phi) &= \frac{1}{\left(1-x_\Phi \right)^{3}}\left[1 + 4x_\Phi - 5x_\Phi^{2} + 2x_\Phi(2+x_\Phi)\log{x_\Phi}\right] \quad  \textrm{(for box, $\chi$ Majorana)} ,
\end{align}
The loop functions satisfy 
\begin{align}
    \{G(x_\Phi), H(x_\Phi), F_{D} (x_\Phi), F_{M} (x_\Phi) \} & \xrightarrow[]{x_\Phi\rightarrow 0} 1,\\ 
    \{G(x_\Phi), H(x_\Phi), F_{D} (x_\Phi), F_{M} (x_\Phi) \} &\xrightarrow[]{x_\Phi\rightarrow 1} \left\{\frac{1}{2},\frac{3}{4},\frac{1}{3}, 0\right\}\,.
\end{align}
The behaviour of the loop functions can also be seen in Fig.~\ref{fig:loop_functions}, which shows that the Majorana loop function $F_M$ vanishes in the degenerate limit.

\subsection{Flavour Observables}
\label{app:FV_observables}
\paragraph{Leptons.}
The class of observables considered in \cref{sec:leptophilic} includes flavour-violating three-body lepton decays, $\mu \to e$ conversion in nuclei, and dipole transitions $\ell_i \to \ell_j \gamma$. The expressions for these observables in terms of the Wilson coefficients listed in \cref{app:matching} are well known in the literature and will not be reported here, see for example \cite{Fernandez-Martinez:2024bxg, Calibbi:2021pyh}. We limit ourselves to a few remarks:

\begin{itemize}
    \item in the spin-independent contribution to $\mu \to e$ conversion, the dipole, four-fermion, and $C_{He}$ operators induce an effective vector coupling only to protons. This follows from the fact that these contributions are dominated by photon exchange or by scalar operators matched onto the electromagnetic current, which does not couple coherently to neutrons. This observable constrains the combination $\sim y^2 / M_\Phi^2$;

    \item dipole transitions $\ell_i \to \ell_j \gamma$ are likewise sensitive to the same parametric combination $\sim y^2 / M_\Phi^2$;

    \item three-body decays receive contributions scaling both as $\sim y^4 / M_\Phi^2$, from four-lepton box diagrams, and as $\sim y^2 / M_\Phi^2$, from dipole and penguin ones. Given the current experimental bounds from $\mu \to e$ conversion and dipole transitions, the bound from three-body decay constraints on the latter combination is subleading. The same also holds when comparing future projections.
\end{itemize}

Note that we compute the observables by matching the SMEFT onto the LEFT, integrating out the heavy $Z$, $W$, and Higgs bosons, and expressing the observables in terms of the resulting LEFT Wilson coefficients. We stress that, in the scenario considered here, this procedure remains valid even when $M_{\Phi,\chi}$ are around or below the electroweak scale. In fact, one could equivalently match directly onto the LEFT and obtain the same result. This follows from two observations. First, the relevant momentum transfer in the processes of interest is $q^2 \lesssim m_\tau^2 \ll m_Z^2$.  Second, the dark sector does not couple directly to the Higgs, so no corrections to \cref{eq:lept_EFT} featuring powers of $|H|^{2n}$ are present, which after electroweak symmetry breaking could invalidate the EFT expansion.

An alternative way to see this is to note that the operators in \cref{eq:lept_EFT} proportional to $g'^2 y^2$ are actually obtained from operators such as $(\bar e \gamma_\mu e) \partial_\nu B^{\mu\nu}$ after applying the equation of motion $\partial_\nu B^{\mu\nu} = J_Y^\mu$, reducing them to Higgs-fermion current and four-fermion interactions. The same operators with a photon or $Z$ fields could be matched directly after electroweak symmetry breaking, and would not get corrected by powers of $m^2_Z/m^2_{\Phi, \chi}$.

\paragraph{Quarks.}
Unlike in the LFV case, where the SM prediction for flavour violation is effectively zero, in the quark sector, the SM uncertainties must be determined precisely. However, in most cases, the SM predictions are dominated by hadronic and CKM-related uncertainties, which propagate into the observables and typically exceed the experimental errors. In the remainder of this section, we summarise the inputs used to compute both the SM and NP contributions to the various flavour observables.

Firstly, since our NP contributions enter through meson mixing, these processes cannot be used to determine the CKM matrix. Instead, we adopt the PDG~\cite{ParticleDataGroup:2024cfk} determination of the CKM elements obtained exclusively from tree-level observables, which our model does not modify. This tree-level determination comes with larger uncertainties, as $B$-meson mixing is otherwise very precisely measured. We have assessed the impact of using different CKM inputs, particularly the dependence on the $V_{cb}$ determination from inclusive versus exclusive decays. We find that adopting the inclusive value leads to better agreement with the SM predictions, especially for $\epsilon_K$, consistent with Ref.~\cite{Buras:2024mnq}, and fully compatible with the use of the tree-level CKM. In our analysis, we have also investigated the effect of performing a simultaneous fit of the CKM parameters, including both meson mixing observables and potential NP contributions. This approach yields results similar to those obtained using the tree-level CKM, and we therefore employ meson mixing primarily to extract bounds on NP. Consequently, we find that using the PDG tree-level CKM determination provides a straightforward and robust implementation, consistent within the corresponding confidence intervals.

For meson mixing, we use the recent Kaon bag parameter from Ref.~\cite{Gorbahn:2024qpe}, and for the $B$-mesons we adopt the bag parameters and decay constants from Ref.~\cite{Dowdall:2019bea}. The detailed procedure and explicit expressions for the QCD running and the computation of the hadronic matrix elements can be found in Appendix C of Ref.~\cite{DiLuzio:2023ndz}, which follows the analyses of the UTfit collaboration~\cite{Ciuchini:1998ix,Becirevic:2001jj,UTfit:2007eik}. In our numerical evaluation, we make use of the Wilson-coefficient bag parameters from Refs.~\cite{Dowdall:2019bea,Carrasco:2014uya,FlavourLatticeAveragingGroupFLAG:2024oxs}.

Finally, to compute the relevant observables, we adopt the following approximations. For $K$-meson mixing, the contribution to the mass splitting is given by the real part of the amplitude,
\begin{equation}
\Delta m_K \simeq 2\mathrm{Re}\left(M_{12}^{\mathrm{SM}} + M_{12}^{\mathrm{NP}}\right) \, , 
\end{equation}
where $M_{12}$ denotes the (short-distance) dispersive part of the mixing Hamiltonian. Currently, no reliable estimate of the long-distance effects is available. Therefore, we simply require the NP contribution not to exceed the SM short-distance contribution~\cite{Brod:2011ty}, ensuring compatibility with experimental values. In the case of $D$-mesons, long-distance effects are expected to dominate; thus, our conservative approach is to demand that the NP contribution does not overshoot the experimental measurements~\cite{ParticleDataGroup:2024cfk}.

However, for CP-violating observables such as $\epsilon_K$ and $x_{12}^{\mathrm{Im},D}$, the SM predictions are better understood. In particular, there exists a precise SM prediction for $\epsilon_K$~\cite{Brod:2019rzc}, and it is well established—both theoretically and experimentally—that CP violation in $D$-meson mixing is extremely small. Consequently, any sizable CP-violating NP contribution could easily exceed the SM expectation.

Finally, for $B_q-$meson mixing, we can make use of SM predictions and compute both SM and NP contributions to meson mixing, using the approximation
\begin{align}
        \Delta m_{B_q} &\simeq 2\left| M_{12}^{\mathrm{SM},q}+M_{12}^{\mathrm{NP},q}\right| \, , \\
        \quad \phi_q &= \arg{\left( M_{12}^{\mathrm{SM},q}+M_{12}^{\mathrm{NP},q}\right)} \, .
\end{align}
The phases can be extracted from time-dependent asymmetries in $B$-meson decays, such as $B_s \to J/\psi\,\phi$ and $B \to J/\psi \, K_S$, yielding $\phi_s$ and $\phi_d$, respectively. Other decay modes can also be used to determine these phases; in our analysis, we adopt the averages provided by HFLAV~\cite{HeavyFlavorAveragingGroupHFLAV:2024ctg}. We note that possible penguin pollution effects are neglected in this work.

For $b\to s \gamma$ we used \texttt{Flavio}~\cite{Straub:2018kue} to extract the bounds on the Wilson Coefficient. However, as we can see in Fig.~\ref{fig:QFV_scale}, these bounds do not provide a competitive probe for the masses considered in the relic density computation.

Other significant $\Delta F=1$ operators include the semileptonic operators, such as $\mathcal{O}_{qe}$ and $\mathcal{O}_{q\ell}^{(1,3)}$, which contribute to well-measured meson decays like $B_{s}\to \mu \mu$ and $K\to \mu\mu$. However, to evaluate the contributions from these operators, one must work within the Low Energy EFT (LEFT) by integrating out the electroweak degrees of freedom. In this basis, the operators relevant to these processes are identified as $\mathcal{O}_9$ and $\mathcal{O}_{10}$, corresponding to the following vector and axial-vector operators:
\begin{equation}
      \mathcal{O}_9 = 
    \frac{4G_F}{\sqrt{2}}\frac{ e^2 \,}{(4 \pi)^2} V_{tq}V_{tq'}^* [\bar{q}' \gamma_\mu  P_L q][\bar{\ell} \gamma^\mu \ell], \quad
  \mathcal{O}_{10}  = 
   \frac{4G_F}{\sqrt{2}}\frac{ e^2 \,}{(4 \pi)^2} V_{tq}V_{tq'}^* [\bar{q}' \gamma_\mu  P_L q][\bar{\ell} \gamma^\mu \gamma_5 \ell] \, .  
\end{equation}
The matching from the SMEFT to these operators can be obtained using~\cite{Alonso:2014csa},
\begin{align}
    C_9=&\frac{4\pi^2}{e^2 V_{tq}V_{tq'}^*} \frac{v^2}{\Lambda^2}\left[C_{qe}+C_{\ell q}^{(1)}+C_{\ell q}^{(3)}-\left(1-4s_W^2\right)\left(C_{Hq}^{(1)}+C_{Hq}^{(3)}\right) \right] =\frac{4\pi^2}{V_{tq}V_{tq'}^*} \frac{v^2}{M_\Phi^2} \frac{y_q^* y_{q'}}{432\pi^2} H(x_\Phi) \,, \\ 
C_{10}=&\frac{4\pi^2}{e^2 V_{tq}V_{tq'}^*}\frac{v^2}{\Lambda^2}\left[C_{qe}-C_{\ell q}^{(1)}-C_{\ell q}^{(3)}+\left(C_{Hq}^{(1)}+C_{Hq}^{(3)}\right) \right] = 0 \, ,
\end{align}
and upon substituting the expressions for our model, we find that $C_{10}$ vanishes exactly. Furthermore, this cancellation occurs within independent sets of operators: the singlet combination $C_{qe}-C_{\ell q}^{(1)}+C_{Hq}^{(1)} = 0$ and the triplet combination $C_{Hq}^{(3)}-C_{\ell q}^{(3)}$ cancel independently. A similar effect was noted in Ref.~\cite{Davighi:2023evx}, where the different contributions to $C_{10}$ were found to be proportional to the hypercharge, i.e., $C_{10}\sim (Y_e-Y_\ell-Y_H)=0$, vanishing due to the conservation of hypercharge in the Yukawa interactions $\bar \ell e H$. We observe the same phenomenon here; similarly to the leptophilic case, after integrating out the new heavy degrees of freedom, we obtain the redundant operator $\bar q' \gamma_\mu (\tau^i) q \partial_\nu B^{\mu\nu} (D_\nu W^{i,\mu\nu})$, where $\tau_i$ are the generators of $\mathrm{SU}(2)_L$. When applying the equations of motion, the derivative acting on the field strength tensor is replaced by the conserved currents, e.g.~$\partial_\nu B^{\mu\nu} =J^\mu_Y$. Consequently, all leptons couple proportionally to the hypercharge, explaining the cancellation in $C_{10}$, and analogously for the $\mathrm{SU}(2)_L$ contribution.

The phenomenological consequence of this cancellation is that purely leptonic decays such as $B_s, K_L \to \mu\mu$ are insensitive to this type of NP. Therefore, we must rely on constraints from the $C_9$ operator, where $\Delta F=1$ processes such as $B\to K^* \mu \mu $ provide the leading constraint. These observables currently exhibit experimental anomalies. Using Ref.~\cite{greljo:2024ytg}, for $b\to s$ transitions we obtain:
\begin{equation}
    \frac{M_\Phi}{\sqrt{|y_{2}^* y_{3}| H(x_\Phi)}} = 126 \pm 12 \, \mathrm{GeV}\, .
\end{equation}
Such a low scale is already ruled out by other observables (see \cref{sec:quarkphilic}); hence, we do not consider this type of operator in our final analysis.

\section{Flavour Reparametrization}
\label{app:LFV_directions}

\begin{figure*}[!tb]
    \centering
    \includegraphics[width=0.9\linewidth]{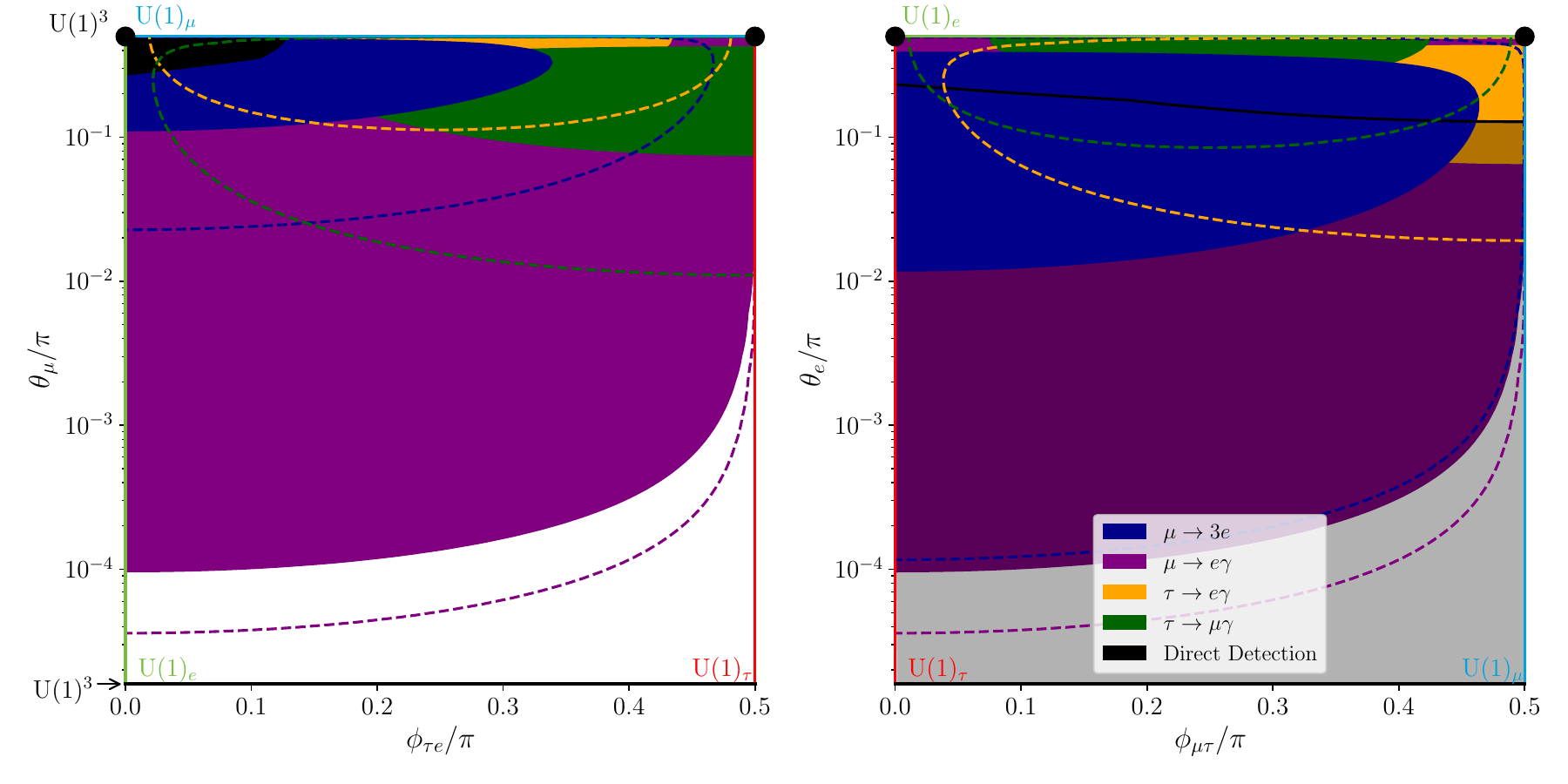}
    \caption{Same bounds as in Fig.~\ref{fig:Majorana_RH_lept}, but redefining the definition of the angles. In particular, the angle redefinition corresponds to aligning the $\mu$- ($e$-)specific flavour to the north-pole of the sphere for the left (right) plot. In this case, we only consider the Majorana case with $\delta_\text{DM} =0.14$ and $M_\chi=170\,$GeV. }
    \label{fig:basis_reparametrization}
\end{figure*}

As shown in Figs.~\ref{fig:Majorana_RH_lept}, \ref{fig:Dirac_RH_lept} the $\mathrm{U}(1)^3$ limit is preserved by flavour observables. This is not the case in the quark sector, where analogous regions are ruled out by the misaligned flavour contributions to $D-\bar{D}$ or $K-\bar{K}$ mixing; see, for instance, Figs.~\ref{fig:Majorana_LH_quark}, \ref{fig:Dirac_LH_quark} near $\phi \sim 0,  \pm\pi/2, \pm \pi$.

For the leptophilic case, it is therefore interesting to quantify how closely one must align to each of these points in order to satisfy the bounds. For this comparison, it is useful to note that the sphere in Fig.~\ref{fig:sphere} can be parametrised with any of the three lepton flavours aligned at the north pole. This allows a visual comparison of the three different $\mathrm{U}(1)^3$ points, since the north pole of the sphere plays a special role in this parametrisation.

In Fig.~\ref{fig:basis_reparametrization}, we show the results for the reparametrisation for the Majorana case corresponding to the right-top plot in Fig.~\ref{fig:Majorana_RH_lept}. We see that the bounds in terms of the angles $\theta_{\mu,e}$, which represent the angular deviation from the north pole, can be directly compared to $\theta_\tau$. This not only confirms that alignment to any of the three $\mathrm{U}(1)^3$ directions is consistent with LFV observables, but also allows us to quantify the degree of alignment required. In fact, alignment to the third family is less constrained than alignment to the lighter families by approximately two orders of magnitude. This can be understood from the parametric suppression of lower-family observables such as $\mu \to e \gamma$, which scales as $\sin^4 \theta_\tau$ in the $\tau$-basis, while in the electron and muon bases it behaves as $\sin^2\theta_{\mu,e}\cos^2\theta_{\mu,e}$. This shows that aligning to the $\tau$ direction leads to a faster decoupling of these observables than aligning to the other families.

\section{Cross sections for dark matter freeze-out}
\label{app:Cross_section_DM}
Many of the cross sections employed in our analysis are taken from earlier studies, as detailed in the main text. In this Appendix, we collect only the cross sections that we have computed, including finite fermion mass effects for processes involving the Yukawa interaction. For completeness, we also present the corresponding expressions in the massless fermion limit. Throughout this section, we denote the SM fermion mass by $m$ and expand the cross sections up to order $\vrel^2$. Since the only relevant fermion mass is that of the top quark, the cross sections are computed either with a single massive fermion or with two massive fermions of equal mass in the final state. We do not provide expressions for the most general case of different fermion masses in the final state in order to avoid unnecessarily lengthy formulas. For the lepton case, one may simply put $N_c=1$ in all expressions below. 

The Majorana and Dirac DM pair annihilation into a massless fermion–antifermion pair reads, respectively,
\begin{align}
    &\sigma_{\chi \chi}^{m=0} \vrel =  \frac{y^4 N_c}{48 \pi}
  \frac{M_\chi^2 (M_\chi^4+M_\Phi^4)}{(M_\chi^2+M_\Phi^2)^4} \vrel^2 \, ,
  \label{sigma_chi_chi_mass0_Maj}
  \\
  &\sigma_{\chi \bar{\chi}}^{m=0} \vrel = \frac{y^4 N_c}{32 \pi}
  \frac{M_\chi^2 }{(M_\chi^2+M_\Phi^2)^2}  \left( 1 + \frac{(M_\Phi^4-3M_\Phi^2 M_\chi^2-M_\Phi^4)}{3(M_\chi^2+M_\Phi^2)^2}  \vrel^2 \right)  \, .
\label{sigma_chi_chi_mass0_Dir}
\end{align}
The corresponding cross sections for one massive quark in the final state read 
{\small
\begin{align}
 &\sigma_{\chi \chi}^{m} \vrel =
 \frac{y^4 N_c}{256 \pi}
 \frac{(4 M_\chi^2-m^2)}{M_\chi^4(2(M_\chi^2+M_\Phi^2)-m^2)^2}
 \bigg\lbrace
 \vphantom{\frac{\vrel^2}{(2(M_\chi^2+M_\Phi^2)-m^2)^2}}
 m^2(4 M_\chi^2-m^2)
 + \frac{\vrel^2}{3(2(M_\chi^2+M_\Phi^2)-m^2)^2}
 \left[
 m^8 - 7 m^6 (M_\Phi^2 + M_\chi^2)
 \right.
 \nonumber \\
 &\qquad
\left.  + m^4 \left( 7 M_\Phi^4 + 38 M_\Phi^2 M_\chi^2 + 27 M_\chi^4 \right)
 - 4 m^2 M_\chi^2 \!\left( 5 M_\Phi^4 + 22 M_\Phi^2 M_\chi^2 + 13 M_\chi^4 \right)
 + 64 M_\chi^4 \!\left( M_\Phi^4 + M_\chi^4 \right) \right]
 \bigg\rbrace ,
 \label{sigma_chi_chi_mass1_Maj}
\end{align}}
\!for Majorana DM, whereas for Dirac DM, one finds 
 {\small
\begin{align}
 &\sigma_{\chi \bar{\chi}}^{m} \vrel =
 \frac{y^4 N_c}{512 \pi}
 \frac{(4 M_\chi^2-m^2)}{M_\chi^4(2(M_\chi^2+M_\Phi^2)-m^2)^2}
 \bigg\lbrace
 \vphantom{\frac{\vrel^2}{(2(M_\chi^2+M_\Phi^2)-m^2)^2}}
 16 M_\chi^4-m^4
 + \frac{\vrel^2}{3(2(M_\chi^2+M_\Phi^2)-m^2)^2}
 \left[
 m^8 - m^6 (7 M_\Phi^2 + 2 M_\chi^2)
 \right.
 \nonumber \\
 &\qquad
 \left. + m^4 \left( 7 M_\Phi^4 + 14 M_\Phi^2 M_\chi^2 -9 M_\chi^4 \right)
 + 4 m^2 M_\chi^2 \!\left( M_\Phi^4 + 2 M_\Phi^2 M_\chi^2 + 17 M_\chi^4 \right)
 + 64 M_\chi^4 \!\left( M_\Phi^4 -3 M_\chi^2 M_\Phi^2+M_\chi^4 \right) \right]
 \bigg\rbrace .
 \label{sigma_chi_chi_mass1_Dir}
\end{align}
}
The cross sections for two massive quarks in the final state read, for Majorana and Dirac DM respectively, as follows 
{\small
\begin{align}
 &\sigma_{\chi \chi}^{2m} \vrel =
 \frac{y^4 N_c}{32 \pi}
 \frac{\sqrt{1-m^2/M_\chi^2}}{(M_\Phi^2+M_\Phi^2-m^2)^2}
 \bigg\lbrace
 \vphantom{\frac{\vrel^2}{(2(M_\chi^2+M_\Phi^2)-m^2)^2}}
 m^2
 + \frac{\vrel^2}{24(M_\Phi^2+M_\Phi^2-m^2)^2(M_\chi^2-m^2)}
 \left[
 13 m^8 - 2 m^6 (13 M_\Phi^2 + 18 M_\chi^2)
 \right.
 \nonumber \\
 &\qquad
 \left. + m^4 \left( 13 M_\Phi^4 + 70 M_\Phi^2 M_\chi^2 + 49 M_\chi^4 \right)
 - 2 m^2 M_\chi^2 \!\left( 13 M_\Phi^4 + 22 M_\Phi^2 M_\chi^2 + 21 M_\chi^4 \right)
 + 16 M_\chi^4 \!\left( M_\Phi^4 + M_\chi^4 \right) \right]
 \bigg\rbrace \, ,
 \label{sigma_chi_chi_mass2_Maj}
\\
 &\sigma_{\chi \bar{\chi}}^{2m} \vrel =
 \frac{y^4 N_c}{32 \pi}
 \frac{\sqrt{1-m^2/M_\chi^2}}{(M_\Phi^2+M_\Phi^2-m^2)^2}
 \bigg\lbrace
 \vphantom{\frac{\vrel^2}{(2(M_\chi^2+M_\Phi^2)-m^2)^2}}
 M_\chi^2
 + \frac{\vrel^2}{24(M_\Phi^2+M_\Phi^2-m^2)^2(M_\chi^2-m^2)}
 \left[
 2 m^8 -  m^6 (4 M_\Phi^2 - 5 M_\chi^2)
 \right.
 \nonumber \\
 &\qquad
\left. + 2 m^4 \left( M_\Phi^4 +  M_\Phi^2 M_\chi^2 -12 M_\chi^4 \right)
 -  m^2 M_\chi^2 \!\left( 7 M_\Phi^4 - 26 M_\Phi^2 M_\chi^2 - 25 M_\chi^4 \right)
 + 8 M_\chi^4 \!\left( M_\Phi^4 -3 M_\Phi^2 M_\chi^2 - M_\chi^4 \right) \right]
 \bigg\rbrace .
 \label{sigma_chi_chi_mass2_Dir}
\end{align}
}
The cross sections with a finite fermion mass in Eqs.~\eqref{sigma_chi_chi_mass1_Maj}-\eqref{sigma_chi_chi_mass2_Dir} reproduce the ones given in Eqs.~\eqref{sigma_chi_chi_mass0_Maj} and \eqref{sigma_chi_chi_mass0_Dir} for $m \to 0$.

Only in the case of Majorana dark matter a Yukawa-driven annihilation channel for the mediator into two fermions exist, namely $\Phi \Phi \to e_i e_j$ and $\Phi \Phi \to q_i q_j$, together with their complex-conjugate processes, for the leptophilic and quarkphilic scenarios, respectively. As in the case of dark matter annihilations, only interactions involving quarks induce a flavour dependence in the total cross section. In the following, we therefore list the relevant cross sections with one massive quark and with two massive quarks in the final state, and provide the massless limit as a reference. The expressions read as follows
{\small 
\begin{align}
    &\sigma_{\Phi \Phi}^{m=0} \vrel =  \frac{y^4 M_\chi^2}{6 \pi (M_\chi^2+M_\Phi^2)^2}
  \left[ 1 - \frac{M_\Phi^2(6 M_\chi^2+M_\Phi^2)}{6(M_\chi^2+M_\Phi^2)^2 }\vrel^2 \right] \, ,
  \\
  &\sigma_{\Phi \Phi}^{m} \vrel =  \frac{y^4 M_\chi^2 (4M_\Phi^2-m^2) }{768 \pi M_\Phi^4 (M_\chi^2+M_\phi^2)^2(M_\Phi^2+M_\chi^2-m^2)^2} \bigg\lbrace
 \vphantom{\frac{\vrel^2}{(M_\Phi^2+M_\Phi^2-m^2)^2}} \left[ 8 (M_\Phi^2+M_\chi^2)^2-8m^2 (M_\chi^2+M_\Phi^2)+3m^4 \right](4M_\Phi^2-m^2) 
 \nonumber \\
 &\qquad
  - \frac{\vrel^2 }{48(M_\Phi^2+M_\Phi^2-m^2)^2(M_\chi^2+M_\Phi^2)^2}
 \left[
 9  m^{14} -2 m^{12} (19 M_\chi^2  + 73 M_\Phi^2) + 2 m^{10} (142 M_\chi^2 M_\Phi^2 + 365 M_\Phi^4 -7 M_\chi^4)  
 \right.  \nonumber
 \\
 &\qquad \left. -8 m^8(200 M_\Phi^6 + 63 M_\Phi^4 M_\chi^2 -102 M_\Phi^2 M_\chi^4 -37 M_\chi^6) \right. 
 \nonumber 
 \\
 &\qquad \left. +4 m^6 (379 M_\Phi^8 -308 M_\Phi^6 M_\chi^2 -1302 M_\Phi^4 M_\chi^4 -772 M_\Phi^2 M_\chi^6 -157 M_\chi^8) 
 \right. 
 \nonumber
 \\
  &\qquad \left. -16 m^4 (M_\Phi^2+M_\chi^2)^2 (7 M_\Phi^6 -342 M_\chi^2 M_\Phi^4 -153 M_\chi^4 M_\Phi^2 -36 M_\chi^6) 
  \right. 
 \nonumber
 \\
  &\qquad \left. - 64\, m^2 \left(M_\chi^2 + M_\Phi^2\right)^3
\left(3 M_\chi^6 + 14 M_\Phi^6 + 66 M_\chi^2 M_\Phi^4 + 15 M_\chi^4 M_\Phi^2
\right) + 256\, M_\Phi^4 \left(M_\chi^2 + M_\Phi^2\right)^4
\left(6 M_\chi^2 + M_\Phi^2\right)
 \right]
 \bigg\rbrace \, ,
  \\
   &\sigma_{\Phi \Phi}^{2m} \vrel = \frac{y^4}{12 \pi}
 \frac{M_\chi^2}{M_\Phi^3(M_\Phi^2+M_\Phi^2-m^2)^2\sqrt{M_\Phi^2-m^2}}
 \bigg\lbrace
 \vphantom{\frac{\vrel^2}{(M_\Phi^2+M_\Phi^2-m^2)^2}}
 2 M_\Phi^4 - 3 m^2 M_\Phi^2 + m^4 
 + \frac{\vrel^2}{24(M_\Phi^2+M_\Phi^2-m^2)^2}
 \left[
 3 m^2 M_\chi^4 (4 M_\Phi^2-3m^2)
 \right.
 \nonumber \\
 &\qquad
 \left. - 6 M_\chi^2 \left( M_\Phi^2-m^2\right)\left( 3m^4-8m^2M_\Phi^2+8M_\Phi^4\right)
 - (M_\Phi^2-m^2)^2(9m^4-16m^2 M_\Phi^2 + 8M_\Phi^4 ) \right]
 \bigg\rbrace .
\end{align}}

\bibliographystyle{JHEP}
\bibliography{refs.bib}

@article{Greljo:2023adz,
    author = "Greljo, Admir and Palavri{\'c}, Ajdin",
    title = "{Leading directions in the SMEFT}",
    eprint = "2305.08898",
    archivePrefix = "arXiv",
    primaryClass = "hep-ph",
    doi = "10.1007/JHEP09(2023)009",
    journal = "JHEP",
    volume = "09",
    pages = "009",
    year = "2023"
}

@article{Biondini:2025gpg,
    author = "Biondini, Simone and Tiberi, Lorenzo and Panella, Orlando",
    title = "{Connecting t-channel dark matter models to the Standard Model Effective Field Theory}",
    eprint = "2507.00925",
    archivePrefix = "arXiv",
    primaryClass = "hep-ph",
    doi = "10.1007/JHEP10(2025)060",
    journal = "JHEP",
    volume = "10",
    pages = "060",
    year = "2025"
}

@article{Calibbi:2017uvl,
    author = "Calibbi, Lorenzo and Signorelli, Giovanni",
    title = "{Charged Lepton Flavour Violation: An Experimental and Theoretical Introduction}",
    eprint = "1709.00294",
    archivePrefix = "arXiv",
    primaryClass = "hep-ph",
    doi = "10.1393/ncr/i2018-10144-0",
    journal = "Riv. Nuovo Cim.",
    volume = "41",
    number = "2",
    pages = "71--174",
    year = "2018"
}

@article{Kahlhoefer:2015bea,
    author = "Kahlhoefer, Felix and Schmidt-Hoberg, Kai and Schwetz, Thomas and Vogl, Stefan",
    title = "{Implications of unitarity and gauge invariance for simplified dark matter models}",
    eprint = "1510.02110",
    archivePrefix = "arXiv",
    primaryClass = "hep-ph",
    reportNumber = "DESY-15-182",
    doi = "10.1007/JHEP02(2016)016",
    journal = "JHEP",
    volume = "02",
    pages = "016",
    year = "2016"
}

@article{deBlas:2017xtg,
    author = "de Blas, J. and Criado, J. C. and Perez-Victoria, M. and Santiago, J.",
    title = "{Effective description of general extensions of the Standard Model: the complete tree-level dictionary}",
    eprint = "1711.10391",
    archivePrefix = "arXiv",
    primaryClass = "hep-ph",
    reportNumber = "CERN-TH-2017-251",
    doi = "10.1007/JHEP03(2018)109",
    journal = "JHEP",
    volume = "03",
    pages = "109",
    year = "2018"
}

@article{Moritsu:2022lem,
    author = "Moritsu, Manabu",
    collaboration = "COMET",
    title = "{Search for Muon-to-Electron Conversion with the COMET Experiment {\textdagger}}",
    eprint = "2203.06365",
    archivePrefix = "arXiv",
    primaryClass = "hep-ex",
    doi = "10.3390/universe8040196",
    journal = "Universe",
    volume = "8",
    number = "4",
    pages = "196",
    year = "2022"
}

@article{Bernstein:2019fyh,
    author = "Bernstein, R. H.",
    collaboration = "Mu2e",
    title = "{The Mu2e Experiment}",
    eprint = "1901.11099",
    archivePrefix = "arXiv",
    primaryClass = "physics.ins-det",
    reportNumber = "FERMILAB-PUB-18-424-PPD",
    doi = "10.3389/fphy.2019.00001",
    journal = "Front. in Phys.",
    volume = "7",
    pages = "1",
    year = "2019"
}

@article{Demetriou:2025ewa,
    author = "Demetriou, Georgios and Isidori, Gino and Piazza, Gioacchino and Pinsard, Emanuelle",
    title = "{The third-generation-philic WIMP: an EFT analysis}",
    eprint = "2505.04708",
    archivePrefix = "arXiv",
    primaryClass = "hep-ph",
    doi = "10.1140/epjc/s10052-025-14580-5",
    journal = "Eur. Phys. J. C",
    volume = "85",
    number = "8",
    pages = "865",
    year = "2025"
}

@article{Buckley:2014fba,
    author = "Buckley, Matthew R. and Feld, David and Goncalves, Dorival",
    title = "{Scalar Simplified Models for Dark Matter}",
    eprint = "1410.6497",
    archivePrefix = "arXiv",
    primaryClass = "hep-ph",
    reportNumber = "IPPP-14-92, DCPT-14-184",
    doi = "10.1103/PhysRevD.91.015017",
    journal = "Phys. Rev. D",
    volume = "91",
    pages = "015017",
    year = "2015"
}

@article{Bell:2016ekl,
    author = "Bell, Nicole F. and Busoni, Giorgio and Sanderson, Isaac W.",
    title = "{Self-consistent Dark Matter Simplified Models with an s-channel scalar mediator}",
    eprint = "1612.03475",
    archivePrefix = "arXiv",
    primaryClass = "hep-ph",
    doi = "10.1088/1475-7516/2017/03/015",
    journal = "JCAP",
    volume = "03",
    pages = "015",
    year = "2017"
}

@article{Albert:2016osu,
    author = "Albert, Andreas and others",
    title = "{Towards the next generation of simplified Dark Matter models}",
    eprint = "1607.06680",
    archivePrefix = "arXiv",
    primaryClass = "hep-ex",
    doi = "10.1016/j.dark.2017.02.002",
    journal = "Phys. Dark Univ.",
    volume = "16",
    pages = "49--70",
    year = "2017"
}

@article{Agrawal:2015kje,
    author = "Agrawal, Prateek and Chacko, Zackaria and Fortes, Elaine C. F. S. and Kilic, Can",
    title = "{Skew-Flavored Dark Matter}",
    eprint = "1511.06293",
    archivePrefix = "arXiv",
    primaryClass = "hep-ph",
    reportNumber = "UTTG-20-15-, TCC-009-15, FERMILAB-PUB-15-622-T",
    doi = "10.1103/PhysRevD.93.103510",
    journal = "Phys. Rev. D",
    volume = "93",
    number = "10",
    pages = "103510",
    year = "2016"
}

@article{Englert:2016joy,
    author = "Englert, Christoph and McCullough, Matthew and Spannowsky, Michael",
    title = "{S-Channel Dark Matter Simplified Models and Unitarity}",
    eprint = "1604.07975",
    archivePrefix = "arXiv",
    primaryClass = "hep-ph",
    reportNumber = "DCPT-16-52, IPPP-16-26, CERN-TH-2016-094",
    doi = "10.1016/j.dark.2016.09.002",
    journal = "Phys. Dark Univ.",
    volume = "14",
    pages = "48--56",
    year = "2016"
}

@article{Goncalves:2016iyg,
    author = "Goncalves, Dorival and Machado, Pedro A. N. and No, Jose Miguel",
    title = "{Simplified Models for Dark Matter Face their Consistent Completions}",
    eprint = "1611.04593",
    archivePrefix = "arXiv",
    primaryClass = "hep-ph",
    reportNumber = "IPPP-16-114, PITT-PACC-1613, FERMILAB-PUB-16-510-T, KCL-PH-TH-2016-61",
    doi = "10.1103/PhysRevD.95.055027",
    journal = "Phys. Rev. D",
    volume = "95",
    number = "5",
    pages = "055027",
    year = "2017"
}

@article{DeSimone:2016fbz,
    author = "De Simone, Andrea and Jacques, Thomas",
    title = "{Simplified models vs. effective field theory approaches in dark matter searches}",
    eprint = "1603.08002",
    archivePrefix = "arXiv",
    primaryClass = "hep-ph",
    reportNumber = "SISSA-21-2016-FISI",
    doi = "10.1140/epjc/s10052-016-4208-4",
    journal = "Eur. Phys. J. C",
    volume = "76",
    number = "7",
    pages = "367",
    year = "2016"
}

@article{Boveia:2016mrp,
    author = "Boveia, Antonio and others",
    editor = "Buchmueller, Oliver and Doglioni, Caterina and Hahn, Kristian and Haisch, Ulrich and Kahlhoefer, Felix and Mangano, Michelangelo and McCabe, Christopher and Tait, Tim M. P.",
    title = "{Recommendations on presenting LHC searches for missing transverse energy signals using simplified $s$-channel models of dark matter}",
    eprint = "1603.04156",
    archivePrefix = "arXiv",
    primaryClass = "hep-ex",
    reportNumber = "CERN-LPCC-2016-001",
    doi = "10.1016/j.dark.2019.100365",
    journal = "Phys. Dark Univ.",
    volume = "27",
    pages = "100365",
    year = "2020"
}

@article{MEGII:2025gzr,
    author = "Afanaciev, K. and others",
    collaboration = "MEG II",
    title = "{New limit on the ${\mu ^+ \rightarrow e^+ \gamma }$ decay with the MEG II experiment}",
    eprint = "2504.15711",
    archivePrefix = "arXiv",
    primaryClass = "hep-ex",
    doi = "10.1140/epjc/s10052-025-14906-3",
    journal = "Eur. Phys. J. C",
    volume = "85",
    number = "10",
    pages = "1177",
    year = "2025",
    note = "[Erratum: Eur.Phys.J.C 85, 1317 (2025)]"
}

@article{Greljo:2025ljr,
    author = "Greljo, Admir and Palavri{\'c}, Ajdin and Tunja, Mirsad and Zupan, Jure",
    title = "{Expanding the Landscape of Exotic Muon Decays}",
    eprint = "2510.08674",
    archivePrefix = "arXiv",
    primaryClass = "hep-ph",
    month = "10",
    year = "2025"
}

@article{Garny:2014waa,
    author = "Garny, Mathias and Ibarra, Alejandro and Rydbeck, Sara and Vogl, Stefan",
    title = "{Majorana Dark Matter with a Coloured Mediator: Collider vs Direct and Indirect Searches}",
    eprint = "1403.4634",
    archivePrefix = "arXiv",
    primaryClass = "hep-ph",
    reportNumber = "DESY-14-029, TUM-HEP-935-14, CERN-PH-TH-2014-041",
    doi = "10.1007/JHEP06(2014)169",
    journal = "JHEP",
    volume = "06",
    pages = "169",
    year = "2014"
}

@article{Allwicher:2022gkm,
    author = "Allwicher, Lukas and Faroughy, Darius A. and Jaffredo, Florentin and Sumensari, Olcyr and Wilsch, Felix",
    title = "{Drell-Yan tails beyond the Standard Model}",
    eprint = "2207.10714",
    archivePrefix = "arXiv",
    primaryClass = "hep-ph",
    doi = "10.1007/JHEP03(2023)064",
    journal = "JHEP",
    volume = "03",
    pages = "064",
    year = "2023"
}

@article{Alonso:2014csa,
    author = "Alonso, Rodrigo and Grinstein, Benjamin and Martin Camalich, Jorge",
    title = "{$SU(2)\times U(1)$ gauge invariance and the shape of new physics in rare $B$ decays}",
    eprint = "1407.7044",
    archivePrefix = "arXiv",
    primaryClass = "hep-ph",
    doi = "10.1103/PhysRevLett.113.241802",
    journal = "Phys. Rev. Lett.",
    volume = "113",
    pages = "241802",
    year = "2014"
}

@article{Greljo:2024ytg,
    author = "Greljo, Admir and Tiblom, Hector and Valenti, Alessandro",
    title = "{New physics through flavor tagging at FCC-ee}",
    eprint = "2411.02485",
    archivePrefix = "arXiv",
    primaryClass = "hep-ph",
    doi = "10.21468/SciPostPhys.18.5.152",
    journal = "SciPost Phys.",
    volume = "18",
    number = "5",
    pages = "152",
    year = "2025"
}

@article{Greljo:2022jac,
    author = "Greljo, Admir and Salko, Jakub and Smolkovi{\v{c}}, Aleks and Stangl, Peter",
    title = "{Rare b decays meet high-mass Drell-Yan}",
    eprint = "2212.10497",
    archivePrefix = "arXiv",
    primaryClass = "hep-ph",
    reportNumber = "CERN-TH-2023-037",
    doi = "10.1007/JHEP05(2023)087",
    journal = "JHEP",
    volume = "05",
    pages = "087",
    year = "2023"
}

@article{Glioti:2024hye,
    author = "Glioti, Alfredo and Rattazzi, Riccardo and Ricci, Lorenzo and Vecchi, Luca",
    title = "{Exploring the flavor symmetry landscape}",
    eprint = "2402.09503",
    archivePrefix = "arXiv",
    primaryClass = "hep-ph",
    doi = "10.21468/SciPostPhys.18.6.201",
    journal = "SciPost Phys.",
    volume = "18",
    number = "6",
    pages = "201",
    year = "2025"
}

@article{Greljo:2017vvb,
    author = "Greljo, Admir and Marzocca, David",
    title = "{High-$p_T$ dilepton tails and flavor physics}",
    eprint = "1704.09015",
    archivePrefix = "arXiv",
    primaryClass = "hep-ph",
    reportNumber = "ZU-TH-12-17",
    doi = "10.1140/epjc/s10052-017-5119-8",
    journal = "Eur. Phys. J. C",
    volume = "77",
    number = "8",
    pages = "548",
    year = "2017"
}

@article{Fuentes-Martin:2019mun,
    author = "Fuentes-Mart{\'\i}n, Javier and Isidori, Gino and Pag{\`e}s, Julie and Yamamoto, Kei",
    title = "{With or without U(2)? Probing non-standard flavor and helicity structures in semileptonic B decays}",
    eprint = "1909.02519",
    archivePrefix = "arXiv",
    primaryClass = "hep-ph",
    reportNumber = "ZU-TH-42/19",
    doi = "10.1016/j.physletb.2019.135080",
    journal = "Phys. Lett. B",
    volume = "800",
    pages = "135080",
    year = "2020"
}

@article{Agrawal:2014una,
    author = "Agrawal, Prateek and Batell, Brian and Hooper, Dan and Lin, Tongyan",
    title = "{Flavored Dark Matter and the Galactic Center Gamma-Ray Excess}",
    eprint = "1404.1373",
    archivePrefix = "arXiv",
    primaryClass = "hep-ph",
    reportNumber = "FERMILAB-PUB-14-069-A-T",
    doi = "10.1103/PhysRevD.90.063512",
    journal = "Phys. Rev. D",
    volume = "90",
    number = "6",
    pages = "063512",
    year = "2014"
}

@article{Agrawal:2014aoa,
    author = "Agrawal, Prateek and Blanke, Monika and Gemmler, Katrin",
    title = "{Flavored Dark Matter beyond Minimal Flavor Violation}",
    eprint = "1405.6709",
    archivePrefix = "arXiv",
    primaryClass = "hep-ph",
    reportNumber = "CERN-PH-TH-2014-098, FERMILAB-PUB-14-141-T",
    doi = "10.1007/JHEP10(2014)072",
    journal = "JHEP",
    volume = "10",
    pages = "072",
    year = "2014"
}

@article{Bishara:2015mha,
    author = "Bishara, Fady and Greljo, Admir and Kamenik, Jernej F. and Stamou, Emmanuel and Zupan, Jure",
    title = "{Dark Matter and Gauged Flavor Symmetries}",
    eprint = "1505.03862",
    archivePrefix = "arXiv",
    primaryClass = "hep-ph",
    reportNumber = "CERN-PH-TH-2015-076, ZU-TH-08-15, FERMILAB-PUB-15-137-T",
    doi = "10.1007/JHEP12(2015)130",
    journal = "JHEP",
    volume = "12",
    pages = "130",
    year = "2015"
}

@article{Acaroglu:2022hrm,
    author = "Acaro{\u{g}}lu, Harun and Agrawal, Prateek and Blanke, Monika",
    title = "{Lepton-flavoured scalar dark matter in Dark Minimal Flavour Violation}",
    eprint = "2211.03809",
    archivePrefix = "arXiv",
    primaryClass = "hep-ph",
    reportNumber = "TTP22-063; P3H-22-104",
    doi = "10.1007/JHEP05(2023)106",
    journal = "JHEP",
    volume = "05",
    pages = "106",
    year = "2023"
}

@article{Kile:2014jea,
    author = "Kile, Jennifer and Kobach, Andrew and Soni, Amarjit",
    title = "{Lepton-Flavored Dark Matter}",
    eprint = "1411.1407",
    archivePrefix = "arXiv",
    primaryClass = "hep-ph",
    reportNumber = "NUHEP-14-03",
    doi = "10.1016/j.physletb.2015.04.005",
    journal = "Phys. Lett. B",
    volume = "744",
    pages = "330--338",
    year = "2015"
}

@article{Chen:2015jkt,
    author = "Chen, Mu-Chun and Huang, Jinrui and Takhistov, Volodymyr",
    title = "{Beyond Minimal Lepton Flavored Dark Matter}",
    eprint = "1510.04694",
    archivePrefix = "arXiv",
    primaryClass = "hep-ph",
    reportNumber = "UCI-TR-2015-17, LA-UR-15-27938",
    doi = "10.1007/JHEP02(2016)060",
    journal = "JHEP",
    volume = "02",
    pages = "060",
    year = "2016"
}

@article{Blanke:2017tnb,
    author = "Blanke, Monika and Kast, Simon",
    title = "{Top-Flavoured Dark Matter in Dark Minimal Flavour Violation}",
    eprint = "1702.08457",
    archivePrefix = "arXiv",
    primaryClass = "hep-ph",
    reportNumber = "TTP17-008",
    doi = "10.1007/JHEP05(2017)162",
    journal = "JHEP",
    volume = "05",
    pages = "162",
    year = "2017"
}

@article{Mescia:2024rki,
    author = "Mescia, Federico and Okawa, Shohei and Wu, Keyun",
    title = "{Multi-component dark matter from Minimal Flavor Violation}",
    eprint = "2408.16812",
    archivePrefix = "arXiv",
    primaryClass = "hep-ph",
    reportNumber = "KEK-TH-2649",
    doi = "10.1007/JHEP11(2024)114",
    journal = "JHEP",
    volume = "11",
    pages = "114",
    year = "2024"
}

@article{Belfatto:2025ids,
    author = {Belfatto, Benedetta and Blanke, Monika and Heisig, Jan and Kr{\"a}mer, Michael and Rathmann, Lena and Wilsch, Felix},
    title = "{Toward a Comprehensive Exploration of Flavored Dark Matter Models}",
    eprint = "2511.10490",
    archivePrefix = "arXiv",
    primaryClass = "hep-ph",
    reportNumber = "P3H-25-091, TTP25-043, TTK-25-37",
    month = "11",
    year = "2025"
}

@article{Kile:2011mn,
    author = "Kile, Jennifer and Soni, Amarjit",
    title = "{Flavored Dark Matter in Direct Detection Experiments and at LHC}",
    eprint = "1104.5239",
    archivePrefix = "arXiv",
    primaryClass = "hep-ph",
    reportNumber = "NUHEP-TH-11-01",
    doi = "10.1103/PhysRevD.84.035016",
    journal = "Phys. Rev. D",
    volume = "84",
    pages = "035016",
    year = "2011"
}

@article{An:2013xka,
    author = "An, Haipeng and Wang, Lian-Tao and Zhang, Hao",
    title = "{Dark matter with $t$-channel mediator: a simple step beyond contact interaction}",
    eprint = "1308.0592",
    archivePrefix = "arXiv",
    primaryClass = "hep-ph",
    reportNumber = "IIT-CAPP-13-06, ANL-HEP-PR-13-38",
    doi = "10.1103/PhysRevD.89.115014",
    journal = "Phys. Rev. D",
    volume = "89",
    number = "11",
    pages = "115014",
    year = "2014"
}

@article{DiFranzo:2013vra,
    author = "DiFranzo, Anthony and Nagao, Keiko I. and Rajaraman, Arvind and Tait, Tim M. P.",
    title = "{Simplified Models for Dark Matter Interacting with Quarks}",
    eprint = "1308.2679",
    archivePrefix = "arXiv",
    primaryClass = "hep-ph",
    reportNumber = "UCI-HEP-TR-2013-17, KEK-TH-1659",
    doi = "10.1007/JHEP11(2013)014",
    journal = "JHEP",
    volume = "11",
    pages = "014",
    year = "2013",
    note = "[Erratum: JHEP 01, 162 (2014)]"
}

@article{Bai:2013iqa,
    author = "Bai, Yang and Berger, Joshua",
    title = "{Fermion Portal Dark Matter}",
    eprint = "1308.0612",
    archivePrefix = "arXiv",
    primaryClass = "hep-ph",
    reportNumber = "SLAC-PUB-15704",
    doi = "10.1007/JHEP11(2013)171",
    journal = "JHEP",
    volume = "11",
    pages = "171",
    year = "2013"
}

@article{Cahill-Rowley:2015aea,
    author = "Cahill-Rowley, Matthew and El Hedri, Sonia and Shepherd, William and Walker, Devin G. E.",
    title = "{Perturbative Unitarity Constraints on Charged/Colored Portals}",
    eprint = "1501.03153",
    archivePrefix = "arXiv",
    primaryClass = "hep-ph",
    reportNumber = "SLAC-PUB-16190, MITP-14-104, SCIPP-15-01",
    doi = "10.1016/j.dark.2018.04.003",
    journal = "Phys. Dark Univ.",
    volume = "22",
    pages = "48--59",
    year = "2018"
}

@article{Garny:2011ii,
    author = "Garny, Mathias and Ibarra, Alejandro and Vogl, Stefan",
    title = "{Dark matter annihilations into two light fermions and one gauge boson: General analysis and antiproton constraints}",
    eprint = "1112.5155",
    archivePrefix = "arXiv",
    primaryClass = "hep-ph",
    reportNumber = "DESY-11-257, TUM-HEP-823-11",
    doi = "10.1088/1475-7516/2012/04/033",
    journal = "JCAP",
    volume = "04",
    pages = "033",
    year = "2012"
}

@article{Lopez-Honorez:2013wla,
    author = "Lopez-Honorez, Laura and Merlo, Luca",
    title = "{Dark matter within the minimal flavour violation ansatz}",
    eprint = "1303.1087",
    archivePrefix = "arXiv",
    primaryClass = "hep-ph",
    reportNumber = "FTUAM-13-129, IFT-UAM-CSIC-13-018, CERN-PH-TH-2013-034",
    doi = "10.1016/j.physletb.2013.04.015",
    journal = "Phys. Lett. B",
    volume = "722",
    pages = "135--143",
    year = "2013"
}

@article{Altmannshofer:2025rxc,
    author = "Altmannshofer, Wolfgang and Stangl, Peter",
    title = "{Flavour Physics Beyond the Standard Model}",
    eprint = "2508.03950",
    archivePrefix = "arXiv",
    primaryClass = "hep-ph",
    reportNumber = "CERN-TH-2025-156",
    month = "8",
    year = "2025"
}

@article{Nir:2020jtr,
    author = "Nir, Y.",
    editor = "Mulders, M. and Tr{\^a}n Thanh V{\^a}n, J.",
    title = "{Flavour physics and CP violation}",
    doi = "10.23730/CYRSP-2020-005.79",
    journal = "CERN Yellow Rep. School Proc.",
    volume = "5",
    pages = "79--128",
    year = "2020"
}

@article{Isidori:2025iyu,
    author = "Isidori, Gino",
    title = "{Flavour Physics and CP Violation}",
    eprint = "2503.14042",
    archivePrefix = "arXiv",
    primaryClass = "hep-ph",
    month = "3",
    year = "2025"
}

@article{Cirelli:2024ssz,
    author = "Cirelli, Marco and Strumia, Alessandro and Zupan, Jure",
    title = "{Dark Matter}",
    eprint = "2406.01705",
    archivePrefix = "arXiv",
    primaryClass = "hep-ph",
    month = "6",
    year = "2024"
}

@article{Edsjo:1997bg,
    author = "Edsjo, Joakim and Gondolo, Paolo",
    title = "{Neutralino relic density including coannihilations}",
    eprint = "hep-ph/9704361",
    archivePrefix = "arXiv",
    reportNumber = "UUITP-11-97, MPI-PHT-97-27",
    doi = "10.1103/PhysRevD.56.1879",
    journal = "Phys. Rev. D",
    volume = "56",
    pages = "1879--1894",
    year = "1997"
}

@article{Baker:2015qna,
    author = "Baker, Michael J. and others",
    title = "{The Coannihilation Codex}",
    eprint = "1510.03434",
    archivePrefix = "arXiv",
    primaryClass = "hep-ph",
    reportNumber = "MITP-15-078",
    doi = "10.1007/JHEP12(2015)120",
    journal = "JHEP",
    volume = "12",
    pages = "120",
    year = "2015"
}

@article{Griest:1990kh,
    author = "Griest, Kim and Seckel, David",
    title = "{Three exceptions in the calculation of relic abundances}",
    reportNumber = "CFPA-TH-90-001A, BA-90-79",
    doi = "10.1103/PhysRevD.43.3191",
    journal = "Phys. Rev. D",
    volume = "43",
    pages = "3191--3203",
    year = "1991"
}

@article{Harz:2019rro,
    author = "Harz, Julia and Petraki, Kalliopi",
    title = "{Higgs-mediated bound states in dark-matter models}",
    eprint = "1901.10030",
    archivePrefix = "arXiv",
    primaryClass = "hep-ph",
    reportNumber = "TUM-HEP-1186-19; Nikhef-2019-004",
    doi = "10.1007/JHEP04(2019)130",
    journal = "JHEP",
    volume = "04",
    pages = "130",
    year = "2019"
}

@article{Greljo:2022cah,
    author = "Greljo, Admir and Palavri{\'c}, Ajdin and Thomsen, Anders Eller",
    title = "{Adding Flavor to the SMEFT}",
    eprint = "2203.09561",
    archivePrefix = "arXiv",
    primaryClass = "hep-ph",
    doi = "10.1007/JHEP10(2022)005",
    journal = "JHEP",
    volume = "10",
    pages = "010",
    year = "2022"
}

@article{Altmannshofer:2024hmr,
    author = "Altmannshofer, Wolfgang and Greljo, Admir",
    title = "{Recent Progress in Flavor Model Building}",
    eprint = "2412.04549",
    archivePrefix = "arXiv",
    primaryClass = "hep-ph",
    doi = "10.1146/annurev-nucl-121423-100950",
    month = "12",
    year = "2024"
}

@article{Faroughy:2020ina,
    author = "Faroughy, Darius A. and Isidori, Gino and Wilsch, Felix and Yamamoto, Kei",
    title = "{Flavour symmetries in the SMEFT}",
    eprint = "2005.05366",
    archivePrefix = "arXiv",
    primaryClass = "hep-ph",
    doi = "10.1007/JHEP08(2020)166",
    journal = "JHEP",
    volume = "08",
    pages = "166",
    year = "2020"
}

@article{Arcadi:2021cwg,
    author = "Arcadi, Giorgio and Calibbi, Lorenzo and Fedele, Marco and Mescia, Federico",
    title = "{Muon $g-2$ and $B$-anomalies from Dark Matter}",
    eprint = "2104.03228",
    archivePrefix = "arXiv",
    primaryClass = "hep-ph",
    reportNumber = "TTP21-009, P3H-21-025",
    doi = "10.1103/PhysRevLett.127.061802",
    journal = "Phys. Rev. Lett.",
    volume = "127",
    number = "6",
    pages = "061802",
    year = "2021"
}

@article{Arcadi:2021glq,
    author = "Arcadi, Giorgio and Calibbi, Lorenzo and Fedele, Marco and Mescia, Federico",
    title = "{Systematic approach to B-physics anomalies and t-channel dark matter}",
    eprint = "2103.09835",
    archivePrefix = "arXiv",
    primaryClass = "hep-ph",
    reportNumber = "TTP21-007, P3H-21-017",
    doi = "10.1103/PhysRevD.104.115012",
    journal = "Phys. Rev. D",
    volume = "104",
    number = "11",
    pages = "115012",
    year = "2021"
}

@article{Marzocca:2024hua,
    author = "Marzocca, David and Nardecchia, Marco and Stanzione, Alfredo and Toni, Claudio",
    title = "{Implications of $B \rightarrow K \nu {\bar{\nu }}$ under rank-one flavor violation hypothesis}",
    eprint = "2404.06533",
    archivePrefix = "arXiv",
    primaryClass = "hep-ph",
    doi = "10.1140/epjc/s10052-024-13534-7",
    journal = "Eur. Phys. J. C",
    volume = "84",
    number = "11",
    pages = "1217",
    year = "2024"
}

@article{Gherardi:2019zil,
    author = "Gherardi, Valerio and Marzocca, David and Nardecchia, Marco and Romanino, Andrea",
    title = "{Rank-One Flavor Violation and B-meson anomalies}",
    eprint = "1903.10954",
    archivePrefix = "arXiv",
    primaryClass = "hep-ph",
    doi = "10.1007/JHEP10(2019)112",
    journal = "JHEP",
    volume = "10",
    pages = "112",
    year = "2019"
}

@article{ATLAS:2020syg,
    author = "Aad, Georges and others",
    collaboration = "ATLAS",
    title = "{Search for squarks and gluinos in final states with jets and missing transverse momentum using 139 fb$^{-1}$ of $\sqrt{s}$ =13 TeV $pp$ collision data with the ATLAS detector}",
    eprint = "2010.14293",
    archivePrefix = "arXiv",
    primaryClass = "hep-ex",
    reportNumber = "CERN-EP-2020-166",
    doi = "10.1007/JHEP02(2021)143",
    journal = "JHEP",
    volume = "02",
    pages = "143",
    year = "2021"
}

@article{ParticleDataGroup:2024cfk,
    author = "Navas, S. and others",
    collaboration = "Particle Data Group",
    title = "{Review of particle physics}",
    doi = "10.1103/PhysRevD.110.030001",
    journal = "Phys. Rev. D",
    volume = "110",
    number = "3",
    pages = "030001",
    year = "2024"
}

@article{Dowdall:2019bea,
    author = "Dowdall, R. J. and Davies, C. T. H. and Horgan, R. R. and Lepage, G. P. and Monahan, C. J. and Shigemitsu, J. and Wingate, M.",
    title = "{Neutral B-meson mixing from full lattice QCD at the physical point}",
    eprint = "1907.01025",
    archivePrefix = "arXiv",
    primaryClass = "hep-lat",
    reportNumber = "INT-PUB-19-031, JLAB-THY-19-3068",
    doi = "10.1103/PhysRevD.100.094508",
    journal = "Phys. Rev. D",
    volume = "100",
    number = "9",
    pages = "094508",
    year = "2019"
}

@article{FlavourLatticeAveragingGroupFLAG:2024oxs,
    author = "Aoki, Y. and others",
    collaboration = "Flavour Lattice Averaging Group (FLAG)",
    title = "{FLAG Review 2024}",
    eprint = "2411.04268",
    archivePrefix = "arXiv",
    primaryClass = "hep-lat",
    reportNumber = "CERN-TH-2024-192, FERMILAB-PUB-24-0785-T",
    month = "11",
    year = "2024"
}

@article{Barbieri:2012uh,
    author = "Barbieri, Riccardo and Buttazzo, Dario and Sala, Filippo and Straub, David M.",
    title = "{Flavour physics from an approximate $U(2)^3$ symmetry}",
    eprint = "1203.4218",
    archivePrefix = "arXiv",
    primaryClass = "hep-ph",
    doi = "10.1007/JHEP07(2012)181",
    journal = "JHEP",
    volume = "07",
    pages = "181",
    year = "2012"
}

@article{ATLAS:2024fub,
    author = "Aad, Georges and others",
    collaboration = "ATLAS",
    title = "{Search for electroweak production of supersymmetric particles in final states with two {\ensuremath{\tau}}-leptons in $ \sqrt{s} $ = 13 TeV pp collisions with the ATLAS detector}",
    eprint = "2402.00603",
    archivePrefix = "arXiv",
    primaryClass = "hep-ex",
    reportNumber = "CERN-EP-2023-295",
    doi = "10.1007/JHEP05(2024)150",
    journal = "JHEP",
    volume = "05",
    pages = "150",
    year = "2024"
}

@article{ATLAS:2019lff,
    author = "Aad, Georges and others",
    collaboration = "ATLAS",
    title = "{Search for electroweak production of charginos and sleptons decaying into final states with two leptons and missing transverse momentum in $\sqrt{s}=13$ TeV $pp$ collisions using the ATLAS detector}",
    eprint = "1908.08215",
    archivePrefix = "arXiv",
    primaryClass = "hep-ex",
    reportNumber = "CERN-EP-2019-106",
    doi = "10.1140/epjc/s10052-019-7594-6",
    journal = "Eur. Phys. J. C",
    volume = "80",
    number = "2",
    pages = "123",
    year = "2020"
}

@article{ATLAS:2019lng,
    author = "Aad, Georges and others",
    collaboration = "ATLAS",
    title = "{Searches for electroweak production of supersymmetric particles with compressed mass spectra in $\sqrt{s}=$ 13 TeV $pp$ collisions with the ATLAS detector}",
    eprint = "1911.12606",
    archivePrefix = "arXiv",
    primaryClass = "hep-ex",
    reportNumber = "CERN-EP-2019-242",
    doi = "10.1103/PhysRevD.101.052005",
    journal = "Phys. Rev. D",
    volume = "101",
    number = "5",
    pages = "052005",
    year = "2020"
}

@article{ATLAS:2022hbt,
    author = "Aad, Georges and others",
    collaboration = "ATLAS",
    title = "{Search for direct pair production of sleptons and charginos decaying to two leptons and neutralinos with mass splittings near the W-boson mass in $ \sqrt{s} $ = 13 TeV pp collisions with the ATLAS detector}",
    eprint = "2209.13935",
    archivePrefix = "arXiv",
    primaryClass = "hep-ex",
    reportNumber = "CERN-EP-2022-132",
    doi = "10.1007/JHEP06(2023)031",
    journal = "JHEP",
    volume = "06",
    pages = "031",
    year = "2023"
}

@article{CMS:2025ttk,
    author = "Chekhovsky, Vladimir and others",
    collaboration = "CMS",
    title = "{A general search for supersymmetric particles in scenarios with compressed mass spectra using proton-proton collisions at $\sqrt{s}$ = 13 TeV}",
    eprint = "2508.13900",
    archivePrefix = "arXiv",
    primaryClass = "hep-ex",
    reportNumber = "CMS-SUS-23-003, CERN-EP-2025-154",
    month = "8",
    year = "2025"
}

@article{CMS:2019zmn,
    author = "Sirunyan, Albert M and others",
    collaboration = "CMS",
    title = "{Search for Supersymmetry with a Compressed Mass Spectrum in Events with a Soft $\tau$ Lepton, a Highly Energetic Jet, and Large Missing Transverse Momentum in Proton-Proton Collisions at $\sqrt{s}=$  TeV}",
    eprint = "1910.01185",
    archivePrefix = "arXiv",
    primaryClass = "hep-ex",
    reportNumber = "CMS-SUS-19-002, CERN-EP-2019-196",
    doi = "10.1103/PhysRevLett.124.041803",
    journal = "Phys. Rev. Lett.",
    volume = "124",
    number = "4",
    pages = "041803",
    year = "2020"
}

@article{ATLAS:2021yij,
    author = "Aad, Georges and others",
    collaboration = "ATLAS",
    title = "{Search for new phenomena in final states with $b$-jets and missing transverse momentum in $\sqrt{s}=13$ TeV $pp$ collisions with the ATLAS detector}",
    eprint = "2101.12527",
    archivePrefix = "arXiv",
    primaryClass = "hep-ex",
    reportNumber = "CERN-EP-2021-001",
    doi = "10.1007/JHEP05(2021)093",
    journal = "JHEP",
    volume = "05",
    pages = "093",
    year = "2021"
}

@article{CMS:2019ybf,
    author = "Sirunyan, Albert M and others",
    collaboration = "CMS",
    title = "{Searches for physics beyond the standard model with the $M_\mathrm{T2}$ variable in hadronic final states with and without disappearing tracks in proton-proton collisions at $\sqrt{s}=$ 13 TeV}",
    eprint = "1909.03460",
    archivePrefix = "arXiv",
    primaryClass = "hep-ex",
    reportNumber = "CMS-SUS-19-005, CERN-EP-2019-180",
    doi = "10.1140/epjc/s10052-019-7493-x",
    journal = "Eur. Phys. J. C",
    volume = "80",
    number = "1",
    pages = "3",
    year = "2020"
}

@article{ATLAS:2020dsf,
    author = "Aad, Georges and others",
    collaboration = "ATLAS",
    title = "{Search for a scalar partner of the top quark in the all-hadronic $t{\bar{t}}$ plus missing transverse momentum final state at $\sqrt{s}=13$ TeV with the ATLAS detector}",
    eprint = "2004.14060",
    archivePrefix = "arXiv",
    primaryClass = "hep-ex",
    reportNumber = "CERN-EP-2020-044",
    doi = "10.1140/epjc/s10052-020-8102-8",
    journal = "Eur. Phys. J. C",
    volume = "80",
    number = "8",
    pages = "737",
    year = "2020"
}

@article{ATLAS:2020xzu,
    author = "Aad, Georges and others",
    collaboration = "ATLAS",
    title = "{Search for new phenomena with top quark pairs in final states with one lepton, jets, and missing transverse momentum in $pp$ collisions at $ \sqrt{s} $ = 13 TeV with the ATLAS detector}",
    eprint = "2012.03799",
    archivePrefix = "arXiv",
    primaryClass = "hep-ex",
    reportNumber = "CERN-EP-2020-177",
    doi = "10.1007/JHEP04(2021)174",
    journal = "JHEP",
    volume = "04",
    pages = "174",
    year = "2021"
}

@article{ATLAS:2024rcx,
    author = "Aad, Georges and others",
    collaboration = "ATLAS",
    title = "{Search for new phenomena with top-quark pairs and large missing transverse momentum using 140 fb$^{-1}$ of pp collision data at $ \sqrt{s} $ = 13 TeV with the ATLAS detector}",
    eprint = "2401.13430",
    archivePrefix = "arXiv",
    primaryClass = "hep-ex",
    reportNumber = "CERN-EP-2024-003",
    doi = "10.1007/JHEP03(2024)139",
    journal = "JHEP",
    volume = "03",
    pages = "139",
    year = "2024"
}

@article{ATLAS:2024lda,
    author = "Aad, Georges and others",
    collaboration = "ATLAS",
    title = "{The quest to discover supersymmetry at the ATLAS experiment}",
    eprint = "2403.02455",
    archivePrefix = "arXiv",
    primaryClass = "hep-ex",
    reportNumber = "CERN-EP-2024-056",
    doi = "10.1016/j.physrep.2024.09.010",
    journal = "Phys. Rept.",
    volume = "1116",
    pages = "261--300",
    year = "2025"
}

@article{CMS:2019zmd,
    author = "Collaboration, The Cms and others",
    collaboration = "CMS",
    title = "{Search for supersymmetry in proton-proton collisions at 13 TeV in final states with jets and missing transverse momentum}",
    eprint = "1908.04722",
    archivePrefix = "arXiv",
    primaryClass = "hep-ex",
    reportNumber = "CMS-SUS-19-006, CERN-EP-2019-152",
    doi = "10.1007/JHEP10(2019)244",
    journal = "JHEP",
    volume = "10",
    pages = "244",
    year = "2019"
}

@article{ATLAS:2017bfj,
    author = "Aaboud, Morad and others",
    collaboration = "ATLAS",
    title = "{Search for dark matter and other new phenomena in events with an energetic jet and large missing transverse momentum using the ATLAS detector}",
    eprint = "1711.03301",
    archivePrefix = "arXiv",
    primaryClass = "hep-ex",
    reportNumber = "CERN-EP-2017-230",
    doi = "10.1007/JHEP01(2018)126",
    journal = "JHEP",
    volume = "01",
    pages = "126",
    year = "2018"
}

@article{ATLAS:2021kxv,
    author = "Aad, Georges and others",
    collaboration = "ATLAS",
    title = "{Search for new phenomena in events with an energetic jet and missing transverse momentum in $pp$ collisions at $\sqrt {s}$ =13  TeV with the ATLAS detector}",
    eprint = "2102.10874",
    archivePrefix = "arXiv",
    primaryClass = "hep-ex",
    reportNumber = "CERN-EP-2020-238",
    doi = "10.1103/PhysRevD.103.112006",
    journal = "Phys. Rev. D",
    volume = "103",
    number = "11",
    pages = "112006",
    year = "2021"
}

@article{Kopp:2014tsa,
    author = "Kopp, Joachim and Michaels, Lisa and Smirnov, Juri",
    title = "{Loopy Constraints on Leptophilic Dark Matter and Internal Bremsstrahlung}",
    eprint = "1401.6457",
    archivePrefix = "arXiv",
    primaryClass = "hep-ph",
    doi = "10.1088/1475-7516/2014/04/022",
    journal = "JCAP",
    volume = "04",
    pages = "022",
    year = "2014"
}

@article{OPAL:2002lje,
    author = "Abbiendi, G. and others",
    collaboration = "OPAL",
    title = "{Search for nearly mass degenerate charginos and neutralinos at LEP}",
    eprint = "hep-ex/0210043",
    archivePrefix = "arXiv",
    reportNumber = "CERN-EP-2002-063",
    doi = "10.1140/epjc/s2003-01237-x",
    journal = "Eur. Phys. J. C",
    volume = "29",
    pages = "479--489",
    year = "2003"
}

@article{Chivukula:1987py,
    author = "Chivukula, R. Sekhar and Georgi, Howard",
    title = "{Composite Technicolor Standard Model}",
    reportNumber = "BUHEP-87-2, HUTP-87/A003",
    doi = "10.1016/0370-2693(87)90713-1",
    journal = "Phys. Lett. B",
    volume = "188",
    pages = "99--104",
    year = "1987"
}

@article{DAmbrosio:2002vsn,
    author = "D'Ambrosio, G. and Giudice, G. F. and Isidori, G. and Strumia, A.",
    title = "{Minimal flavor violation: An Effective field theory approach}",
    eprint = "hep-ph/0207036",
    archivePrefix = "arXiv",
    reportNumber = "CERN-TH-2002-147, IFUP-TH-2002-17",
    doi = "10.1016/S0550-3213(02)00836-2",
    journal = "Nucl. Phys. B",
    volume = "645",
    pages = "155--187",
    year = "2002"
}

@article{Barbieri:2011ci,
    author = "Barbieri, Riccardo and Isidori, Gino and Jones-Perez, Joel and Lodone, Paolo and Straub, David M.",
    title = "{$U(2)$ and Minimal Flavour Violation in Supersymmetry}",
    eprint = "1105.2296",
    archivePrefix = "arXiv",
    primaryClass = "hep-ph",
    doi = "10.1140/epjc/s10052-011-1725-z",
    journal = "Eur. Phys. J. C",
    volume = "71",
    pages = "1725",
    year = "2011"
}

@article{Allwicher:2023shc,
    author = "Allwicher, Lukas and Cornella, Claudia and Isidori, Gino and Stefanek, Ben A.",
    title = "{New physics in the third generation. A comprehensive SMEFT analysis and future prospects}",
    eprint = "2311.00020",
    archivePrefix = "arXiv",
    primaryClass = "hep-ph",
    reportNumber = "ZU-TH 71/23, MITP-23-060, KCL-PH-TH/2023-59",
    doi = "10.1007/JHEP03(2024)049",
    journal = "JHEP",
    volume = "03",
    pages = "049",
    year = "2024"
}

@article{Martin:1997ns,
    author = "Martin, Stephen P.",
    editor = "Kane, Gordon L.",
    title = "{A Supersymmetry primer}",
    eprint = "hep-ph/9709356",
    archivePrefix = "arXiv",
    reportNumber = "FERMILAB-PUB-97-425-T",
    doi = "10.1142/9789812839657_0001",
    journal = "Adv. Ser. Direct. High Energy Phys.",
    volume = "18",
    pages = "1--98",
    year = "1998"
}

@article{DELPHI:2003uqw,
    author = "Abdallah, J. and others",
    collaboration = "DELPHI",
    title = "{Searches for supersymmetric particles in e+ e- collisions up to 208-GeV and interpretation of the results within the MSSM}",
    eprint = "hep-ex/0311019",
    archivePrefix = "arXiv",
    reportNumber = "CERN-EP-2003-007",
    doi = "10.1140/epjc/s2003-01355-5",
    journal = "Eur. Phys. J. C",
    volume = "31",
    pages = "421--479",
    year = "2003"
}

@article{Blondel:2013ia,
    author = "Blondel, A. and others",
    title = "{Research Proposal for an Experiment to Search for the Decay $\mu \to eee$}",
    eprint = "1301.6113",
    archivePrefix = "arXiv",
    primaryClass = "physics.ins-det",
    month = "1",
    year = "2013"
}

@article{Belle-II:2018jsg,
    author = "Altmannshofer, W. and others",
    editor = "Kou, E. and Urquijo, P.",
    collaboration = "Belle-II",
    title = "{The Belle II Physics Book}",
    eprint = "1808.10567",
    archivePrefix = "arXiv",
    primaryClass = "hep-ex",
    reportNumber = "KEK Preprint 2018-27, BELLE2-PUB-PH-2018-001, FERMILAB-PUB-18-398-T, JLAB-THY-18-2780, INT-PUB-18-047, UWThPh 2018-26",
    doi = "10.1093/ptep/ptz106",
    journal = "PTEP",
    volume = "2019",
    number = "12",
    pages = "123C01",
    year = "2019",
    note = "[Erratum: PTEP 2020, 029201 (2020)]"
}

@article{Banerjee:2022xuw,
    author = "Banerjee, Swagato and others",
    title = "{Snowmass 2021 White Paper: Charged lepton flavor violation in the tau sector}",
    eprint = "2203.14919",
    archivePrefix = "arXiv",
    primaryClass = "hep-ph",
    month = "3",
    year = "2022"
}

@article{deBlas:2025gyz,
    author = "de Blas, Jorge and others",
    title = "{Physics Briefing Book: Input for the 2026 update of the European Strategy for Particle Physics}",
    eprint = "2511.03883",
    archivePrefix = "arXiv",
    primaryClass = "hep-ex",
    reportNumber = "CERN-ESU-2025-001, CERN-ESU-2025-001",
    doi = "10.17181/CERN.35CH.2O2P",
    month = "11",
    year = "2025"
}

@article{MEGII:2018kmf,
    author = "Baldini, A. M. and others",
    collaboration = "MEG II",
    title = "{The design of the MEG II experiment}",
    eprint = "1801.04688",
    archivePrefix = "arXiv",
    primaryClass = "physics.ins-det",
    doi = "10.1140/epjc/s10052-018-5845-6",
    journal = "Eur. Phys. J. C",
    volume = "78",
    number = "5",
    pages = "380",
    year = "2018"
}

@article{MEGII:2023ltw,
    author = "Afanaciev, K. and others",
    collaboration = "MEG II",
    title = "{A search for $\mu ^+ \rightarrow \textrm{e}^+ \gamma $ with the first dataset of the MEG~II experiment}",
    eprint = "2310.12614",
    archivePrefix = "arXiv",
    primaryClass = "hep-ex",
    doi = "10.1140/epjc/s10052-024-12416-2",
    journal = "Eur. Phys. J. C",
    volume = "84",
    number = "3",
    pages = "216",
    year = "2024",
    note = "[Erratum: Eur.Phys.J.C 84, 1042 (2024)]"
}

@article{Ibarra:2014qma,
    author = "Ibarra, Alejandro and Toma, Takashi and Totzauer, Maximilian and Wild, Sebastian",
    title = "{Sharp Gamma-ray Spectral Features from Scalar Dark Matter Annihilations}",
    eprint = "1405.6917",
    archivePrefix = "arXiv",
    primaryClass = "hep-ph",
    reportNumber = "TUM-HEP-946-14, IPPP-14-48, DCPT-14-96",
    doi = "10.1103/PhysRevD.90.043526",
    journal = "Phys. Rev. D",
    volume = "90",
    number = "4",
    pages = "043526",
    year = "2014"
}

@article{Liu:2021mhn,
    author = "Liu, Jia and Wang, Xiao-Ping and Xie, Ke-Pan",
    title = "{Searching for lepton portal dark matter with colliders and gravitational waves}",
    eprint = "2104.06421",
    archivePrefix = "arXiv",
    primaryClass = "hep-ph",
    doi = "10.1007/JHEP06(2021)149",
    journal = "JHEP",
    volume = "06",
    pages = "149",
    year = "2021"
}

@article{BaBar:2009hkt,
    author = "Aubert, Bernard and others",
    collaboration = "BaBar",
    title = "{Searches for Lepton Flavor Violation in the Decays $\tau^\pm \to e^\pm \gamma$ and $\tau^\pm \to \mu^\pm \gamma$}",
    eprint = "0908.2381",
    archivePrefix = "arXiv",
    primaryClass = "hep-ex",
    reportNumber = "SLAC-PUB-13753, BABAR-PUB-09-026",
    doi = "10.1103/PhysRevLett.104.021802",
    journal = "Phys. Rev. Lett.",
    volume = "104",
    pages = "021802",
    year = "2010"
}

@article{Belle:2021ysv,
    author = "Abdesselam, A. and others",
    collaboration = "Belle",
    title = "{Search for lepton-flavor-violating tau-lepton decays to $\ell\gamma$ at Belle}",
    eprint = "2103.12994",
    archivePrefix = "arXiv",
    primaryClass = "hep-ex",
    doi = "10.1007/JHEP10(2021)019",
    journal = "JHEP",
    volume = "10",
    pages = "19",
    year = "2021"
}

@article{Fernandez-Martinez:2024bxg,
    author = "Fern{\'a}ndez-Mart{\'\i}nez, Enrique and Marcano, Xabier and Naredo-Tuero, Daniel",
    title = "{Global lepton flavour violating constraints on new physics}",
    eprint = "2403.09772",
    archivePrefix = "arXiv",
    primaryClass = "hep-ph",
    reportNumber = "IFT-UAM/CSIC-24-39",
    doi = "10.1140/epjc/s10052-024-12973-6",
    journal = "Eur. Phys. J. C",
    volume = "84",
    number = "7",
    pages = "666",
    year = "2024"
}

@article{Calibbi:2022ddo,
    author = "Calibbi, Lorenzo and Li, Tong and Marcano, Xabier and Schmidt, Michael A.",
    title = "{Indirect constraints on lepton-flavor-violating quarkonium decays}",
    eprint = "2207.10913",
    archivePrefix = "arXiv",
    primaryClass = "hep-ph",
    reportNumber = "CPPC-2022-08, IFT-UAM/CSIC-22-82",
    doi = "10.1103/PhysRevD.106.115039",
    journal = "Phys. Rev. D",
    volume = "106",
    number = "11",
    pages = "115039",
    year = "2022"
}

@article{ElHedri:2016onc,
    author = "El Hedri, Sonia and Kaminska, Anna and de Vries, Maikel",
    title = "{A Sommerfeld Toolbox for Colored Dark Sectors}",
    eprint = "1612.02825",
    archivePrefix = "arXiv",
    primaryClass = "hep-ph",
    reportNumber = "MITP-16-135",
    doi = "10.1140/epjc/s10052-017-5168-z",
    journal = "Eur. Phys. J. C",
    volume = "77",
    number = "9",
    pages = "622",
    year = "2017"
}

@article{Becker:2025vgq,
    author = "Becker, Mathias and Copello, Emanuele and Harz, Julia and Napetschnig, Martin",
    title = "{Manual for SE+BSF4DM -- A micrOMEGAs package for Sommerfeld Effect and Bound State Formation in colored Dark Sectors}",
    eprint = "2512.02155",
    archivePrefix = "arXiv",
    primaryClass = "hep-ph",
    reportNumber = "TUM-HEP-1572/25, MITP-25-077",
    month = "12",
    year = "2025"
}

@article{DeLaTorreLuque:2023fyg,
    author = "De La Torre Luque, Pedro and Smirnov, Juri and Linden, Tim",
    title = "{Gamma-ray lines in 15~years of Fermi-LAT data: New constraints on Higgs portal dark matter}",
    eprint = "2309.03281",
    archivePrefix = "arXiv",
    primaryClass = "hep-ph",
    doi = "10.1103/PhysRevD.109.L041301",
    journal = "Phys. Rev. D",
    volume = "109",
    number = "4",
    pages = "L041301",
    year = "2024"
}

@article{Garny:2017rxs,
    author = {Garny, Mathias and Heisig, Jan and L{\"u}lf, Benedikt and Vogl, Stefan},
    title = "{Coannihilation without chemical equilibrium}",
    eprint = "1705.09292",
    archivePrefix = "arXiv",
    primaryClass = "hep-ph",
    reportNumber = "TUM-HEP-1085-17, TTK-17-18",
    doi = "10.1103/PhysRevD.96.103521",
    journal = "Phys. Rev. D",
    volume = "96",
    number = "10",
    pages = "103521",
    year = "2017"
}

@article{Fermi-LAT:2022byn,
    author = "Abdollahi, Soheila and others",
    collaboration = "Fermi-LAT",
    title = "{Incremental Fermi Large Area Telescope Fourth Source Catalog}",
    eprint = "2201.11184",
    archivePrefix = "arXiv",
    primaryClass = "astro-ph.HE",
    doi = "10.3847/1538-4365/ac6751",
    journal = "Astrophys. J. Supp.",
    volume = "260",
    number = "2",
    pages = "53",
    year = "2022"
}

@article{HESS:2018cbt,
    author = "Abdallah, H. and others",
    collaboration = "HESS",
    title = "{Search for $\gamma$-Ray Line Signals from Dark Matter Annihilations in the Inner Galactic Halo from 10 Years of Observations with H.E.S.S.}",
    eprint = "1805.05741",
    archivePrefix = "arXiv",
    primaryClass = "astro-ph.HE",
    doi = "10.1103/PhysRevLett.120.201101",
    journal = "Phys. Rev. Lett.",
    volume = "120",
    number = "20",
    pages = "201101",
    year = "2018"
}

@article{Fuentes-Martin:2022jrf,
    author = {Fuentes-Mart{\'\i}n, Javier and K{\"o}nig, Matthias and Pag{\`e}s, Julie and Thomsen, Anders Eller and Wilsch, Felix},
    title = "{A proof of concept for matchete: an automated tool for matching effective theories}",
    eprint = "2212.04510",
    archivePrefix = "arXiv",
    primaryClass = "hep-ph",
    reportNumber = "MITP-22-105, TUM-HEP-1443/22, ZU-TH-58/22",
    doi = "10.1140/epjc/s10052-023-11726-1",
    journal = "Eur. Phys. J. C",
    volume = "83",
    number = "7",
    pages = "662",
    year = "2023"
}

@article{Dorsner:2018ynv,
    author = "Dor{\v{s}}ner, Ilja and Greljo, Admir",
    title = "{Leptoquark toolbox for precision collider studies}",
    eprint = "1801.07641",
    archivePrefix = "arXiv",
    primaryClass = "hep-ph",
    reportNumber = "MITP-18-005",
    doi = "10.1007/JHEP05(2018)126",
    journal = "JHEP",
    volume = "05",
    pages = "126",
    year = "2018"
}

@article{Buras:2024mnq,
    author = "Buras, Andrzej J. and Stangl, Peter",
    title = "{On the interplay of constraints from $B_{s},$D,~ and K meson mixing in $Z^\prime $ models with implications for $b\!\rightarrow \! s \nu {\bar{\nu }}$ transitions}",
    eprint = "2412.14254",
    archivePrefix = "arXiv",
    primaryClass = "hep-ph",
    reportNumber = "AJB-24-3, CERN-TH-2024-218",
    doi = "10.1140/epjc/s10052-025-14168-z",
    journal = "Eur. Phys. J. C",
    volume = "85",
    number = "5",
    pages = "519",
    year = "2025"
}

@article{Gorbahn:2024qpe,
    author = {Gorbahn, Martin and J{\"a}ger, Sebastian and Kvedarait{\.{e}}, Sandra},
    title = "{RI-(S)MOM to $ \overline{\textrm{MS}} $ conversion for B$_{K}$ at two-loop order}",
    eprint = "2411.19861",
    archivePrefix = "arXiv",
    primaryClass = "hep-ph",
    doi = "10.1007/JHEP09(2025)011",
    journal = "JHEP",
    volume = "09",
    pages = "011",
    year = "2025"
}

@article{DiLuzio:2023ndz,
    author = "Di Luzio, Luca and Guerrera, Alfredo Walter Mario and D{\'\i}az, Xavier Ponce and Rigolin, Stefano",
    title = "{On the IR/UV flavour connection in non-universal axion models}",
    eprint = "2304.04643",
    archivePrefix = "arXiv",
    primaryClass = "hep-ph",
    doi = "10.1007/JHEP06(2023)046",
    journal = "JHEP",
    volume = "06",
    pages = "046",
    year = "2023"
}

@article{Ciuchini:1998ix,
    author = "Ciuchini, Marco and others",
    title = "{Delta M(K) and epsilon(K) in SUSY at the next-to-leading order}",
    eprint = "hep-ph/9808328",
    archivePrefix = "arXiv",
    reportNumber = "TUM-HEP-320-98, ROME1-1217-98, EDINBURGH-98-14, ROM2F-98-33",
    doi = "10.1088/1126-6708/1998/10/008",
    journal = "JHEP",
    volume = "10",
    pages = "008",
    year = "1998"
}

@article{Becirevic:2001jj,
    author = "Becirevic, D. and Ciuchini, Marco and Franco, E. and Gimenez, V. and Martinelli, G. and Masiero, A. and Papinutto, M. and Reyes, J. and Silvestrini, L.",
    title = "{$B_d - \bar{B}_d$ mixing and the $B_d \to J/\psi K_s$ asymmetry in general SUSY models}",
    eprint = "hep-ph/0112303",
    archivePrefix = "arXiv",
    reportNumber = "DFPD-01-TH-55, FTUV-IFIC-01-1129, RM3-TH-01-16, ROMA-1328-01",
    doi = "10.1016/S0550-3213(02)00291-2",
    journal = "Nucl. Phys. B",
    volume = "634",
    pages = "105--119",
    year = "2002"
}

@article{UTfit:2007eik,
    author = "Bona, M. and others",
    collaboration = "UTfit",
    title = "{Model-independent constraints on $\Delta F=2$ operators and the scale of new physics}",
    eprint = "0707.0636",
    archivePrefix = "arXiv",
    primaryClass = "hep-ph",
    doi = "10.1088/1126-6708/2008/03/049",
    journal = "JHEP",
    volume = "03",
    pages = "049",
    year = "2008"
}

@article{Carrasco:2014uya,
    author = "Carrasco, N. and others",
    title = "{$D^0-\bar{D}^0$ mixing in the standard model and beyond from $N_f$ =2 twisted mass QCD}",
    eprint = "1403.7302",
    archivePrefix = "arXiv",
    primaryClass = "hep-lat",
    doi = "10.1103/PhysRevD.90.014502",
    journal = "Phys. Rev. D",
    volume = "90",
    number = "1",
    pages = "014502",
    year = "2014"
}

@article{Brod:2011ty,
    author = "Brod, Joachim and Gorbahn, Martin",
    title = "{Next-to-Next-to-Leading-Order Charm-Quark Contribution to the $CP$ Violation Parameter $\epsilon_K$ and $\Delta M_K$}",
    eprint = "1108.2036",
    archivePrefix = "arXiv",
    primaryClass = "hep-ph",
    doi = "10.1103/PhysRevLett.108.121801",
    journal = "Phys. Rev. Lett.",
    volume = "108",
    pages = "121801",
    year = "2012"
}

@article{Vairo:2003gh,
    author = "Vairo, Antonio",
    title = "{A Theoretical review of heavy quarkonium inclusive decays}",
    eprint = "hep-ph/0311303",
    archivePrefix = "arXiv",
    reportNumber = "IFUM-778-FT",
    doi = "10.1142/S0217732304012927",
    journal = "Mod. Phys. Lett. A",
    volume = "19",
    pages = "253--269",
    year = "2004"
}

@article{Bodwin:1994jh,
    author = "Bodwin, Geoffrey T. and Braaten, Eric and Lepage, G. Peter",
    title = "{Rigorous QCD analysis of inclusive annihilation and production of heavy quarkonium}",
    eprint = "hep-ph/9407339",
    archivePrefix = "arXiv",
    reportNumber = "ANL-HEP-PR-94-24, FERMILAB-PUB-94-073-T, NUHEP-TH-94-5",
    doi = "10.1103/PhysRevD.55.5853",
    journal = "Phys. Rev. D",
    volume = "51",
    pages = "1125--1171",
    year = "1995",
    note = "[Erratum: Phys.Rev.D 55, 5853 (1997)]"
}

@article{Binder:2021vfo,
    author = "Binder, Tobias and Filimonova, Anastasiia and Petraki, Kalliopi and White, Graham",
    title = "{Saha equilibrium for metastable bound states and dark matter freeze-out}",
    eprint = "2112.00042",
    archivePrefix = "arXiv",
    primaryClass = "hep-ph",
    doi = "10.1016/j.physletb.2022.137323",
    journal = "Phys. Lett. B",
    volume = "833",
    pages = "137323",
    year = "2022"
}

@article{Mitridate:2017izz,
    author = "Mitridate, Andrea and Redi, Michele and Smirnov, Juri and Strumia, Alessandro",
    title = "{Cosmological Implications of Dark Matter Bound States}",
    eprint = "1702.01141",
    archivePrefix = "arXiv",
    primaryClass = "hep-ph",
    reportNumber = "CERN-TH-2017-030, IFUP-TH-2017",
    doi = "10.1088/1475-7516/2017/05/006",
    journal = "JCAP",
    volume = "05",
    pages = "006",
    year = "2017"
}

@article{Brod:2019rzc,
    author = "Brod, Joachim and Gorbahn, Martin and Stamou, Emmanuel",
    title = "{Standard-Model Prediction of $\epsilon_K$ with Manifest Quark-Mixing Unitarity}",
    eprint = "1911.06822",
    archivePrefix = "arXiv",
    primaryClass = "hep-ph",
    doi = "10.1103/PhysRevLett.125.171803",
    journal = "Phys. Rev. Lett.",
    volume = "125",
    number = "17",
    pages = "171803",
    year = "2020"
}

@article{Arcadi:2023imv,
    author = "Arcadi, Giorgio and Cabo-Almeida, David and Mescia, Federico and Virto, Javier",
    title = "{Dark Matter Direct Detection in {\ensuremath{\mathsf{t}}}-channel mediator models}",
    eprint = "2309.07896",
    archivePrefix = "arXiv",
    primaryClass = "hep-ph",
    doi = "10.1088/1475-7516/2024/02/005",
    journal = "JCAP",
    volume = "02",
    pages = "005",
    year = "2024"
}

@article{Cirelli:2005uq,
    author = "Cirelli, Marco and Fornengo, Nicolao and Strumia, Alessandro",
    title = "{Minimal dark matter}",
    eprint = "hep-ph/0512090",
    archivePrefix = "arXiv",
    reportNumber = "DFTT40-2005, IFUP-TH-2005-34",
    doi = "10.1016/j.nuclphysb.2006.07.012",
    journal = "Nucl. Phys. B",
    volume = "753",
    pages = "178--194",
    year = "2006"
}

@article{Chang:2013oia,
    author = "Chang, Spencer and Edezhath, Ralph and Hutchinson, Jeffrey and Luty, Markus",
    title = "{Effective WIMPs}",
    eprint = "1307.8120",
    archivePrefix = "arXiv",
    primaryClass = "hep-ph",
    doi = "10.1103/PhysRevD.89.015011",
    journal = "Phys. Rev. D",
    volume = "89",
    number = "1",
    pages = "015011",
    year = "2014"
}

@article{Arcadi:2024ukq,
    author = "Arcadi, Giorgio and Cabo-Almeida, David and Dutra, Ma{\'\i}ra and Ghosh, Pradipta and Lindner, Manfred and Mambrini, Yann and Neto, Jacinto P. and Pierre, Mathias and Profumo, Stefano and Queiroz, Farinaldo S.",
    title = "{The Waning of the WIMP: Endgame?}",
    eprint = "2403.15860",
    archivePrefix = "arXiv",
    primaryClass = "hep-ph",
    doi = "10.1140/epjc/s10052-024-13672-y",
    journal = "Eur. Phys. J. C",
    volume = "85",
    number = "2",
    pages = "152",
    year = "2025"
}

@article{Hisano:2018bpz,
    author = "Hisano, Junji and Nagai, Ryo and Nagata, Natsumi",
    title = "{Singlet Dirac Fermion Dark Matter with Mediators at Loop}",
    eprint = "1808.06301",
    archivePrefix = "arXiv",
    primaryClass = "hep-ph",
    reportNumber = "IPMU18-0085, TU-1071, UT-18-16",
    doi = "10.1007/JHEP12(2018)059",
    journal = "JHEP",
    volume = "12",
    pages = "059",
    year = "2018"
}

@article{Ibarra:2015fqa,
    author = "Ibarra, Alejandro and Wild, Sebastian",
    title = "{Dirac dark matter with a charged mediator: a comprehensive one-loop analysis of the direct detection phenomenology}",
    eprint = "1503.03382",
    archivePrefix = "arXiv",
    primaryClass = "hep-ph",
    doi = "10.1088/1475-7516/2015/05/047",
    journal = "JCAP",
    volume = "05",
    pages = "047",
    year = "2015"
}

@article{Arina:2025zpi,
    author = "Arina, Chiara and others",
    title = "{t-channel dark matter models {\textendash} a whitepaper}",
    eprint = "2504.10597",
    archivePrefix = "arXiv",
    primaryClass = "hep-ph",
    reportNumber = "CERN-LPCC-2025-001, IRMP-CP3-25-07, TTK-25-07",
    doi = "10.1140/epjc/s10052-025-14635-7",
    journal = "Eur. Phys. J. C",
    volume = "85",
    pages = "975",
    year = "2025",
    note = "[Erratum: Eur.Phys.J.C 85, 1105 (2025)]"
}

@article{Ibarra:2024mpq,
    author = "Ibarra, Alejandro and Reichard, Merlin and Tomar, Gaurav",
    title = "{Probing dark matter electromagnetic properties in direct detection experiments}",
    eprint = "2408.15760",
    archivePrefix = "arXiv",
    primaryClass = "hep-ph",
    doi = "10.1088/1475-7516/2025/02/072",
    journal = "JCAP",
    volume = "02",
    pages = "072",
    year = "2025"
}

@article{DAgnolo:2017dbv,
    author = "D'Agnolo, Raffaele Tito and Pappadopulo, Duccio and Ruderman, Joshua T.",
    title = "{Fourth Exception in the Calculation of Relic Abundances}",
    eprint = "1705.08450",
    archivePrefix = "arXiv",
    primaryClass = "hep-ph",
    doi = "10.1103/PhysRevLett.119.061102",
    journal = "Phys. Rev. Lett.",
    volume = "119",
    number = "6",
    pages = "061102",
    year = "2017"
}

@article{HeavyFlavorAveragingGroupHFLAV:2024ctg,
    author = "Banerjee, Swagato and others",
    collaboration = "Heavy Flavor Averaging Group (HFLAV)",
    title = "{Averages of $b$-hadron, $c$-hadron, and $\tau$-lepton properties as of 2023}",
    eprint = "2411.18639",
    archivePrefix = "arXiv",
    primaryClass = "hep-ex",
    month = "11",
    year = "2024"
}

@article{Kayser:1983wm,
    author = "Kayser, Boris and Goldhaber, Alfred S.",
    title = "{{CPT} and {CP} Properties of Majorana Particles, and the Consequences}",
    reportNumber = "Print-83-0630 (NSF)",
    doi = "10.1103/PhysRevD.28.2341",
    journal = "Phys. Rev. D",
    volume = "28",
    pages = "2341",
    year = "1983"
}

@article{Radescu:1985wf,
    author = "Radescu, E. E.",
    title = "{Comments on the Electromagnetic Properties of Majorana Fermions}",
    reportNumber = "JINR-E2-85-341",
    doi = "10.1103/PhysRevD.32.1266",
    journal = "Phys. Rev. D",
    volume = "32",
    pages = "1266",
    year = "1985"
}

@article{Batell:2011tc,
    author = "Batell, Brian and Pradler, Josef and Spannowsky, Michael",
    title = "{Dark Matter from Minimal Flavor Violation}",
    eprint = "1105.1781",
    archivePrefix = "arXiv",
    primaryClass = "hep-ph",
    doi = "10.1007/JHEP08(2011)038",
    journal = "JHEP",
    volume = "08",
    pages = "038",
    year = "2011"
}

@article{Kamenik:2011nb,
    author = "Kamenik, Jernej F. and Zupan, Jure",
    title = "{Discovering Dark Matter Through Flavor Violation at the LHC}",
    eprint = "1107.0623",
    archivePrefix = "arXiv",
    primaryClass = "hep-ph",
    doi = "10.1103/PhysRevD.84.111502",
    journal = "Phys. Rev. D",
    volume = "84",
    pages = "111502",
    year = "2011"
}

@article{Agrawal:2011ze,
    author = "Agrawal, Prateek and Blanchet, Steve and Chacko, Zackaria and Kilic, Can",
    title = "{Flavored Dark Matter, and Its Implications for Direct Detection and Colliders}",
    eprint = "1109.3516",
    archivePrefix = "arXiv",
    primaryClass = "hep-ph",
    reportNumber = "UMD-PP-011-014, RUNHETC-2011-17, UTTG-19-11, TCC-020-11",
    doi = "10.1103/PhysRevD.86.055002",
    journal = "Phys. Rev. D",
    volume = "86",
    pages = "055002",
    year = "2012"
}

@article{Straub:2018kue,
    author = "Straub, David M.",
    title = "{flavio: a Python package for flavour and precision phenomenology in the Standard Model and beyond}",
    eprint = "1810.08132",
    archivePrefix = "arXiv",
    primaryClass = "hep-ph",
    month = "10",
    year = "2018"
}

@article{Gaunt:2025peq,
    author = "Gaunt, Jonathan R. and Owen, Adam",
    title = "{FeynCraft: A Game of Feynman Diagrams}",
    eprint = "2510.14082",
    archivePrefix = "arXiv",
    primaryClass = "physics.ed-ph",
    month = "10",
    year = "2025"
}

@article{Gondolo:1990dk,
    author = "Gondolo, Paolo and Gelmini, Graciela",
    title = "{Cosmic abundances of stable particles: Improved analysis}",
    reportNumber = "UCLA-90-TEP-68",
    doi = "10.1016/0550-3213(91)90438-4",
    journal = "Nucl. Phys. B",
    volume = "360",
    pages = "145--179",
    year = "1991"
}

@article{Tsai:2019eqi,
    author = "Tsai, Yue-Lin Sming and Lu, Chih-Ting and Tran, Van Que",
    title = "{Confronting dark matter co-annihilation of Inert two Higgs Doublet Model with a compressed mass spectrum}",
    eprint = "1912.08875",
    archivePrefix = "arXiv",
    primaryClass = "hep-ph",
    doi = "10.1007/JHEP06(2020)033",
    journal = "JHEP",
    volume = "06",
    pages = "033",
    year = "2020"
}

@article{Garny:2015wea,
    author = "Garny, Mathias and Ibarra, Alejandro and Vogl, Stefan",
    title = "{Signatures of Majorana dark matter with t-channel mediators}",
    eprint = "1503.01500",
    archivePrefix = "arXiv",
    primaryClass = "hep-ph",
    reportNumber = "CERN-PH-TH-2015-036, TUM-HEP-985-15",
    doi = "10.1142/S0218271815300190",
    journal = "Int. J. Mod. Phys. D",
    volume = "24",
    number = "07",
    pages = "1530019",
    year = "2015"
}

@article{Biondini:2017ufr,
    author = "Biondini, S. and Laine, M.",
    title = "{Re-derived overclosure bound for the inert doublet model}",
    eprint = "1706.01894",
    archivePrefix = "arXiv",
    primaryClass = "hep-ph",
    doi = "10.1007/JHEP08(2017)047",
    journal = "JHEP",
    volume = "08",
    pages = "047",
    year = "2017"
}

@article{Biondini:2018pwp,
      author         = "Biondini, S. and Laine, M.",
      title          = "{Thermal dark matter co-annihilating with a strongly
                        interacting scalar}",
      journal        = "JHEP",
      volume         = "04",
      year           = "2018",
      pages          = "072",
      eprint         = "1801.05821",
      archivePrefix  = "arXiv",
      primaryClass   = "hep-ph",
      SLACcitation   = "%%CITATION = ARXIV:1801.05821;%%"
}

@article{Garny:2018ali,
    author = "Garny, Mathias and Heisig, Jan",
    title = "{Interplay of super-WIMP and freeze-in production of dark matter}",
    eprint = "1809.10135",
    archivePrefix = "arXiv",
    primaryClass = "hep-ph",
    reportNumber = "TUM-HEP 1166/18, TTK-18-39",
    doi = "10.1103/PhysRevD.98.095031",
    journal = "Phys. Rev. D",
    volume = "98",
    number = "9",
    pages = "095031",
    year = "2018"
}

@article{Becker:2022iso,
    author = "Becker, Mathias and Copello, Emanuele and Harz, Julia and Mohan, Kirtimaan A. and Sengupta, Dipan",
    title = "{Impact of Sommerfeld effect and bound state formation in simplified t-channel dark matter models}",
    eprint = "2203.04326",
    archivePrefix = "arXiv",
    primaryClass = "hep-ph",
    reportNumber = "ADP-22-6/T1177, MSUHEP-22-002, TUM-HEP-1387-22",
    doi = "10.1007/JHEP08(2022)145",
    journal = "JHEP",
    volume = "08",
    pages = "145",
    year = "2022"
}

@article{Hisano:2004ds,
    author = "Hisano, Junji and Matsumoto, Shigeki. and Nojiri, Mihoko M. and Saito, Osamu",
    title = "{Non-perturbative effect on dark matter annihilation and gamma ray signature from galactic center}",
    eprint = "hep-ph/0412403",
    archivePrefix = "arXiv",
    reportNumber = "ICRR-REPORT-513-2004-11, YITP-04-73",
    doi = "10.1103/PhysRevD.71.063528",
    journal = "Phys. Rev. D",
    volume = "71",
    pages = "063528",
    year = "2005"
}

@article{Iengo:2009ni,
    author = "Iengo, Roberto",
    title = "{Sommerfeld enhancement: General results from field theory diagrams}",
    eprint = "0902.0688",
    archivePrefix = "arXiv",
    primaryClass = "hep-ph",
    doi = "10.1088/1126-6708/2009/05/024",
    journal = "JHEP",
    volume = "05",
    pages = "024",
    year = "2009"
}

@article{Feng:2010zp,
    author = "Feng, Jonathan L. and Kaplinghat, Manoj and Yu, Hai-Bo",
    title = "{Sommerfeld Enhancements for Thermal Relic Dark Matter}",
    eprint = "1005.4678",
    archivePrefix = "arXiv",
    primaryClass = "hep-ph",
    reportNumber = "UCI-TR-2010-06",
    doi = "10.1103/PhysRevD.82.083525",
    journal = "Phys. Rev. D",
    volume = "82",
    pages = "083525",
    year = "2010"
}

@article{Detmold:2014qqa,
    author = "Detmold, William and McCullough, Matthew and Pochinsky, Andrew",
    title = "{Dark Nuclei I: Cosmology and Indirect Detection}",
    eprint = "1406.2276",
    archivePrefix = "arXiv",
    primaryClass = "hep-ph",
    reportNumber = "MIT-CTP-4554",
    doi = "10.1103/PhysRevD.90.115013",
    journal = "Phys. Rev. D",
    volume = "90",
    number = "11",
    pages = "115013",
    year = "2014"
}

@article{vonHarling:2014kha,
    author = "von Harling, Benedict and Petraki, Kalliopi",
    title = "{Bound-state formation for thermal relic dark matter and unitarity}",
    eprint = "1407.7874",
    archivePrefix = "arXiv",
    primaryClass = "hep-ph",
    reportNumber = "NIKHEF-2014-018",
    doi = "10.1088/1475-7516/2014/12/033",
    journal = "JCAP",
    volume = "12",
    pages = "033",
    year = "2014"
}

@article{Petraki:2015hla,
    author = "Petraki, Kalliopi and Postma, Marieke and Wiechers, Michael",
    title = "{Dark-matter bound states from Feynman diagrams}",
    eprint = "1505.00109",
    archivePrefix = "arXiv",
    primaryClass = "hep-ph",
    reportNumber = "NIKHEF-2015-013",
    doi = "10.1007/JHEP06(2015)128",
    journal = "JHEP",
    volume = "06",
    pages = "128",
    year = "2015"
}

@article{Biondini:2018ovz,
      author         = "Biondini, S. and Vogl, Stefan",
      title          = "{Coloured coannihilations: Dark matter phenomenology
                        meets non-relativistic EFTs}",
      journal        = "JHEP",
      volume         = "02",
      year           = "2019",
      pages          = "016",
      doi            = "10.1007/JHEP02(2019)016",
      eprint         = "1811.02581",
      archivePrefix  = "arXiv",
      primaryClass   = "hep-ph",
      SLACcitation   = "%%CITATION = ARXIV:1811.02581;%%"
}

@article{Biondini:2019int,
      author         = "Biondini, Simone and Vogl, Stefan",
      title          = "{Scalar dark matter coannihilating with a coloured
                        fermion}",
      journal        = "JHEP",
      volume         = "11",
      year           = "2019",
      pages          = "147",
      doi            = "10.1007/JHEP11(2019)147",
      eprint         = "1907.05766",
      archivePrefix  = "arXiv",
      primaryClass   = "hep-ph",
      SLACcitation   = "%%CITATION = ARXIV:1907.05766;%%"
}

@article{Bollig:2021psb,
    author = "Bollig, Julian and Vogl, Stefan",
    title = "{Impact of bound states on non-thermal dark matter production}",
    eprint = "2112.01491",
    archivePrefix = "arXiv",
    primaryClass = "hep-ph",
    doi = "10.1088/1475-7516/2022/10/031",
    journal = "JCAP",
    volume = "10",
    pages = "031",
    year = "2022"
}

@article{Biondini:2022ggt,
    author = "Biondini, Simone and Schicho, Philipp and Tenkanen, Tuomas V. I.",
    title = "{Strong electroweak phase transition in t-channel simplified dark matter models}",
    eprint = "2207.12207",
    archivePrefix = "arXiv",
    primaryClass = "hep-ph",
    reportNumber = "HIP-2022-19/TH, NORDITA 2022-050",
    doi = "10.1088/1475-7516/2022/10/044",
    journal = "JCAP",
    volume = "10",
    pages = "044",
    year = "2022"
}

@article{Planck:2018vyg,
    author = "Aghanim, N. and others",
    collaboration = "Planck",
    title = "{Planck 2018 results. VI. Cosmological parameters}",
    eprint = "1807.06209",
    archivePrefix = "arXiv",
    primaryClass = "astro-ph.CO",
    doi = "10.1051/0004-6361/201833910",
    journal = "Astron. Astrophys.",
    volume = "641",
    pages = "A6",
    year = "2020",
    note = "[Erratum: Astron.Astrophys. 652, C4 (2021)]"
}

@article{Acaroglu:2021qae,
    author = "Acaro{\u{g}}lu, Harun and Blanke, Monika",
    title = "{Tasting flavoured Majorana dark matter}",
    eprint = "2109.10357",
    archivePrefix = "arXiv",
    primaryClass = "hep-ph",
    reportNumber = "TTP21-031; P3H-21-065, TTP21-031, P3H-21-065",
    doi = "10.1007/JHEP05(2022)086",
    journal = "JHEP",
    volume = "05",
    pages = "086",
    year = "2022"
}

@article{Blanke:2017fum,
    author = "Blanke, Monika and Das, Satrajit and Kast, Simon",
    title = "{Flavoured Dark Matter Moving Left}",
    eprint = "1711.10493",
    archivePrefix = "arXiv",
    primaryClass = "hep-ph",
    reportNumber = "TTP17-049",
    doi = "10.1007/JHEP02(2018)105",
    journal = "JHEP",
    volume = "02",
    pages = "105",
    year = "2018"
}

@article{LZ:2024zvo,
    author = "Aalbers, J. and others",
    collaboration = "LZ",
    title = "{Dark Matter Search Results from 4.2{\,}{\,}Tonne-Years of Exposure of the LUX-ZEPLIN (LZ) Experiment}",
    eprint = "2410.17036",
    archivePrefix = "arXiv",
    primaryClass = "hep-ex",
    reportNumber = "FERMILAB-PUB-24-0796-V",
    doi = "10.1103/4dyc-z8zf",
    journal = "Phys. Rev. Lett.",
    volume = "135",
    number = "1",
    pages = "011802",
    year = "2025"
}

@article{Gondolo:2013wwa,
    author = "Gondolo, Paolo and Scopel, Stefano",
    title = "{On the sbottom resonance in dark matter scattering}",
    eprint = "1307.4481",
    archivePrefix = "arXiv",
    primaryClass = "hep-ph",
    reportNumber = "CETUP2013-008",
    doi = "10.1088/1475-7516/2013/10/032",
    journal = "JCAP",
    volume = "10",
    pages = "032",
    year = "2013"
}

@article{Olgoso:2025jot,
    author = "Olgoso, Pablo and Paradisi, Paride and Selimovic, Nudzeim",
    title = "{The Dark Side of a Tera-Z Factory}",
    eprint = "2507.17803",
    archivePrefix = "arXiv",
    primaryClass = "hep-ph",
    month = "7",
    year = "2025"
}

@article{Bramante:2016rdh,
    author = "Bramante, Joseph and Fox, Patrick J. and Kribs, Graham D. and Martin, Adam",
    title = "{Inelastic frontier: Discovering dark matter at high recoil energy}",
    eprint = "1608.02662",
    archivePrefix = "arXiv",
    primaryClass = "hep-ph",
    reportNumber = "FERMILAB-PUB-16-301-T",
    doi = "10.1103/PhysRevD.94.115026",
    journal = "Phys. Rev. D",
    volume = "94",
    number = "11",
    pages = "115026",
    year = "2016"
}

@article{Biondini:2025ihi,
    author = "Biondini, S. and Eriksson, M. and Laine, M.",
    title = "{Computing singlet scalar freeze-out with plasmon and plasmino states}",
    eprint = "2505.05206",
    archivePrefix = "arXiv",
    primaryClass = "hep-ph",
    doi = "10.1007/JHEP08(2025)197",
    journal = "JHEP",
    volume = "08",
    pages = "197",
    year = "2025"
}

@article{DeSimone:2014qkh,
    author = "De Simone, Andrea and Giudice, Gian Francesco and Strumia, Alessandro",
    title = "{Benchmarks for Dark Matter Searches at the LHC}",
    eprint = "1402.6287",
    archivePrefix = "arXiv",
    primaryClass = "hep-ph",
    reportNumber = "CERN-PH-TH-2014-008, SISSA-01-2014-FISI",
    doi = "10.1007/JHEP06(2014)081",
    journal = "JHEP",
    volume = "06",
    pages = "081",
    year = "2014"
}

@article{Ellis:2014ipa,
    author = "Ellis, John and Olive, Keith A. and Zheng, Jiaming",
    title = "{The Extent of the Stop Coannihilation Strip}",
    eprint = "1404.5571",
    archivePrefix = "arXiv",
    primaryClass = "hep-ph",
    reportNumber = "KCL-PH-TH-2014-17, LCTS-2014-16, CERN-PH-TH-2014-067, FTPI-MINN-14-11, UMN-TH-3333-14",
    doi = "10.1140/epjc/s10052-014-2947-7",
    journal = "Eur. Phys. J. C",
    volume = "74",
    pages = "2947",
    year = "2014"
}

@article{Ibarra:2015nca,
    author = "Ibarra, A. and Pierce, A. and Shah, N. R. and Vogl, S.",
    editor = "Tecchio, Monica and Levin, Daniel",
    title = "{Anatomy of Coannihilation with a Scalar Top Partner}",
    eprint = "1501.03164",
    archivePrefix = "arXiv",
    primaryClass = "hep-ph",
    reportNumber = "MCTP-15-04, TUM-HEP-974-15",
    doi = "10.1103/PhysRevD.91.095018",
    journal = "Phys. Rev. D",
    volume = "91",
    number = "9",
    pages = "095018",
    year = "2015"
}

@article{ElHedri:2017nny,
    author = "El Hedri, Sonia and Kaminska, Anna and de Vries, Maikel and Zurita, Jose",
    title = "{Simplified Phenomenology for Colored Dark Sectors}",
    eprint = "1703.00452",
    archivePrefix = "arXiv",
    primaryClass = "hep-ph",
    reportNumber = "MITP-17-002, TTP17-006",
    doi = "10.1007/JHEP04(2017)118",
    journal = "JHEP",
    volume = "04",
    pages = "118",
    year = "2017"
}

@article{Binder:2023ckj,
    author = "Binder, Tobias and Garny, Mathias and Heisig, Jan and Lederer, Stefan and Urban, Kai",
    title = "{Excited bound states and their role in dark matter production}",
    eprint = "2308.01336",
    archivePrefix = "arXiv",
    primaryClass = "hep-ph",
    reportNumber = "TUM-HEP 1469/23, TTK-23-21",
    doi = "10.1103/PhysRevD.108.095030",
    journal = "Phys. Rev. D",
    volume = "108",
    number = "9",
    pages = "095030",
    year = "2023"
}

@article{Garny:2021qsr,
    author = "Garny, Mathias and Heisig, Jan",
    title = "{Bound-state effects on dark matter coannihilation: Pushing the boundaries of conversion-driven freeze-out}",
    eprint = "2112.01499",
    archivePrefix = "arXiv",
    primaryClass = "hep-ph",
    reportNumber = "TUM-HEP 1379/21, TTK-21-52",
    doi = "10.1103/PhysRevD.105.055004",
    journal = "Phys. Rev. D",
    volume = "105",
    number = "5",
    pages = "055004",
    year = "2022"
}

@article{Harz:2018csl,
    author = "Harz, Julia and Petraki, Kalliopi",
    title = "{Radiative bound-state formation in unbroken perturbative non-Abelian theories and implications for dark matter}",
    eprint = "1805.01200",
    archivePrefix = "arXiv",
    primaryClass = "hep-ph",
    reportNumber = "Nikhef-2018-023",
    doi = "10.1007/JHEP07(2018)096",
    journal = "JHEP",
    volume = "07",
    pages = "096",
    year = "2018"
}

@article{Biondini:2023zcz,
    author = "Biondini, Simone and Brambilla, Nora and Qerimi, Gramos and Vairo, Antonio",
    title = "{Effective field theories for dark matter pairs in the early universe: cross sections and widths}",
    eprint = "2304.00113",
    archivePrefix = "arXiv",
    primaryClass = "hep-ph",
    doi = "10.1007/JHEP07(2023)006",
    journal = "JHEP",
    volume = "07",
    pages = "006",
    year = "2023"
}

@article{Binder:2019erp,
    author = "Binder, Tobias and Mukaida, Kyohei and Petraki, Kalliopi",
    title = "{Rapid bound-state formation of Dark Matter in the Early Universe}",
    eprint = "1910.11288",
    archivePrefix = "arXiv",
    primaryClass = "hep-ph",
    reportNumber = "DESY-19-181, IPMU19-0148",
    doi = "10.1103/PhysRevLett.124.161102",
    journal = "Phys. Rev. Lett.",
    volume = "124",
    number = "16",
    pages = "161102",
    year = "2020"
}

@article{Cirelli:2013hv,
    author = "Cirelli, Marco and Giesen, Gaelle",
    title = "{Antiprotons from Dark Matter: Current constraints and future sensitivities}",
    eprint = "1301.7079",
    archivePrefix = "arXiv",
    primaryClass = "hep-ph",
    doi = "10.1088/1475-7516/2013/04/015",
    journal = "JCAP",
    volume = "04",
    pages = "015",
    year = "2013"
}

@article{Bringmann:2012ez,
    author = "Bringmann, Torsten and Weniger, Christoph",
    title = "{Gamma Ray Signals from Dark Matter: Concepts, Status and Prospects}",
    eprint = "1208.5481",
    archivePrefix = "arXiv",
    primaryClass = "hep-ph",
    doi = "10.1016/j.dark.2012.10.005",
    journal = "Phys. Dark Univ.",
    volume = "1",
    pages = "194--217",
    year = "2012"
}

@article{McDaniel:2023bju,
    author = "McDaniel, Alex and Ajello, Marco and Karwin, Christopher M. and Di Mauro, Mattia and Drlica-Wagner, Alex and S{\'a}nchez-Conde, Miguel A.",
    title = "{Legacy analysis of dark matter annihilation from the Milky~Way dwarf spheroidal galaxies with 14~years of Fermi-LAT data}",
    eprint = "2311.04982",
    archivePrefix = "arXiv",
    primaryClass = "astro-ph.HE",
    reportNumber = "FERMILAB-PUB-23-686-PPD",
    doi = "10.1103/PhysRevD.109.063024",
    journal = "Phys. Rev. D",
    volume = "109",
    number = "6",
    pages = "063024",
    year = "2024"
}

@article{Garny:2018icg,
    author = {Garny, Mathias and Heisig, Jan and Hufnagel, Marco and L{\"u}lf, Benedikt},
    title = "{Top-philic dark matter within and beyond the WIMP paradigm}",
    eprint = "1802.00814",
    archivePrefix = "arXiv",
    primaryClass = "hep-ph",
    reportNumber = "TUM-HEP 1131/18, TTK-18-05, DESY-18-018, TUM-HEP-1131-18",
    doi = "10.1103/PhysRevD.97.075002",
    journal = "Phys. Rev. D",
    volume = "97",
    number = "7",
    pages = "075002",
    year = "2018"
}

@article{Biondini:2025jvp,
    author = "Biondini, Simone and Brambilla, Nora and Dashko, Andrii and Qerimi, Gramos and Vairo, Antonio",
    title = "{Effective field theories for dark matter pairs in the early universe: Debye mass effects}",
    eprint = "2501.03327",
    archivePrefix = "arXiv",
    primaryClass = "hep-ph",
    reportNumber = "DESY-24-154, TUM-EFT 179/23",
    doi = "10.1007/JHEP04(2025)091",
    journal = "JHEP",
    volume = "04",
    pages = "091",
    year = "2025"
}

@article{Arcadi:2017kky,
    author = "Arcadi, Giorgio and Dutra, Ma\'\i{}ra and Ghosh, Pradipta and Lindner, Manfred and Mambrini, Yann and Pierre, Mathias and Profumo, Stefano and Queiroz, Farinaldo S.",
    title = "{The waning of the WIMP? A review of models, searches, and constraints}",
    eprint = "1703.07364",
    archivePrefix = "arXiv",
    primaryClass = "hep-ph",
    doi = "10.1140/epjc/s10052-018-5662-y",
    journal = "Eur. Phys. J. C",
    volume = "78",
    number = "3",
    pages = "203",
    year = "2018"
}

@article{Schumann:2019eaa,
    author = "Schumann, Marc",
    title = "{Direct Detection of WIMP Dark Matter: Concepts and Status}",
    eprint = "1903.03026",
    archivePrefix = "arXiv",
    primaryClass = "astro-ph.CO",
    doi = "10.1088/1361-6471/ab2ea5",
    journal = "J. Phys. G",
    volume = "46",
    number = "10",
    pages = "103003",
    year = "2019"
}

@article{Greljo:2025mwj,
    author = "Greljo, Admir and Palavri{\'c}, Ajdin and Stefanek, Ben A.",
    title = "{Minimal Flavor Protection for TeV-scale New Physics}",
    eprint = "2512.04159",
    archivePrefix = "arXiv",
    primaryClass = "hep-ph",
    month = "12",
    year = "2025"
}

@article{Glashow:1970gm,
    author = "Glashow, S. L. and Iliopoulos, J. and Maiani, L.",
    title = "{Weak Interactions with Lepton-Hadron Symmetry}",
    doi = "10.1103/PhysRevD.2.1285",
    journal = "Phys. Rev. D",
    volume = "2",
    pages = "1285--1292",
    year = "1970"
}

@article{DARWIN:2016hyl,
    author = "Aalbers, J. and others",
    collaboration = "DARWIN",
    title = "{DARWIN: towards the ultimate dark matter detector}",
    eprint = "1606.07001",
    archivePrefix = "arXiv",
    primaryClass = "astro-ph.IM",
    doi = "10.1088/1475-7516/2016/11/017",
    journal = "JCAP",
    volume = "11",
    pages = "017",
    year = "2016"
}

@inproceedings{Alarcon:2022ero,
    author = "Alarcon, Ricardo and others",
    title = "{Electric dipole moments and the search for new physics}",
    booktitle = "{Snowmass 2021}",
    eprint = "2203.08103",
    archivePrefix = "arXiv",
    primaryClass = "hep-ph",
    month = "3",
    year = "2022"
}

@article{n2EDM:2021yah,
    author = "Ayres, N. J. and others",
    collaboration = "n2EDM",
    title = "{The design of the n2EDM experiment: nEDM Collaboration}",
    eprint = "2101.08730",
    archivePrefix = "arXiv",
    primaryClass = "physics.ins-det",
    doi = "10.1140/epjc/s10052-021-09298-z",
    journal = "Eur. Phys. J. C",
    volume = "81",
    number = "6",
    pages = "512",
    year = "2021"
}

@article{pEDM:2022ytu,
    author = "Alexander, Jim and others",
    collaboration = "pEDM",
    title = "{The storage ring proton EDM experiment}",
    eprint = "2205.00830",
    archivePrefix = "arXiv",
    primaryClass = "hep-ph",
    reportNumber = "FERMILAB-PUB-22-611-PPD",
    month = "4",
    year = "2022"
}

@article{Calibbi:2021pyh,
    author = "Calibbi, Lorenzo and Marcano, Xabier and Roy, Joydeep",
    title = "{Z lepton flavour violation as a probe for new physics at future $e^+e^-$ colliders}",
    eprint = "2107.10273",
    archivePrefix = "arXiv",
    primaryClass = "hep-ph",
    reportNumber = "TUM-HEP 1352/21",
    doi = "10.1140/epjc/s10052-021-09777-3",
    journal = "Eur. Phys. J. C",
    volume = "81",
    number = "12",
    pages = "1054",
    year = "2021"
}

@article{Mu2e:2014fns,
    author = "Bartoszek, L. and others",
    collaboration = "Mu2e",
    title = "{Mu2e Technical Design Report}",
    eprint = "1501.05241",
    archivePrefix = "arXiv",
    primaryClass = "physics.ins-det",
    reportNumber = "FERMILAB-TM-2594, FERMILAB-DESIGN-2014-01",
    doi = "10.2172/1172555",
    month = "10",
    year = "2014"
}

@article{COMET:2018auw,
    author = "Abramishvili, R. and others",
    collaboration = "COMET",
    title = "{COMET Phase-I Technical Design Report}",
    eprint = "1812.09018",
    archivePrefix = "arXiv",
    primaryClass = "physics.ins-det",
    doi = "10.1093/ptep/ptz125",
    journal = "PTEP",
    volume = "2020",
    number = "3",
    pages = "033C01",
    year = "2020"
}

@article{Achasov:2023gey,
    author = "Achasov, M. and others",
    title = "{STCF conceptual design report (Volume 1): Physics {\&} detector}",
    eprint = "2303.15790",
    archivePrefix = "arXiv",
    primaryClass = "hep-ex",
    doi = "10.1007/s11467-023-1333-z",
    journal = "Front. Phys. (Beijing)",
    volume = "19",
    number = "1",
    pages = "14701",
    year = "2024"
}

@article{FCC:2025lpp,
    author = "Benedikt, M. and others",
    collaboration = "FCC",
    title = "{Future Circular Collider Feasibility Study Report: Volume 1, Physics, Experiments, Detectors}",
    eprint = "2505.00272",
    archivePrefix = "arXiv",
    primaryClass = "hep-ex",
    reportNumber = "CERN-FCC-PHYS-2025-0002",
    month = "4",
    year = "2025"
}

@article{Davighi:2023evx,
    author = "Davighi, Joe and Stefanek, Ben A.",
    title = "{Deconstructed hypercharge: a natural model of flavour}",
    eprint = "2305.16280",
    archivePrefix = "arXiv",
    primaryClass = "hep-ph",
    reportNumber = "ZU-TH 24/23",
    doi = "10.1007/JHEP11(2023)100",
    journal = "JHEP",
    volume = "11",
    pages = "100",
    year = "2023"
}

@article{Abdallah:2015ter,
    author = "Abdallah, Jalal and others",
    title = "{Simplified Models for Dark Matter Searches at the LHC}",
    eprint = "1506.03116",
    archivePrefix = "arXiv",
    primaryClass = "hep-ph",
    reportNumber = "FERMILAB-PUB-15-283-CD, CERN-PH-TH-2015-139",
    doi = "10.1016/j.dark.2015.08.001",
    journal = "Phys. Dark Univ.",
    volume = "9-10",
    pages = "8--23",
    year = "2015"
}

@article{Greljo:2023bdy,
    author = "Greljo, Admir and Palavri{\'c}, Ajdin and Smolkovi{\v{c}}, Aleks",
    title = "{Leading directions in the SMEFT: Renormalization effects}",
    eprint = "2312.09179",
    archivePrefix = "arXiv",
    primaryClass = "hep-ph",
    doi = "10.1103/PhysRevD.109.075033",
    journal = "Phys. Rev. D",
    volume = "109",
    number = "7",
    pages = "075033",
    year = "2024"
}

\end{document}